\documentclass[twocolumn,superscriptaddress,floatfix,preprintnumbers,aps,nofootinbib,hyperref]{revtex4-2}
\usepackage[utf8]{inputenc}
\usepackage{epsf}
\usepackage{latexsym,amssymb,euscript}
\usepackage[dvips]{graphicx}
\usepackage{amsmath}
\usepackage{nicefrac}
\usepackage{slashed}
\usepackage{booktabs}
\usepackage[linktocpage]{hyperref}
\usepackage{braket}
\usepackage{chngcntr}
\usepackage{bbold}
\usepackage{graphics}
\usepackage{graphicx}
\usepackage{mciteplus}
\usepackage{caption}
\usepackage{subcaption}
\usepackage{adjustbox}
\usepackage{ragged2e}
\usepackage{placeins}
\usepackage[titletoc]{appendix}
\usepackage[normalem]{ulem}
\graphicspath{{./figures/}}
\hypersetup{
 linktocpage = false,
 urlcolor = urlblue,
 colorlinks = true,
 linkcolor = urlblue,
 anchorcolor = citegreen,
 citecolor = citegreen,
 pdfstartview = {XYZ null null 1.25} 
           }
\usepackage[left=2cm, right=2cm]{geometry}
\usepackage{pstricks}
\usepackage{color}
\usepackage{xcolor}
\definecolor{urlblue}{rgb}{0.2,0.4,0.7}
\definecolor{citegreen}{rgb}{0,0.4,0.2}
\definecolor{linkred}{rgb}{0.9,0.2,0.1}
\usepackage{float}
\usepackage{academicons}
\definecolor{orcidlogocol}{HTML}{A6CE39}
\usepackage{changepage}
\usepackage{fancyhdr}
\usepackage{epsfig}
\usepackage{letltxmacro}
\LetLtxMacro{\oldcite}{\cite}
\renewcommand{\cite}[1]{\mbox{\oldcite{#1}}}

\usepackage{orcidlink}

\newcommand{\drv}{{\rm d}}

\newcommand{\as}{\alpha_s}

\newcommand{\MSb}{\overline{\rm MS}}

\newcommand{\LL}{{\rm LL/LO}}

\newcommand{\NLLp}{{\rm NLL/NLO^+}}

\newcommand{\HENLOp}{{\rm HE}\mbox{-}{\rm NLO^+}}

\newcommand{\DY}{\Delta Y}

\newcommand{\F}{{\cal F}}

\newcommand{\Jpsi}{J/\psi}
\newcommand{\Yps}{\Upsilon}
\newcommand{\BCs}{B_c(^1S_0)}
\newcommand{\Bss}{B_c(^3S_1)}

\newcommand{\QXQq}{X_{Q\bar{Q}q\bar{q}}}

\newcommand{\TQQ}{T_{4Q}}
\newcommand{\TQc}{T_{4c}}
\newcommand{\TQcZpp}{T_{4c}(0^{++})}
\newcommand{\TQcOpm}{T_{4c}(1^{+-})}
\newcommand{\TQcTpp}{T_{4c}(2^{++})}
\newcommand{\TQb}{T_{4b}}
\newcommand{\TQbZpp}{T_{4b}(0^{++})}
\newcommand{\TQbOpm}{T_{4b}(1^{+-})}
\newcommand{\TQbTpp}{T_{4b}(2^{++})}

\newcommand{{\HFNRevo}}{\tt HF-NRevo}

\newcommand{{\Jethad}}{\tt JETHAD}
\newcommand{{\symJethad}}{\tt symJETHAD}
\newcommand{{\psymJethad}}{\tt (sym)JETHAD}
\newcommand{{\Hell}}{\tt HELL}
\newcommand{{\RadISH}}{\tt RadISH}
\newcommand{{\Pegasus}}{\tt QCD-PEGASUS}
\newcommand{{\HOPPET}}{\tt HOPPET}
\newcommand{{\QCDNUM}}{\tt QCDNUM}
\newcommand{{\APFEL}}{\tt APFEL}
\newcommand{{\APFELpp}}{\tt APFEL++}
\newcommand{{\APFELppp}}{\tt APFEL(++)}
\newcommand{{\EKO}}{\tt EKO}
\newcommand{{\FeynCalc}}{\tt FeynCalc}

\allowdisplaybreaks

\bibpunct{[}{]}{,}{n}{}{,}

\begin{document}

\title{Fragmentation of fully heavy tetraquarks: The TQ4Q1.1 functions as a case study}

\author{Francesco~Giovanni~Celiberto\,\orcidlink{0000-0003-3299-2203}} 
\email{francesco.celiberto@uah.es}
\affiliation{Universidad de Alcal\'a (UAH), Departamento de F\'isica y Matem\'aticas, E-28805 Alcal\'a de Henares, Madrid, Spain}


\begin{abstract}
We extend the study of exotic matter formation via the {\tt TQ4Q1.1} set of collinear, variable-flavor-number-scheme fragmentation functions for fully charmed or bottomed tetraquarks in three quantum configurations: scalar ($J^{PC} = 0^{++}$), axial vector ($J^{PC} = 1^{+-}$), and tensor ($J^{PC} = 2^{++}$).
We adopt single-parton fragmentation at leading power and implement a nonrelativistic Quantum Chromodynamics (NRQCD) factorization scheme tailored to tetraquark Fock-state configurations.
Short-distance inputs at the initial scale are modeled using updated calculations for both gluon- and heavy-quark-initiated channels.
A threshold-consistent Dokshitzer-Gribov-Lipatov-Altarelli-Parisi (DGLAP) evolution is then applied via the novel Heavy-flavor nonrelativistic-evolution ({\HFNRevo}) hybrid scheme.
We provide the first systematic treatment of uncertainties from nonperturbative color-composite long-distance matrix elements (LDMEs), as well as from perturbative hard-scattering (H-MHOUs) and fragmentation-scale inputs (F-MHOUs), assessed separately and in combination.
To support phenomenology, we compute NLL/NLO$^+$ cross sections for tetraquark-jet systems at the HL-LHC and FCC within the hybrid collinear and high-energy factorization (HyF) as implemented in {\psymJethad}, incorporating angular multiplicities as key observables sensitive to high-energy QCD dynamics. 
We also provide expected event yields based on realistic luminosity scenarios, offering a concrete benchmark for experimental searches.
This work connects the investigation of exotic hadrons with state-of-the-art precision QCD.
\end{abstract}

\maketitle

\setcounter{tocdepth}{3}
\begingroup
\renewcommand{\baselinestretch}{0.75}\footnotesize
\parskip 3.0pt
\tableofcontents
\renewcommand{\baselinestretch}{1.0}\normalsize
\endgroup

\parskip 6pt

\section{Introduction}
\label{sec:introduction}

Understanding how matter organizes itself beyond the conventional meson and baryon structure remains one of the central open challenges in hadron physics.
Exotic configurations such as tetraquarks and pentaquarks---genuine multiquark states---invite us to rethink the underlying mechanisms of color confinement and hadronization, due to their beyond-minimal valence content and possible internal complexity.

High-energy experiments at the Large Hadron Collider (LHC), and future projects like the Electron-Ion Collider (EIC)~\cite{AbdulKhalek:2021gbh,Khalek:2022bzd,Hentschinski:2022xnd,Amoroso:2022eow,Abir:2023fpo,Allaire:2023fgp} and the Future Circular Collider (FCC)~\cite{FCC:2025lpp,FCC:2025uan,FCC:2025jtd}, provide access to unprecedented kinematic regimes where such exotic hadrons can be produced and studied.
These machines offer a clean testing ground for probing multiquark formation, resolving their inner quark-gluon structure, and isolating the dominant production mechanisms.

Recent advances in QCD factorization and the development of all-order resummation techniques have expanded the theoretical reach of precision calculations for exotic-matter observables~\cite{Feng:2020riv,Feng:2020qee,Feng:2023agq,Feng:2023ghc,Bai:2024ezn,Bai:2024flh,Nejad:2021mmp,Celiberto:2023rzw,Celiberto:2024mab,Celiberto:2024beg,Celiberto:2025dfe,Celiberto:2025ipt}.
Accurate predictions of cross sections and differential distributions, validated against experimental measurements, are now possible, enabling a more refined understanding of the interplay between perturbative dynamics and nonperturbative hadron formation.

A comprehensive physics program that combines exotic spectroscopy with state-of-the-art QCD approaches could shed light on the effective degrees of freedom at play in strongly interacting systems and help reveal the organizing dynamics of multiquark states.
Such insights would enhance our grasp of QCD in the nonperturbative regime and contribute to clarifying the mechanisms that drive the formation of bound hadronic matter.

Exotic hadrons are generally classified into two categories: gluon-rich states, such as quark-gluon hybrids~\cite{Kou:2005gt,Braaten:2013boa,Berwein:2015vca} and glueballs~\cite{Minkowski:1998mf,Mathieu:2008me,Chen:2021cjr,Csorgo:2019ewn,D0:2020tig}, and multiquark states, such as tetraquarks, pentaquarks,  and hexaquarks~\cite{Gell-Mann:1964ewy,Jaffe:1976ig,Jaffe:1976ih,Jaffe:1976yi,Ader:1981db,Rosner:1985yh,Pepin:1998ih,Vijande:2011im,Esposito:2016noz,Lebed:2016hpi,Guo:2017jvc,Lucha:2017mof,Ali:2019roi}.
While hybrids and glueballs contain explicit gluonic degrees of freedom, compact multiquarks are modeled as leading Fock-state combinations of four or five quarks, respectively.

The observation of the $X(3872)$ particle by the Belle experiment at KEKB in 2003~\cite{Belle:2003nnu}, later confirmed by several Collaborations~\cite{CDF:2003cab,LHCb:2013kgk,CMS:2021znk,Swanson:2006st}, marked the beginning of what is often referred to as the ``Exotic-matter Revolution'' or Second Quarkonium Revolution''.
It followed the earlier ``Quarkonium Revolution'' of the 1970s, which had been ignited by the historic discovery of the first doubly charmed hadron, the $\Jpsi$ meson~\cite{SLAC-SP-017:1974ind,E598:1974sol,Bacci:1974za}.
The $X(3872)$, identified as a hidden-charm state with a $[c\bar{c}]$ component, is widely interpreted as the first experimentally observed candidate for a hidden-charm tetraquark~\cite{Chen:2016qju,Liu:2019zoy}.
This milestone was followed in 2021 by the LHCb detection of the $X(2900)$~\cite{LHCb:2020bls}, representing the first exotic hadron observed with open-charm flavor and thus expanding the landscape of exotic spectroscopy.

Despite exhibiting conventional quantum numbers, the $X(3872)$ features decay modes that explicitly violate isospin conservation.
Such anomalies point to a structural complexity that cannot be captured by the standard quarkonium picture, motivating a wealth of theoretical interpretations framed within the tetraquark paradigm.

Among the most explored frameworks is the compact diquark model, where the $X(3872)$ is envisioned as a tightly bound diquark-antidiquark configuration, distinct from the standard quark-antiquark composition of mesons~\cite{Maiani:2004vq,tHooft:2008rus,Maiani:2013nmn,Maiani:2014aja,Maiani:2017kyi,Mutuk:2021hmi,Wang:2013vex,Wang:2013exa,Grinstein:2024rcu}.
An alternative view is offered by the meson molecule scenario, which interprets the $X(3872)$ as a loosely bound system of two mesons held together by residual strong interactions, in analogy with molecular binding in atomic physics~\cite{Tornqvist:1993ng,Braaten:2003he,Guo:2013sya,Mutuk:2022ckn,Wang:2013daa,Wang:2014gwa,Esposito:2023mxw,Grinstein:2024rcu}.
A third possibility is the hadroquarkonium picture, where the $X(3872)$ is described as a compact quarkonium core surrounded by a light mesonic cloud, bearing similarities to a hadronic analog of an atomic system~\cite{Dubynskiy:2008mq,Voloshin:2013dpa,Guo:2017jvc,Ferretti:2018ojb,Ferretti:2018tco,Ferretti:2020ewe}.

These models aim to disentangle the subtleties of the $X(3872)$ inner dynamics, and to shed light on the broader class of exotic hadrons sharing similar features.
Further insight into its nature may be gleaned through phenomenological investigations in high-multiplicity proton collisions~\cite{Esposito:2020ywk}, as well as via theoretical analyses of its thermal behavior in a hadronic medium~\cite{Armesto:2024zad}.

In 2021, the LHCb Collaboration reported the first observation of a doubly charmed tetraquark, $T_{cc}^+$~\cite{LHCb:2021vvq,LHCb:2021auc}.
This state is interpreted as a near-threshold $|DD\rangle$ molecular configuration, and its dynamics have been theoretically investigated using the XEFT effective field theory framework~\cite{Fleming:2021wmk,Dai:2023mxm,Hodges:2024awq,Fleming:2007rp,Fleming:2008yn,Braaten:2010mg,Fleming:2011xa,Mehen:2015efa,Braaten:2020iye}.

Until recently, the $X(3872)$ was the only exotic hadron clearly observed in prompt proton-proton collisions.
This picture evolved with the detection of both the $T_{cc}^+$ and a broad resonance in the di-$\Jpsi$ invariant mass spectrum~\cite{LHCb:2020bwg}, now dubbed $X(6900)$ and believed to correspond either to the $0^{++}$ scalar state or, more likely, to the $2^{++}$ tensor excitation of a fully charmed $\TQc$ tetraquark~\cite{Chen:2022asf}.
A recent review by the LHCb Collaboration further highlights these developments and the prospects for future Run-3 searches and beyond~\cite{Nogga:2025qcm}.

The $X(6900)$ resonance has since garnered significant theoretical attention.
Some studies interpret the structure as a dynamically generated resonance stemming from Pomeron exchanges and coupled-channel effects~\cite{Gong:2020bmg}.
Additional analyses using Bethe-Salpeter equations and Regge trajectories propose that $X(6900)$ may correspond to radial or orbital excitations of a compact $|c \bar {c} c \bar{c} \rangle$ tetraquark, possibly accompanied by other fully heavy states in the spectrum.
Finally, recent proposals to decipher the spin-parity structure of $\TQc$ states through photon-photon fusion in ultraperipheral heavy-ion collisions offer promising experimental avenues~\cite{Niu:2022cug}.

The CMS Collaboration has recently advanced our understanding of fully charmed tetraquarks by performing the first spin-parity determination of the states observed in the $[\Jpsi]$ and $[\Jpsi,\psi(2S)]$ channels.
The initial observation of a family of three resonances in the $4m_c < M < 7\,{\rm GeV}$ region was reported in Refs.~\cite{CMS:2023owd} (see also Ref.~\cite{Zhu:2024swp}), and has now been complemented by a new angular analysis~\cite{CMS:2025fpt} favoring a $[J^{PC} = 2^{++}]$ assignment for the leading signal. 
This result disfavors loosely bound molecular interpretations and supports a compact diquark-antidiquark structure, where spin-1 diquarks naturally produce $[J = 2]$ configurations. 
Such spin constraints do not apply to mixed-flavor exotics such as $X(3872)$ or $Z_c(3900)$, which remain consistent with lower spin and molecular descriptions. 
The CMS findings thus provide crucial input for modeling the internal dynamics of fully tetraquarks.

Furthermore, related measurements of quarkonium pair production from ATLAS and CMS provide important empirical knowledge of multiple-parton interaction (MPI) dynamics in a deeper way. 
ATLAS measured prompt $\Jpsi$ pair production in proton-proton collisions at $\sqrt{s}=8$~TeV and extracted effective cross sections of DPS~\cite{ATLAS:2016ydt}. 
CMS conducted a similar study at $\sqrt{s}=7$~TeV, providing total and differential cross sections for double $\Jpsi$ production~\cite{CMS:2014cmt}. 
More recently, CMS reported the first observation of double $\Jpsi$ production in proton-lead collisions at $8.16$~TeV, delineating single-parton and DPS contributions across kinematic regimes~\cite{CMS:2024wgu}.

From a theoretical viewpoint, doubly and fully heavy tetraquarks, $\QXQq$ and $\TQQ$, represent particularly accessible systems for probing the strong force.
In $\QXQq$ states, nonrelativistic heavy quarks interact with light degrees of freedom, often via diquarklike substructures, enabling a clean test of QCD in a mixed-mass regime.
Similarly, $\TQQ$ hadrons consist solely of heavy quarks and antiquarks in a $|Q\bar{Q}Q\bar{Q}\rangle$ configuration, devoid of valence light quarks or active gluon fields, thus resembling a doubled, exotic quarkonium.\footnote{For analogous studies on fully stranged tetraquarks, the ``cousins'' of the $\TQQ$ states, see Refs.~\cite{Ding:2006ya,Ho:2019org,Su:2022eun,Liu:2020lpw,Dong:2022otb,Xi:2023byo,Ma:2024vsi}.}

Given the nonrelativistic dynamics of their constituents and the Fock-space structure, these exotic states can be analyzed using techniques developed for quarkonia.
While charmonia are often described as QCD analogs of hydrogen atoms~\cite{Pineda:2011dg}, $\QXQq$ and $\TQQ$ tetraquarks may be seen as QCD nuclei or molecules, depending on the modeling approach~\cite{Maiani:2019cwl}.

Despite major progress in understanding exotic hadrons' mass spectra and decay modes since the $X(3872)$ discovery, their production mechanisms remain poorly understood. Only a few model-dependent approaches, based on color evaporation~\cite{Maciula:2020wri} or hadron-quark duality~\cite{Berezhnoy:2011xy,Karliner:2016zzc,Becchi:2020mjz}, have been proposed so far.

Complementary efforts explored the role of multiparticle interactions in tetraquark production at hadron colliders~\cite{Carvalho:2015nqf,Abreu:2023wwg} and investigated potential signatures of high-energy dynamics in exotic formation~\cite{Cisek:2022uqx}. 
Additional studies focused on exclusive radiative decays of $\TQQ$ at bottom factories~\cite{Feng:2020qee} and on $\TQQ$ photoproduction in lepton-hadron collisions~\cite{Feng:2023ghc}.
A deep neural network approach to multiquark bound states was recently proposed~\cite{Wu:2025wvv}.

In the bottom sector, our knowledge is still limited. The BELLE Collaboration first reported two charged bottomoniumlike structures in $\Yps(5S)$ decays~\cite{Belle:2011aa}, hinting at exotic contributions. 
However, no $|b\bar{b}b\bar{b}\rangle$ or $|b\bar{b}q\bar{q}\rangle$ tetraquarks have yet been confirmed.

A recent observation by the ANDY Collaboration at RHIC of a resonance near 18.15~GeV in copper-gold collisions~\cite{ANDY:2019bfn} matches predictions for $\TQb$ states~\cite{Vogt:2021lei}. 
On the lattice side, bottom-charmed and doubly bottomed tetraquarks have been investigated in Refs.~\cite{Francis:2018jyb,Padmanath:2023rdu} and~\cite{Bicudo:2015vta,Leskovec:2019ioa,Alexandrou:2024iwi}, respectively.

The unexpectedly high cross sections for $X(3872)$ at large transverse momenta, reported by ATLAS, CMS, and LHCb~\cite{CMS:2013fpt,ATLAS:2016kwu,LHCb:2021ten}, point to a fragmentation-based production mechanism. These findings offer a key opportunity to refine theoretical descriptions rooted in high-energy QCD.

To this end, in Ref.~\cite{Celiberto:2024mab} we introduced the first-generation fragmentation functions (FFs) for fully charmed tetraquarks, the {\tt TQ4Q1.0} sets, focused on scalar ($J^{PC} = 0^{++}$) and tensor ($J^{PC} = 2^{++}$) fully charmed states.
These FFs incorporate initial-scale inputs from both gluon and charm fragmentation, respectively modeled using quark-potential nonrelativistic QCD (NRQCD)~\cite{Caswell:1985ui,Thacker:1990bm,Bodwin:1994jh,Cho:1995vh,Cho:1995ce,Leibovich:1996pa,Bodwin:2005hm} and spin-physics inspired prescriptions~\cite{Suzuki:1977km,Suzuki:1985up,Amiri:1986zv,Nejad:2021mmp}.

The evolution to energy scales covered by experiments is performed using the novel \emph{Heavy-flavor nonrelativistic-evolution} ({\HFNRevo}) method~\cite{Celiberto:2025euy,Celiberto:2024mex,Celiberto:2024bxu,Celiberto:2024rxa}, which enables a DGLAP-consistent treatment of multistep heavy-flavor thresholds, including gluon and charm activation.

The scalar and tensor channels provide valuable benchmarks in the study of fully heavy tetraquarks.
Recent experimental efforts, notably the CMS analysis~\cite{CMS:2023owd,CMS:2025fpt}, have favored a $2^{++}$ assignment for observed all-charm tetraquark peaks.
This supports the picture of a compact configuration with internal spin-1 diquark constituents, where the tensor channel emerges as the natural ground-state candidate due to symmetry and dynamical constraints.
Such findings reinforce the relevance of these channels as targets for phenomenological modeling and experimental comparison.

Subsequently, in Ref.~\cite{Celiberto:2024beg}, the FFs were extended to the {\tt TQ4Q1.1} sets, which represent the successors of the {\tt 1.0} versions and include updated initial-scale inputs derived from NRQCD for both gluon~\cite{Feng:2020riv} and charm~\cite{Bai:2024ezn} channels.
Remarkably, the {\tt TQ4Q1.1} framework includes both fully charmed and fully bottomed tetraquark states, thus generalizing the fragmentation picture across different heavy-flavor sectors.

A further step was achieved in Ref.~\cite{Celiberto:2025dfe}, with a dedicated study of axial-vector ($J^{PC} = 1^{+-}$) tetraquarks, also marking the first time that long-distance matrix-element (LDME) uncertainties were consistently propagated into the FFs.
These states are of particular interest in hadron physics, as they offer a relatively clean window into multiquark dynamics.
Although light $1^{+-}$ tetraquarks can undergo strong mixing with conventional mesons due to chiral symmetry breaking and meson-tetraquark couplings~\cite{Kim:2017yvd,Kim:2018zob}, fully heavy systems behave differently.
Heavy-quark spin symmetry (HQSS)~\cite{Isgur:1991wq,Neubert:1993mb} implies weak spin-dependent interactions among charm and bottom quarks, thus suppressing $[1^{++} \leftrightarrow 1^{+-}]$ mixing~\cite{Weng:2020jao,An:2022qpt}.
This leads to more rigid spin structures and nearly pure physical eigenstates in the heavy sector, allowing for precise theoretical control and a clean separation from molecularlike admixtures.

The axial-vector states are further distinguished by their production characteristics.
Due to Landau-Yang constraints~\cite{Landau:1948kw,Yang:1950rg} and $C$-parity selection rules, they are suppressed at leading order (LO) in gluon fusion, making them excellent probes of nonperturbative QCD effects~\cite{Karliner:2020dta,Becchi:2020mjz}.
Their selective dynamics and reduced mixing enhance their diagnostic value, especially when compared to scalar and tensor configurations, which may suffer from overlapping resonant interpretations.

Taken together, the scalar, tensor, and axial-vector components of the {\tt TQ4Q1.1} sets embody a unified, evolution-consistent framework for modeling the fragmentation of fully heavy tetraquarks.
This modular infrastructure, grounded in QCD and responsive to experimental input, is well-suited for future developments such as \emph{multimodal} modeling and comparative studies across flavor sectors.

In this work, we extend and complete the phenomenological study of collinear fragmentation to fully heavy tetraquarks, presenting a detailed examination of the {\tt TQ4Q1.1} family of functions.
This comprehensive update includes collinear FFs in a variable-flavor number scheme (VFNS)~\cite{Mele:1990cw,Cacciari:1993mq,Buza:1996wv} for both charmed and bottomed tetraquarks, across all three spin-parity configurations: scalar ($J^{PC} = 0^{++}$), axial-vector ($J^{PC} = 1^{+-}$), and tensor ($J^{PC} = 2^{++}$).
As mentioned, the construction of these FFs is grounded in the {\HFNRevo} methodology.

By embedding all spin channels within a unified formalism, the {\tt TQ4Q1.1} framework, we provide a systematic treatment of collinear fragmentation for compact four-quark bound states, allowing for controlled comparisons across quantum numbers and heavy-flavor sectors.

A novel aspect of this study is the implementation of a quantitative treatment of nonperturbative uncertainties.
For the first time, we propagate theoretical errors from LDMEs into the FFs themselves.
These LDMEs are defined on a color-composite basis, consistent with diquark-antidiquark clustering, and are extracted from potential NRQCD calculations for the scalar and tensor channels and from a model-averaged prescription in the axial-vector case.

By incorporating uncertainty bands, {\tt TQ4Q1.1} serves not only as a functional tool for high-energy predictions, but also as a diagnostic instrument to assess the model dependence of exotic hadron formation.
Designed as a case study within a broader strategy, this work lays the foundation for future \emph{multimodal} fragmentation frameworks, where perturbative and nonperturbative ingredients of different nature can be jointly explored and validated against upcoming experimental results.

In summary, the {\tt TQ4Q1.1} sets presented in this work generalize and supersede the first-generation {\tt TQ4Q1.0} FFs~\cite{Celiberto:2024mab}, previously limited to scalar and tensor $\TQc$ states and constructed via a hybrid modeling based on quark-potential NRQCD inputs (for gluon fragmentation) and a Suzuki-inspired kinematic ansatz~\cite{Suzuki:1977km} (for charm fragmentation). 
The {\tt 1.1} sets provide a complete NRQCD-based treatment of the short-distance coefficients, matched to updated LDMEs from potential models~\cite{Feng:2020riv,Bai:2024ezn}. 
They cover all three relevant spin-parity channels ($0^{++}$, $1^{+-}$, $2^{++}$), extend the analysis to both fully charmed and fully bottomed tetraquark states, and include analytic expressions for all partonic channels.

As a test case for our fragmentation framework, we examine the inclusive production of fully heavy tetraquarks accompanied by a jet in proton-proton collisions at energies relevant for the High-Luminosity LHC (HL-LHC) and FCC.
Our reference reaction is studied within a $\NLLp$ hybrid formalism (HyF) that combines next-to-leading order (NLO) collinear factorization with the resummation of high-energy logarithms beyond NLL accuracy.

This setup allows us to access observables that are particularly sensitive high-energy dynamics, including rapidity intervals and azimuthal-angle multiplicities, the latter being novelly introduced as discriminators of energy-flow patterns---resummed versus fixed order---in QCD radiation.
All predictions are obtained using the {\Jethad} numerical infrastructure supplemented by its symbolic extension {\symJethad}~\cite{Celiberto:2020wpk,Celiberto:2022rfj,Celiberto:2023fzz,Celiberto:2024mrq,Celiberto:2024swu}.

This work contributes to the broader effort of establishing a systematic, evolution-consistent framework for the fragmentation of fully heavy tetraquarks, bridging the flexible modeling strategies typical of hadron-structure studies with the most advanced tools from precision QCD.
By addressing multiple spin-parity channels and both charm and bottom flavors, it provides a unified reference for the entire family of compact four-quark states.
It sets the stage for future investigations and cross-validations with potential-model predictions, lattice QCD results, and heavy-flavor phenomenology at current and next-generation colliders.

The structure of this article is as follows.
In Sec.~\ref{sec:FFs}, we begin with a brief overview of the general features of heavy-flavor fragmentation, including how the NRQCD framework---originally developed for quarkonia---can be extended to the case of fully heavy tetraquarks.
We then present the technical foundation underlying the construction of the {\tt TQ4Q1.1} functions for both fully charmed and fully bottomed systems.
In Sec.~\ref{sec:phenomenology}, we discuss our phenomenological analysis for tetraquark-jet systems at HL-LHC and FCC energies and within the HyF framework.
In Sec.~\ref{sec:conclusions} we presents our conclusions and outlook.

\section{Fragmentation of fully heavy tetraquarks}
\label{sec:FFs}

In the first part of this section, we briefly outline the main features of heavy-flavor fragmentation, focusing on heavy-light hadrons, quarkonium states, and exotic bound systems (Sec.~\ref{ssec:FFs_intro}).
We then discuss the application of NRQCD to model initial-scale inputs for gluon and constituent heavy-quark fragmentation channels to $\TQQ$ tetraquarks (Secs.~\ref{ssec:FFs_NRQCD} and~\ref{ssec:FFs_initial_scale}).
Finally, in Sec.~\ref{ssec:FFs_TQ4Q11}, we study the timelike DGLAP evolution within the {\HFNRevo} framework of the complete {\tt TQ4Q1.1} family of FFs, as a case study for a modular and consistent approach to the fragmentation of fully heavy tetraquarks.

\subsection{Heavy-flavor fragmentation: Key concepts}
\label{ssec:FFs_intro}

Contrariwise to their light-flavor counterpart, heavy-flavor hadrons entail a fragmentation mechanism partially rooted in perturbative QCD. 
Unlike light hadrons, heavy-flavor hadronization involves additional complexity due to the large mass of heavy quarks, which places them within the perturbative QCD domain.
As a result, while FFs for light hadrons are entirely nonperturbative, those for heavy hadrons must include perturbative elements at the initial scale.

For singly heavy-flavored hadrons like $D$, $B$, or $\Lambda_{c,b}$, the fragmentation process can be modeled as a two-step mechanism~\cite{Cacciari:1996wr,Cacciari:1993mq,Jaffe:1993ie,Kniehl:2005mk,Helenius:2018uul,Helenius:2023wkn}. 
First, a highly energetic parton $i$ fragments into a heavy quark $Q$, a process calculable in perturbative QCD since the coupling at the heavy-quark mass scale is small. 
This short-distance coefficient (SDC), describing the $[i \to Q]$ subprocess, occurs over a shorter timescale than hadronization. 
The first NLO computations for such SDCs appeared in Refs.~\cite{Mele:1990yq,Mele:1990cw}, with further next-to-NLO developments in Refs.~\cite{Rijken:1996vr,Mitov:2006wy,Blumlein:2006rr,Melnikov:2004bm,Mitov:2004du,Biello:2024zti}.

The second, nonperturbative stage involves the hadronization of the heavy quark into a physical hadron, typically modeled through phenomenological functions~\cite{Kartvelishvili:1977pi,Bowler:1981sb,Peterson:1982ak,Andersson:1983jt,Collins:1984ms,Colangelo:1992kh} or effective theories~\cite{Georgi:1990um,Eichten:1989zv,Grinstein:1992ss,Neubert:1993mb,Jaffe:1993ie}.
To build a full VFNS FF set, one then evolves these initial inputs through DGLAP equations, using numerical methods to incorporate scaling violations at the desired perturbative accuracy.

This two-stage strategy also applies to quarkonium fragmentation. 
However, the presence of a $[Q\bar{Q}]$ pair in the leading Fock state makes the dynamics more involved. 
This is addressed within the framework of NRQCD~\cite{Caswell:1985ui,Thacker:1990bm,Bodwin:1994jh,Cho:1995vh,Cho:1995ce,Leibovich:1996pa,Bodwin:2005hm}, whose foundations and phenomenological implications are extensively reviewed in Refs.~\cite{Grinstein:1998xb,Kramer:2001hh,QuarkoniumWorkingGroup:2004kpm,Lansberg:2005aw,Lansberg:2019adr}.

In NRQCD, heavy quarks and antiquarks are treated as nonrelativistic degrees of freedom, enabling a factorization between SDCs, describing the perturbative production of a $[Q\bar{Q}]$ pair, and LDMEs, which model nonperturbative hadronization.
A physical quarkonium is expressed as a linear combination of Fock states, ordered via a double expansion in $\alpha_s$ and the relative velocity $v_{\cal Q}$ of the constituent quarks.

NRQCD accommodates both low and high transverse-momentum ($p_T$) production.
At low $p_T$ values, the dominant mechanism is the direct production of the $[Q\bar{Q}]$ pair in the hard scattering, followed by nonperturbative hadronization.
At higher $p_T$, fragmentation of a single parton into the hadron plus radiation becomes dominant.
While this single-parton process corresponds to a VFNS governed by DGLAP evolution, the low-$p_T$ short-distance channel aligns with a fixed-flavor number scheme (FFNS)~\cite{Alekhin:2009ni}, involving two-parton fragmentation and higher-power corrections~\cite{Fleming:2012wy,Kang:2014tta,Echevarria:2019ynx,Boer:2023zit,Celiberto:2024mex,Celiberto:2024bxu,Celiberto:2024rxa}.

Early LO calculations for gluon and heavy-quark fragmentation into $S$-wave vector quarkonia in color-singlet states were provided in Refs.~\cite{Braaten:1993rw,Braaten:1993mp}.
NLO order results became available later~\cite{Zheng:2019gnb,Zheng:2021sdo}, enabling the development of the first pioneering VFNS DGLAP-evolved FFs, known as the {\tt ZCW19$^+$} sets~\cite{Celiberto:2022dyf,Celiberto:2023fzz}.
This was extended to $\BCs$ and $\Bss$ mesons via the {\tt ZCFW22} sets~\cite{Celiberto:2022keu,Celiberto:2024omj}.

Predictions using {\tt ZCFW22} aligned with LHCb data~\cite{LHCb:2014iah,LHCb:2016qpe,Celiberto:2024omj}, confirming that $\BCs$ production stays below 0.1\% relative to singly bottomed $B$ mesons~\cite{Celiberto:2024omj}, thereby validating the use of the VFNS scheme at large $p_T$.

Recent studies suggest that the NRQCD factorization also applies to exotic states, such as double $\Jpsi$ signals observed at the LHC~\cite{LHCb:2020bwg,ATLAS:2023bft,CMS:2023owd}, interpreted as compact tetracharms~\cite{Zhang:2020hoh,Zhu:2020xni}.
Here, $\TQc$ formation proceeds from the short-distance production of two charm and two anticharm quarks, at a scale $\sim 1/m_c$.
As in the quarkonium case, fragmentation is modeled as a two-step convolution: a perturbative stage followed by nonperturbative hadronization.

The first NRQCD-based calculation of the initial input for gluon fragmentation into color-singlet $S$-wave $\TQc$ states was carried out in Ref.~\cite{Feng:2020riv}.
Building on that framework, the first VFNS FF sets for heavy-light tetraquarks, {\tt TQHL1.0}, where introduced in Ref.~\cite{Celiberto:2023rzw} (see also the review in~\cite{Celiberto:2024mrq}).
Reference~\cite{Celiberto:2024beg} expanded upon this effort by releasing the successor families {\tt TQ4Q1.1} and {\tt TQHL1.1}, featuring NRQCD-based modeling also for the $[Q \to \TQQ]$ input~\cite{Bai:2024ezn}, improved treatment of doubly-heavy tetraquark fragmentation, and inclusion of bottomoniumlike states.

Then, a prime dedicated study of axial-vector ($1^{+-}$) states was presented in Ref.~\cite{Celiberto:2025dfe}, also marking the first time that composite LDME uncertainties were propagated into the FFs.
As a first application, the {\tt TQ4Q1.1} FFs were used to investigate indirect detection of charmed tetraquarks through Higgs and electroweak decays~\cite{Ma:2025ryo}.

As a direct application of our exploration on exotic-matter production, FF determinations for fully charmed pentaquarks and rare $\Omega$ baryons were presented in Refs.~\cite{Celiberto:2025ipt} and~\cite{Celiberto:2025ogy}, respectively, leading to the release of the {\tt PQ5Q1.0} and {\tt OMG3Q1.0} sets.

\subsection{NRQCD from quarkonia to tetraquarks}
\label{ssec:FFs_NRQCD}

In this subsection, we outline the NRQCD-based approach to collinear fragmentation, detailing how it provides a systematic framework for describing the transition from a partonic state to the physical hadron.
To ensure clarity and consistency, we first recall the structure of NRQCD-based fragmentation in the case of conventional quarkonium.
This provides the necessary background for understanding the subsequent extension to exotic channels.

The derivation of FFs for tetraquark states follows the formalism presented in Refs.~\cite{Feng:2020riv,Bai:2024ezn}, which we adapt to match the specific scope of our study. We refer the reader to those references for technical details on the NRQCD-based calculations and the modeling of both short-distance and long-distance contributions.

According to NRQCD, the FF of a parton $i$ into a physical quarkonium state $\cal Q$ with momentum fraction $z$ at the initial energy scale $\mu_{F,0}$ is given by the following expression
\begin{equation}
\label{FFs_NRQCD_onium}
D_i^{\cal Q}(z, \mu_{F,0}) = \sum_{[n]} {\cal D}_i^{Q\bar{Q}}(z, [n]) \langle {\cal O}^{\cal Q}([n]) \rangle \;.
\end{equation}
Here, ${\cal D}_i^{Q\bar{Q}}(z, [n])$ is the perturbative SDC for the production of the intermediate $[Q \bar Q]$ pair, expandable in $\alpha_s$ and featuring DGLAP-type logarithms for resummation.
The factor $\langle {\cal O}^{\cal Q}([n]) \rangle$ represents the nonperturbative NRQCD LDME, containing suppression effects in the relative velocity $v_{\cal Q}$ of the $[Q \bar Q]$ system.
These LDMEs must be determined through global fits or estimated via lattice or potential models.
The $[n] \equiv \,^{2S+1}L_J^{(c)}$ label denotes the spectroscopic quantum numbers, with $(c)$ indicating the color state, singlet $(1)$ or octet $(8)$.

Equation~\eqref{FFs_NRQCD_onium} reflects the two core assumptions of NRQCD; first, the physical quarkonium is described by a linear superposition of all contributing Fock states labeled by $[n]$, and second, all contributions are organized via a double expansion in $\alpha_s$ and $v_{\cal Q}$.

A comparison between heavy-light-hadron and quarkonium fragmentation reveals both structural analogies and crucial differences.
For singly heavy-flavored hadrons $h_Q$, the $[i \to h_Q]$ FF at the initial scale involves a convolution between a perturbative FF for $[i \to h_Q]$ and a nonperturbative transition function that encodes the probability for $Q$ to hadronize into $h_Q$~\cite{Cacciari:1996wr,Cacciari:1993mq,Jaffe:1993ie}.
This convolution reflects the dynamical sharing of momentum with soft constituents, and leads to an explicit $z$-dependence in the hadronization model.

In contrast, NRQCD expresses the $[i \to {\cal Q}]$ FF as a sum over Fock states $[n]$, each term being a product of a perturbative SDC and a constant LDME.
Since the heavy constituents of the quarkonium originate from the short-distance process itself, hadronization does not involve additional momentum redistribution.
As a result, LDMEs are $z$-independent and represent fixed probabilities for the transition of a $[Q\bar Q]$ pair into the physical state $\cal Q$.
This fundamental difference explains why FFs for open-flavor hadrons require phenomenological $z$-dependent models, while NRQCD-based FFs are built from linear combinations of perturbative functions weighted by scalar LDMEs.

Although LDMEs are not known from first principles, they can be constrained by experimental data or estimated through nonperturbative approaches such as lattice QCD or potential models. However, specific assumptions may simplify their structure.

A notable assumption is the so-called \emph{vacuum saturation approximation} (VSA)~\cite{Shifman:1978bx,Gilman:1979bc,Bodwin:1994jh}, which posits that intermediate states other than the vacuum are suppressed by powers of the relative velocity $v_{\cal Q}$ of the heavy quarks. Under this approximation, the LDME reduces to the product of matrix elements between the vacuum and the lowest Fock state $|Q\bar Q\rangle$. 
Symbolically, the VSA reads
\begin{equation}
\label{VSA}
\hspace{-0.45cm}
\langle 0 | \chi^\dagger \Pi_n \psi \, \mathcal{P}_{\cal Q} \, \psi^\dagger \Pi_n^\prime \chi | 0 \rangle
\simeq
\langle 0 | \chi^\dagger \Pi_n \psi | {\cal Q} \rangle \langle {\cal Q} | \psi^\dagger \Pi_n^\prime \chi | 0 \rangle \;
\end{equation}
Here, $\Pi_n$ and $\Pi_n^\prime$ are spin-color projectors for the relevant NRQCD state, while $\mathcal{P}_{\cal Q}$ denotes the projector onto the physical quarkonium ${\cal Q}$.
The fields $\chi$ and $\psi$ represent nonrelativistic Pauli spinors that annihilate a heavy antiquark and create a heavy quark, respectively, within the NRQCD framework.

In analogy with Eq.~\eqref{FFs_NRQCD_onium}, which describes the fragmentation of a parton $i$ into a quarkonium $\cal Q$ within the NRQCD factorization framework, the initial-scale FF of a parton $i$ into a fully heavy tetraquark $\TQQ$ can also be expressed as a weighted sum over different Fock states.
In this extended scenario, the nonperturbative formation of the physical tetraquark state is described by color-composite LDMEs that encode the probability of hadronization from a perturbatively produced four-quark configuration.
The general structure of the initial-scale FF reads
\begin{equation}
\label{FFs_TQc_general}
D_i^{\TQQ}(z, \mu_{F,0}) = \sum_{[n]} {\cal D}_i^{4Q}(z,[n]) \langle {\cal O}^{\TQQ}([n]) \rangle \;,
\end{equation}
where the role of the ${\cal D}_i^{4Q}(z,[n])$ SDCs is to describe the production of a compact $|Q\bar{Q}Q\bar{Q}\rangle$ configuration in a definite spin, color, and orbital state $[n]$, and $\langle {\cal O}^{\TQQ}([n]) \rangle$ are the color-composite LDMEs encoding the nonperturbative hadronization into the physical tetraquark.

The key distinction with respect to the quarkonium case lies in the complexity of the intermediate state.
Instead of a two-body heavy-quark pair, the tetraquark system originates from a four-body configuration, subject to additional spin-color couplings and internal symmetries.
This richer structure implies a larger number of contributing Fock states and, consequently, a more intricate pattern of LDMEs, reflecting the possible internal color-spin arrangements compatible with a color-singlet final state.

Moreover, while the NRQCD framework continues to apply thanks to the $m_Q \gg \Lambda_{\rm QCD}$ hierarchy, the reduced binding of the four-heavy-quark system and the possible coexistence of tightly and loosely bound configurations may influence the convergence of the velocity expansion and the size of subleading terms.
Nevertheless, for low-lying $S$-wave compact tetraquarks, one expects the factorized structure of Eq.~\eqref{FFs_TQc_general} to remain a valid starting point for constructing perturbative inputs at the initial energy scale, and for matching the evolution to high energies via DGLAP equations.

Considering a fully heavy tetraquark $\TQQ(J^{PC})$, with total angular momentum, parity, and charge $J^{PC} = 0^{++}$, $1^{+-}$, or $2^{++}$, and retaining only the lowest-order terms in the velocity expansion, we can recast Eq.~\eqref{FFs_TQc_general} as follows
\begin{equation}
\begin{split}
 \label{TQQ_FF_initial-scale}
 D^{\TQQ(J^{PC})}_i(z,\mu_{F,0}) \, &= \,
 \frac{1}{m_Q^9}
 \sum_{[n]} 
 \tilde{\cal D}^{(J^{PC})}_i(z,[n]) \\
 &\times \, \langle {\cal O}^{\TQQ(J^{PC})}([n]) \rangle
 \;,
\end{split}
\end{equation}
with $m_Q = m_c = 1.5$~GeV ($m_Q = m_b = 4.9$~GeV) being the charm (bottom) mass.
Furthermore, the composite quantum number $[n]$ runs over the combinations $[3,3]$, $[6,6]$, $[3,6]$, and $[6,3]$.
Throughout this work, we define the SDCs as dimensionless quantities, given by
\begin{equation}
\begin{split}
 \label{TQQ_FF_SDCs_tilde}
 \tilde{\cal D}^{(J^{PC})}_i(z,[n]) \rangle
 \, \equiv \,
 m_Q^9 \,
 {\cal D}^{(J^{PC})}_i(z,[n])
\;.
\end{split}
\end{equation}
Finally, the following symmetry relations hold
\begin{equation}
\begin{split}
 \label{TQQ_FF_initial-scale_symmetry}
 \tilde{\cal D}^{(J^{PC})}_i(z,[3,6]) \, &= \, \tilde{\cal D}^{(J^{PC})}_i(z,[6,3]) \;,
 \\
 \langle {\cal O}^{\TQc(J^{PC})}([3,6]) \rangle \, &= \, \langle {\cal O}^{\TQQ(J^{PC})}([6,3]) \rangle^*
 \;.
\end{split}
\end{equation}

The fragmentation formalism adopted here is expected to dominate in the high-transverse-momentum region, typically defined by the condition $p_T \gtrsim 3 M_{\TQQ}$, where $M_{\TQQ}$ denotes the mass of the produced tetraquark state. 
This threshold is commonly adopted in fragmentation-based studies as a practical criterion for the onset of the collinear regime, where logarithmic terms $\ln(p_T^2/M_{\TQQ}^2)$ become sizable, power corrections of order $\mathcal{O}(M_{\TQQ}^2/p_T^2)$ are suppressed, and nonfragmentation mechanisms are subleading~\cite{Cacciari:1995yt,Artoisenet:2007xi,Ma:2014svb}.

Early studies on heavy quarkonium production~\cite{Cacciari:1994dr,Cacciari:1995yt,Cacciari:1996dg} indicate that gluon fragmentation becomes dominant for $p_T \gtrsim 10 \div 15$ GeV in the case of $\Jpsi$. 
More recent analyses on charmed $B$ mesons~\cite{Kolodziej:1995nv,Artoisenet:2007xi} suggest even higher thresholds, up to 80~GeV. 
In analogy, for tetraquarks with masses $M_{\TQc} \sim 4 m_c \simeq 6.5$ GeV and $M_{\TQb} \sim 4 m_b \simeq 18$ GeV, we conservatively identify the fragmentation-dominated regime as $p_T \gtrsim 20$ GeV and $p_T \gtrsim 50$ GeV, respectively. 
These estimates define the kinematic domain in which the FF sets introduced in this work are expected to be phenomenologically reliable.

An important competing mechanism to fragmentation at moderate and low transverse momentum is the contribution from MPIs, particularly double-parton scattering (DPS). 
This process---where two independent hard scatterings occur in a single proton-proton collision---has been identified as an active channel for producing multiple heavy-quark pairs.

DPS has been extensively studied in a variety of channels, including double-jet production~\cite{Ducloue:2015jba}, where it affects rapidity distributions at low $p_T$, and in multiquarkonium final states such as $[\Jpsi{+}\Jpsi]$~\cite{Lansberg:2014swa,Lansberg:2020rft}, $[\Jpsi{+}\Yps]$~\cite{Lansberg:2020rft}, $[\Jpsi{+}Z]$~\cite{Lansberg:2016rcx}, and $[\Jpsi{+}W]$~\cite{Lansberg:2017chq}, as well as triple-$\Jpsi$ production~\cite{dEnterria:2016ids,Shao:2019qob}. These studies confirm the phenomenological relevance of DPS at low to moderate $p_T$ and motivate its consideration in the context of exotic heavy-flavor final states, including fully heavy tetraquarks.

In this context, early studies have modeled tetraquark formation via a double color-evaporation approach and found cross sections that grow rapidly with collision energy in both proton-proton and proton-ion collisions, with potential enhancements in proton-ion of order the nuclear mass number $A$ or larger~\cite{Carvalho:2015nqf,Abreu:2023wwg}. 
Similarly, single-parton and DPS mechanisms for the production of $|c\bar{c}c\bar{c}\rangle$ states---relevant to $T_{4c}$ formation---have been compared, showing that DPS can exceed single-parton contributions by more than one order of magnitude, depending on the kinematics~\cite{Maciula:2020wri}.

Such evidence indicates that, at low to moderate $p_T$, DPS can represent a competitive or even dominant source for tetraquark production. 
However, it is generally expected that DPS contributions decrease faster with increasing $p_T$ and lack the characteristic logarithmic enhancements that make collinear fragmentation dominant at high $p_T$. 
Given the current focus on the high-$p_T$ regime where $p_T \gtrsim 3 M_{\TQQ}$, fragmentation remains the leading theoretical mechanism in our framework.
DPS effects are therefore expected to be subdominant in the observables considered here and can be safely neglected at leading power, though they should be incorporated in future studies targeting lower-$p_T$ regions.

\subsection{Initial-scale FF inputs}
\label{ssec:FFs_initial_scale}

\begin{figure*}[!t]
\centering
\includegraphics[width=0.475\textwidth]{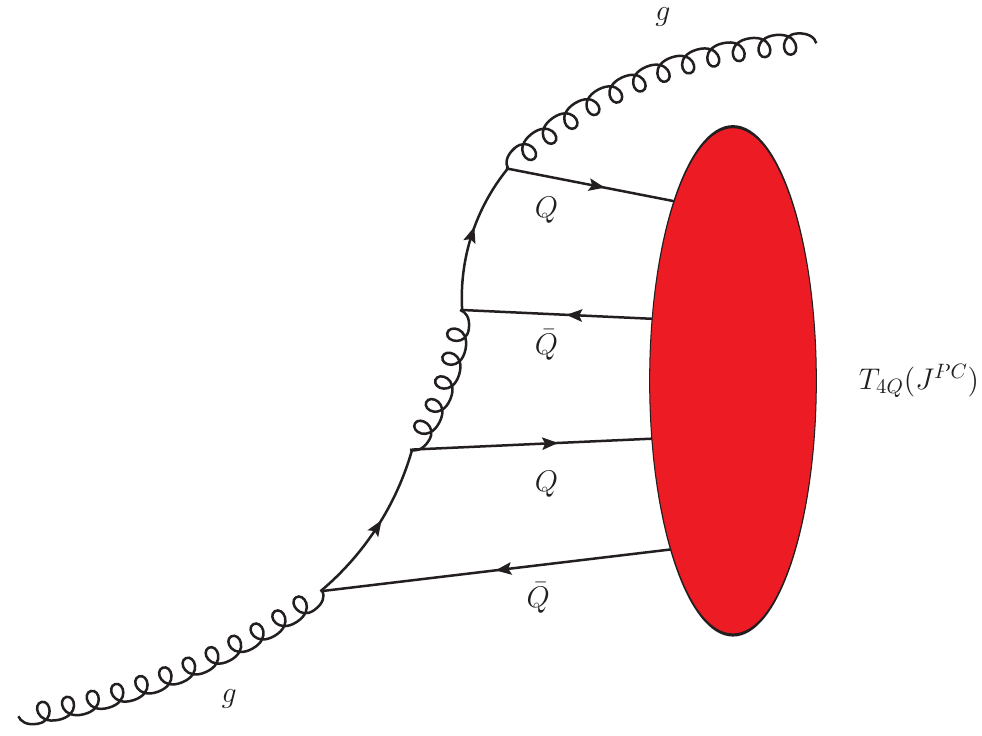}
\hspace{0.40cm}
\includegraphics[width=0.475\textwidth]{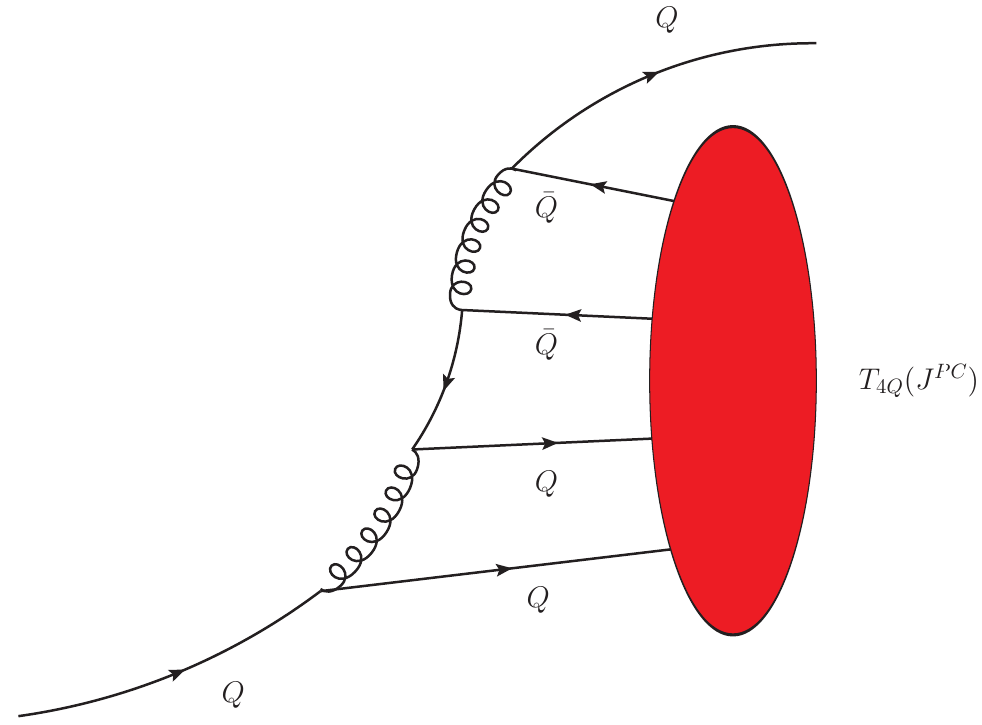}

\caption{
\justifying
\noindent
LO diagrams illustrating the collinear fragmentation of a gluon (left) and a heavy quark (right) into a fully heavy $S$-wave tetraquark in a color-singlet configuration.
Parton interactions on the left-hand side of each diagram depict the perturbative short-distance coefficients (SDCs), while red blobs on the right-hand side portray the nonperturbative long-distance matrix elements (LDMEs) describing the final hadronization step.
}
\label{fig:FF_diagrams}
\end{figure*}

For the sake of clarity, this section is divided into four parts: we first introduce the dimensionless SDCs, then present the general strategy to construct the color-composite LDMEs, discuss the role of color-octet mechanisms, and finally specify the LDME values and their associated uncertainties.

\vspace{1em}
\noindent
\textbf{Dimensionless SDCs.} 
To calculate the SDCs entering Eq.~\eqref{TQQ_FF_initial-scale}, Authors of Refs.~\cite{Feng:2020riv,Bai:2024ezn} followed the standard perturbative matching procedure between pure QCD and NRQCD matrix elements.
Within the standard NRQCD approach, the SCDs are determined by matching perturbative QCD and NRQCD matrix elements computed with fictitious, free multiquark states that carry the same quantum numbers as the physical quarkonium~\cite{Bodwin:1994jh,Petrelli:1997ge}.
At present, analytic calculations for the initial-scale FF inputs of $\TQQ$ states are available only at LO in both $\alpha_s$ and the relative velocity $v_{\cal Q}$. 

As in the quarkonium case, SDCs for $\TQQ$ states are insensitive to the long-distance dynamics of hadronization, and can thus be computed by replacing the physical tetraquark state $\TQQ$ with an artificial, free four-quark configuration $|[QQ][\bar{Q}\bar{Q}]\rangle$ carrying the same quantum numbers.

The matching is performed at LO in $\alpha_s$ and in the relative velocity $v_{\cal Q}$, retaining only the lowest-order NRQCD operators and neglecting derivative contributions.
The tetraquark is modeled as a diquark-antidiquark configuration, and the projection onto definite spin, parity, and color is achieved through the construction of appropriate NRQCD four-quark operators.\footnote{The diquark-antidiquark configuration simplifies the NRQCD matching by reducing the four-body problem to two-body building blocks, making the projection onto spin, parity, and color quantum numbers more tractable. References~\cite{Feng:2020riv,Bai:2024ezn} adopt this approach to construct gauge-invariant operators and extract the corresponding SDCs.}
These operators include both $[3 \otimes \bar{3}]$ and $[6 \otimes \bar{6}]$ color structures. 

However, the allowed quantum numbers depend on the interplay between Fermi-Dirac statistics and the $S$-wave orbital structure of the system. 
Specifically, for scalar ($0^{++}$) and tensor ($2^{++}$) channels, both color configurations are in principle allowed, with the $[6 \otimes \bar{6}]$ component contributing only to spin-0 states.
In contrast, for the axial-vector state ($1^{+-}$), Fermi-Dirac constraints eliminate the $[6 \otimes \bar{6}]$ configuration, leaving only the $[\bar{3} \otimes 3]$ contribution. 
Additionally, in the axial-vector case, the gluon-initiated channel is forbidden at LO due to the Landau-Yang theorem, which prohibits the coupling of a single on-shell gluon to a $1^{+-}$ state. 
As a result, only the constituent heavy-quark fragmentation channel is retained in this case.

The analytic expressions for the dimensionless SDCs of our tetraquarks are reported in Appendix~\hyperlink{app:A}{A}.
The gluon and heavy-quark NRQCD channels, first calculated in Refs.~\cite{Feng:2020riv} and~\cite{Bai:2024ezn}, and later independently rederived using the {\psymJethad}~\cite{Celiberto:2020wpk,Celiberto:2022rfj,Celiberto:2023fzz,Celiberto:2024mrq,Celiberto:2024swu} symbolic engine, are valid at LO accuracy and for color-singlet configurations (see Fig.~\ref{fig:FF_diagrams}).

\vspace{1em}
\noindent
\textbf{Building color-composite LDMEs.} 
As mentioned above, our approach provides analytic expressions for the SDCs at the initial energy scale, which are then matched to the corresponding NRQCD LDMEs.
These matrix elements encode the nonperturbative transition of the $| [QQ][\bar{Q}\bar{Q}] \rangle$ configuration into the physical tetraquark state and must be either extracted from experimental data or modeled through phenomenological assumptions.

In analogy with the quarkonium case, a commonly adopted simplification is provided by the VSA, which reduces the LDMEs to the product of matrix elements between the vacuum and the lowest Fock component of the physical state.
For tetraquarks, this structure is more intricate due to the presence of two color-connected diquark and antidiquark subsystems.
Following Refs.~\cite{Feng:2020riv,Bai:2024ezn}, we adopt the so-called color-composite LDMEs, where the physical tetraquark state is expanded in a diquark-antidiquark basis and projected onto definite color-singlet operators.

This decomposition enables the classification of the long-distance matrix elements according to the color configuration of the constituent clusters, namely $[3,3]$, $[6,6]$, and the mixed $[3,6]$/$[6,3]$ components, as previously introduced in Eq.~\eqref{TQQ_FF_initial-scale}.
These LDMEs describe the probability amplitude for hadronizing a perturbatively produced four-quark state into the physical bound state, and must be specified for each spin and color configuration.
For completeness, here we report the color composite operators associated with the $\TQQ$ LDMEs, and refer the reader to Sec.~IV of Ref.~\cite{Feng:2020riv} for technical details.
We write
\begin{align}
\label{TQQ_FF_LDMEs_operators}
{\cal O}^{(0)}_{\bar{3} \otimes 3} &= - \frac{1}{\sqrt{3}} 
\left[
\psi^T_a (i \sigma^2) \sigma^i \psi_b \right]
\left[
\chi^\dagger_c \sigma^i (i \sigma^2) \chi^\ast_d 
\right]
{\cal C}^{ab;cd}_{\bar{3} \otimes 3} \;,
\nonumber \\[0.20cm]
{\cal O}^{\alpha\beta;(2)}_{\bar{3} \otimes 3} &= 
\left[
\psi^T_a (i \sigma^2) \sigma^m \psi_b \right]
\left[
\chi^\dagger_c \sigma^n (i \sigma^2) \chi^\ast_d 
\right]
\Gamma^{\alpha\beta;mn}
{\cal C}^{ab;cd}_{\bar{3} \otimes 3} \;,
\nonumber \\[0.20cm]
{\cal O}^{(0)}_{6 \otimes \bar{6}} &= 
\left[
\psi^T_a (i \sigma^2) \psi_b \right]
\left[
\chi^\dagger_c (i \sigma^2) \chi^\ast_d 
\right]
{\cal C}^{ab;cd}_{6 \otimes \bar{6}} \;,
\end{align}
with $\sigma^2$ standing for the second Pauli matrix, and $\psi$ and $\chi$ representing the standard NRQCD fields, as in Eq.~\eqref{VSA}.
The rank-4 Lorentz tensor is defined as
\begin{equation}
\label{Gamma_klmn_rank4}
 \Gamma^{kl;mn} \equiv \frac{1}{2} \left( \delta^{km} \delta^{ln} + \delta^{kn} \delta^{lm} - \frac{2}{3} \delta^{kl} \delta^{mn} \right) \,,
\end{equation}
and the rank-4 color tensors take the form
\begin{equation}
\begin{split}
\label{C_abcd}
 C_{\bar{3} \otimes 3}^{ab;cd} &= \frac{1}{(\sqrt{2})^2} \, \epsilon^{abm} \epsilon^{cdn} \, \frac{\delta^{mn}}{\sqrt{N_c}} \\
 &=\, \frac{1}{2\sqrt{3}} \left( \delta^{ac} \delta^{bd} - \delta^{ad} \delta^{bc} \right) \;, \\[0.20cm]
 C_{6 \otimes \bar{6}}^{ab;cd} &= \frac{1}{2\sqrt{6}} \left( \delta^{ac} \delta^{bd} + \delta^{ad} \delta^{bc} \right) \;,
\end{split}
\end{equation}
where we have introduced the rank-3 Levi-Civita tensors $\epsilon^{ijk}$ and the Kronecker delta $\delta^{mn}$ to construct color-singlet combinations.

We note that the adopted diquark-antidiquark basis is formally equivalent, via Fierz rearrangements, to a molecularlike decomposition with color structure $[1 \otimes 1]$ or $[8 \otimes 8]$.
However, this choice is not meant to imply a literal bound-state interpretation of the diquark-antidiquark configuration.
Rather, it reflects the structural assumptions of our proxy model, where the tetraquark formation is driven by dominant color and spin correlations between tightly coupled diquark and antidiquark pairs.
Such a representation is fully consistent with the nature of a fragmentation production inherently based on the direct formation of compact, multiquark bound states.

\vspace{1em}
\noindent
\textbf{On the role of color-octet mechanisms.}
The inclusion of color-octet channels in the NRQCD modeling of initial-scale FFs remains a challenging task, both conceptually and phenomenologically. 
In the present work, following the strategy adopted in Refs.~\cite{Feng:2020riv,Bai:2024ezn}, we restrict our analysis to color-singlet configurations only. 
This choice reflects the current status of reliable input for the LDMEs associated with color-octet channels, which are notoriously difficult to constrain. 
As discussed in those studies, a consistent treatment of such contributions will require dedicated efforts to obtain trustworthy estimates of octet LDMEs, possibly through lattice simulations or advanced potential models. 
Until such inputs become available, any attempt to include color-octet terms would rely on uncontrolled modeling assumptions.

However, it is instructive to assess the potential relevance of color-octet mechanisms by drawing an analogy with the well-known ``quarkonium production puzzle'' in (semi-)inclusive reactions (see Refs.~\cite{Brambilla:2010cs,Andronic:2015wma} and references therein). 
In the case of $\Jpsi$ production, extensive phenomenological evidence has shown that color-singlet mechanisms alone fail to describe high-$p_T$ data at hadron colliders. 
Prediction within the dominant singlet contribution, proceeding through ${}^3S_1^{(1)}$ configurations, underestimate the observed rates. 
In contrast, the inclusion of fragmentation processes involving gluons or heavy quarks into color-octet $|c\bar{c}\rangle$ states, (particularly ${}^3S_1^{(8)}$, ${}^1S_0^{(8)}$, and ${}^3P_J^{(8)}$) yields a more accurate description~\cite{Braaten:1993rw,Cho:1995vh,Cho:1995ce,Beneke:1996tk, Butenschoen:2010rq,Chao:2012iv, Gong:2012ug}. 
These observations have led to the widely accepted conclusion that a complete description of vector charmonia requires the inclusion of color-octet channels.

For other quarkonia, however, the situation is different. 
For pseudoscalar states like $\eta_c$ and $\eta_b$, NRQCD predicts the dominance of the ${}^1S_0^{(1)}$ singlet channel, with octet contributions strongly suppressed by symmetry constraints~\cite{Braaten:1993rw,Han:2014jya}. 
The available data (especially for $\eta_c$~\cite{LHCb:2014oii,LHCb:2019zaj}) can be reproduced by singlet-only approaches, confirming this prediction. 
A similar pattern holds for $\Upsilon(nS)$ states~\cite{LHCb:2022byt}, where the larger mass scale and reduced soft-gluon sensitivity limit the impact of octet mechanisms~\cite{Artoisenet:2008fc}, although minor corrections may still be allowed at high transverse momentum~\cite{Gong:2010bk}.

Turning our attention to fully heavy tetraquarks, the applicability of a singlet-only approximation is highly sensitive to the quantum numbers $J^{PC}$ of the bound state. 
In the scalar channel ($0^{++}$), the internal color-spin wave function receives contributions from multiple diquark-antidiquark configurations, most notably the $[3,3]$ and $[6,6]$ color combinations. 
These components can interfere and mix, leading to a fragmentation structure that is inherently model dependent. 
In this case, the interplay of color configurations and their sensitivity to the assumed potential model may render color-octet mechanisms necessary to account for transitions that cannot be captured by the singlet component alone. 
As a consequence, predictions for scalar tetraquark fragmentation could be particularly vulnerable to theoretical uncertainties.

The situation is structurally simpler in the tensor channel ($2^{++}$), where only the $[3,3]$ configuration contributes. This channel is $S$-wave and fully symmetric in spin, and Fermi-Dirac statistics permit just one allowed color-spin coupling. 
Despite the absence of internal mixing, the resulting predictions still retain a moderate level of model dependence, especially through the specific shape and normalization of the bound-state wave function. 
In addition, the lack of strict symmetry constraints on the spinor structure leaves more room for ambiguity in modeling the wave function, thereby increasing the sensitivity to potential-model assumptions. 
Moreover, while a singlet-only treatment may be justified in first approximation, the possibility of subleading octet contributions cannot be fully excluded at the quantitative level.

A qualitatively different scenario emerges in the axial-vector channel ($1^{+-}$), which stands out for its structural clarity. Here, the $S$-wave nature of the bound state, combined with the antisymmetric requirements of Fermi-Dirac statistics, constrains the internal color-spin wave function to a single configuration: the $[3,3]$ component. 
This eliminates the possibility of interference between different color structures and leads to a much cleaner factorization scheme. 
The resulting fragmentation dynamics are largely insensitive to the details of the potential model, and the singlet-only approximation proves to be not only sufficient but also phenomenologically robust. 
This is supported by the observed stability of predictions across different modeling frameworks, as reported in Refs.~\cite{Bai:2024ezn,Celiberto:2025dfe}.

It is tempting to draw a parallel between the axial-vector tetraquark and the $\Jpsi$, both being spin-1 states. However, such a comparison can be misleading. The $\Jpsi$ is a vector $^3S_1$ quarkonium with well-understood spin and parity assignments, while the $1^{+-}$ tetraquark is an exotic multiquark system with different symmetry properties and a different color-spin composition. Unlike the $\Jpsi$, whose singlet contribution is insufficient at high $p_T$, the $1^{+-}$ tetraquark can, in many cases, be well-described by the singlet term alone, particularly in gluon fragmentation.

From a structural viewpoint, the difference stems from the dominance of a single color configuration in the diquark-antidiquark basis adopted within the NRQCD color-decomposition framework. 
The fragmentation process itself further enhances this compactness, favoring the formation of correlated subsystems with minimal color rearrangement. 
In this context, the role of color-octet contributions becomes increasingly model-specific: relevant in some configurations, negligible in others.

Although SDCs and LDMEs for color-octet configurations are currently unavailable for fully heavy tetraquarks, an order-of-magnitude estimate of their impact can be derived using NRQCD scaling arguments. 
In analogy with pseudoscalar quarkonia such as the $\eta_c$ and $\eta_b$, the leading color-singlet contribution to the FFs typically scales as $v_{\cal Q}^3$~\cite{Bodwin:1994jh,Lucha:1991vn,Eichten:1995ch}, while octet contributions scale as $v_{\cal Q}^7$ or beyond, and involve an additional $\alpha_s$ suppression at the hard-scattering level~\cite{Braaten:1994vv}. 
A conservative estimate for the ratio between octet and singlet contributions thus reads
\begin{equation*}
\frac{\text{octet}}{\text{singlet}} \sim \alpha_s \, v_{\cal Q}^4 \;.
\end{equation*}
For $T_{4c}$ states, using typical values $v_{{\cal Q}_c}^2 \sim 0.3$~\cite{Bodwin:1994jh} and $\alpha_s \sim 0.2$, this yields
\begin{equation}
\alpha_s \, v_{{\cal Q}_c}^4 \sim 0.2 \times (0.3)^2 \simeq 0.02 \;,
\end{equation}
suggesting a suppression by at least two orders of magnitude.

For fully bottomed tetraquarks, the suppression is expected to be even stronger. The relative velocity among the constituent quarks is smaller, with $v_{{\cal Q}_b}^2 \sim 0.1$~\cite{Brambilla:2010cs}, leading to
\begin{equation}
\alpha_s \, v_{{\cal Q}_b}^4 \sim 0.15 \times (0.1)^2 \simeq 0.0015 \;,
\end{equation}
\emph{i.e.}, a suppression of roughly three orders of magnitude. 
This effect is further enhanced by the reduced phase space available for soft-gluon emissions in bottom systems, making color-octet transitions even less likely at leading power.
These estimates reinforce the validity of the color-singlet approximation adopted in this work, particularly in the axial-vector channel, while also delineating the path for future improvements once the necessary nonperturbative inputs become available.

In light of this analysis, our choice to restrict the present implementation of FFs to color-singlet channels is well-motivated, particularly for the axial-vector channel. At the same time, our discussion outlines the path for future improvements: once reliable determinations of color-octet LDMEs become available, their inclusion will allow for a more comprehensive and quantitatively accurate modeling of tetraquark production across all quantum number sectors.

\vspace{1em}
\noindent
\textbf{LDME values and uncertainties.}
As mentioned, the LDMEs $\langle {\cal O}^{\TQQ(J^{PC})}([n]) \rangle$ represent the genuinely nonperturbative inputs to our initial-scale FFs. 
As in our previous works~\cite{Celiberto:2024mab,Celiberto:2024beg}, and in the absence of experimental data or lattice inputs for fully heavy tetraquarks, we extract estimates of these matrix elements from quark potential models.

This approach follows the strategy introduced in Ref.~\cite{Feng:2020riv}, where the radial wave functions at the origin are computed for fully charmed states by solving the Schr\"odinger equation with a Cornell-like potential~\cite{Eichten:1974af,Eichten:1978tg}, and then related to LDMEs through the VSA. 
In that work, three models were proposed~\cite{Zhao:2020nwy,Lu:2020cns,liu:2020eha}: the first and third adopt nonrelativistic quark fields, while the second includes relativistic corrections.

The first model tends to significantly overestimate the cross section when compared to CMS data on $\Jpsi$ production at $\sqrt{s} = 13$~TeV~\cite{CMS:2017dju}, which are in any case expected to exceed the actual $\TQc$ yield. 
Moreover, numerical tests (not shown here) revealed that the third model leads to severe instabilities; FFs constructed with its LDMEs exhibit large numerical fluctuations under minimal parameter variations, on the order of 0.1\%. 
For these reasons, in Refs.~\cite{Celiberto:2024mab,Celiberto:2024beg} we discarded both the first and third models and selected the second model~\cite{Lu:2020cns} as default reference.

A further extension was introduced in Ref.~\cite{Bai:2024ezn}, where two additional LDME models, denoted as Models~IV and~V in that work, were incorporated into the NRQCD-based construction of heavy-quark FFs.
From the inspection of Table~1 of Ref.~\cite{Bai:2024ezn}, it clearly emerges that Model~IV~\cite{Yu:2022lak} predicts values for the $1S$ axial-vector channel that are numerically close to those in~\cite{Lu:2020cns}, whereas Model~V~\cite{Wang:2019rdo} yields results that are suppressed by roughly one order of magnitude. 
In light of this results, and aiming to reduce model dependence while avoiding overconservative suppression, in Ref.~\cite{Celiberto:2025dfe} we based our analysis of the $\TQcOpm$ case on the average of the LDMEs from Refs.~\cite{Lu:2020cns} and~\cite{Yu:2022lak}, taking the spread between them as a measure of the theoretical uncertainty.

The axial-vector channel stands out as a particularly clean benchmark for this purpose. 
As already explained, due to Fermi-Dirac statistics and the $S$-wave nature of the state, its internal color-spin wave function is dominated by a single configuration, $[3,3]$, with no interference or mixing among color structures. 
This leads to a much cleaner factorization structure, reduces sensitivity to model assumptions, and enhances prediction stability. 
As observed in Ref.~\cite{Celiberto:2025dfe}, the $1^{+-}$ channel displays remarkable robustness under variations of the LDME model---particularly for Models I, II, and IV in Ref.~\cite{Bai:2024ezn}---unlike the scalar and tensor channels.

In contrast, the scalar ($0^{++}$) and tensor ($2^{++}$) states suffer from stronger model dependence. 
As discussed in Ref.~\cite{Feng:2020riv} and revisited in Refs.~\cite{Celiberto:2024mab,Celiberto:2024beg}, the $0^{++}$ state receives contributions from multiple interfering components in the diquark-antidiquark basis, notably the $[3,3]$, $[6,6]$, and mixed $[3,6]$ channels. 
These induce destructive or constructive interference patterns that vary with the potential model, making the resulting LDMEs unstable. 
The $2^{++}$ channel is structurally simpler, since it involves only a $[3,3]$ component, but still displays significant sensitivity to the shape and normalization of the radial wave function.

Given these considerations, in the present study we adopt a conservative prescription for estimating uncertainties on the scalar and tensor LDMEs. 
Since no direct uncertainty estimate is available for these channels, we take the relative uncertainty extracted in Ref.~\cite{Celiberto:2025dfe} for the axial-vector LDME as a reference and double it to reflect the enhanced model dependence.

Then, in the absence of first-principles determinations for the LDMEs of fully bottomed tetraquarks, we rely on a phenomenological \emph{Ansatz} inspired by physical arguments.
We consider the $\TQb$ system to behave as a compact diquark-antidiquark bound state, whose internal dynamics are predominantly governed by attractive color-Coulomb interactions.
Under this assumption, a dimensional analysis allows us to relate the four-body Schr\"odinger wave function at the origin for $\TQb$ and $\TQc$ states.
This strategy, originally proposed in Ref.~\cite{Feng:2023agq}, leads to the following scaling relation
\begin{equation}
\label{sLDMEs_T4b}
\hspace{-0.40cm}
 \frac{\langle {\cal O}^{\TQb(J^{PC})}([n]) \rangle}{\langle {\cal O}^{\TQc(J^{PC})}([n]) \rangle} 
 = 
 \frac{{\langle \cal O}^{\TQb}_{\rm [C]} \rangle}{{\langle \cal O}^{\TQc}_{\rm [C]}\rangle} 
 \simeq
 \left( \frac{m_b \as^{[b]}}{m_c \as^{[c]}} \right)^9 
 \simeq \,  
 400
 \;.
\end{equation}
In this expression, $\as^{[Q=c,b]}$ represents the strong coupling constant evaluated at the characteristic scale $m_Q v_{\cal Q}$, with $v_{\cal Q}$ denoting the typical relative velocity between the two constituent heavy quarks.
The `${\rm [C]}$' label specifies that the associated LDME has been obtained using a potential model dominated by Coulomb-like interactions within the diquark subsystem.

In this framework, the LDME scales as the square of the four-body wave function at the origin, with $|{\cal R}_{\cal Q}(0)|^2 \propto (\alpha_s m_Q)^3$ for each diquark pair, leading to an overall exponent of 9. 
This scaling mirrors the behavior found in heavy quarkonium systems, where the nonrelativistic wave function at the origin is governed by the Bohr radius of the bound state, $r_{B,{\cal Q}} \sim 1/(\alpha_s m_Q)$. 
Thus, as mentioned, the Coulombic approximation yields $|{\cal R}_{\cal Q}(0)|^2 \propto 1/r_{B,{\cal Q}}^3 \propto (\alpha_s m_Q)^3$, a result supported by early potential model studies~\cite{Bodwin:1994jh,Eichten:2019gig}.

In the case of a compact $\TQQ$ system treated as two weakly bound heavy diquarks, this scaling law applies independently to each diquark-antidiquark interaction, resulting in the overall cubic dependence squared in the four-body LDME. 
The use of Coulomb-like dynamics is further justified by the large masses and small relative velocities of the heavy quarks, which suppress nonperturbative effects and allow for a short-distance, weak-coupling treatment. 
This makes the Coulombic approximation a reasonable starting point for estimating the hierarchy of LDMEs across different flavor sectors.

To estimate the theoretical uncertainty associated with our Coulomb-based scaling of $\TQb$ LDMEs, we consider the impact of adopting alternative potential models, such as the Cornell~\cite{Eichten:1978tg} or Buchm{\"u}ller-Tye~\cite{Buchmuller:1980su} potentials, on the four-body wave function at the origin. 
Studies of heavy quarkonium systems show that the inclusion of a confining term in the interquark potential typically reduces the radial wave function at the origin by up to 25\% with respect to the purely Coulombic case.
Since our LDMEs scale as the square of the diquark wave function squared, \emph{i.e.}, as $|{\cal R}_{\cal Q}(0)|^4$ per diquark pair, this reduction translates into a relative variation of the LDME of order

\begin{equation}
\label{sLDMEs_T4b_Delta}
 \left( \frac{\Delta_{\rm model} {\cal R}_{\cal Q}(0)}{{\cal R}_{\cal Q}(0)} \right)^4 \simeq (0.25)^4 = 0.4\% \;.
\end{equation}

Although it would be formally appropriate to add this model-dependent uncertainty in quadrature with the propagated error from the $\TQc$ sector, we find that its impact is numerically negligible.
Indeed, since the LDME scaling factor is approximately 400, the relative uncertainty introduced by switching to a confining potential, estimated to be around 0.4\%, remains subleading compared to the 20--25\% uncertainty already inherited from the $\TQc$ values.
We therefore retain only the $\TQc$-propagated uncertainty in our $\TQb$ analysis, which remains the dominant source in our error budget.

We collect the numerical values of the color-composite LDMEs employed in our analysis in Tables~\ref{tab:T4c_LDMEs} and~\ref{tab:T4b_LDMEs}, for the $\TQc(J^{PC})$ and $\TQb(J^{PC})$ states, respectively. 
The quoted uncertainties reflect the model-dependent prescriptions detailed above.

\begin{table}[t]
\centering
\begin{tabular}{c|c|c|c}
\toprule
$[n]$ & $\TQcZpp$ [GeV$^9$] & $\TQcOpm$ [GeV$^9$] & $\TQcTpp$ [GeV$^9$] \\
\midrule
$[3,3]$ & $0.0347 \pm 0.0076$ & $0.0878 \pm 0.0098$ & $0.0720 \pm 0.0158$ \\
$[6,6]$ & $0.0128 \pm 0.0028$ & $0$                 & $0$                 \\
$[3,6]$ & $0.0211 \pm 0.0046$ & $0$                 & $0$                 \\
\bottomrule
\end{tabular}
\caption{
\justifying
\noindent
Color-composite LDMEs $\langle {\cal O}^{\TQc(J^{PC})}([n]) \rangle$ for the $\TQcZpp$, $\TQcOpm$, and $\TQcTpp$ states. The uncertainties reflect the strategy described in the main text.
}
\label{tab:T4c_LDMEs}
\end{table}

\begin{table}[t]
\centering
\begin{tabular}{c|c|c|c}
\toprule
$[n]$ & $\TQbZpp$ [GeV$^9$] & $\TQbOpm$ [GeV$^9$] & $\TQbTpp$ [GeV$^9$] \\
\midrule
$[3,3]$ & $13.88 \pm 3.05$ & $35.1 \pm 3.9$ & $28.80 \pm 6.34$ \\
$[6,6]$ & $5.12 \pm 1.13$  & $0$            & $0$             \\
$[3,6]$ & $8.44 \pm 1.86$  & $0$            & $0$             \\
\bottomrule
\end{tabular}
\caption{
\justifying
\noindent
Color-composite LDMEs $\langle {\cal O}^{\TQb(J^{PC})}([n]) \rangle$ for the $\TQbZpp$, $\TQbOpm$, and $\TQbTpp$ states at the initial scale. The values are obtained via color-Coulomb scaling from the corresponding $\TQc$ entries, as explained in the main text.
}
\label{tab:T4b_LDMEs}
\end{table}

\subsection{{\tt TQ4Q1.1} functions from {\HFNRevo}}
\label{ssec:FFs_TQ4Q11}

The final ingredient in constructing the {\tt TQ4Q1.1} collinear FFs for fully heavy tetraquarks is the proper application of DGLAP evolution to the initial-scale inputs discussed in the previous subsection.
Unlike the case of light-hadron fragmentation, a key feature here is that both the heavy-quark ($Q$ or $\bar{Q}$) and the gluon channels exhibit nontrivial evolution thresholds.
This behavior arises directly from the kinematics of the perturbative splittings $[g \to (Q\bar{Q}Q\bar{Q})]$ and $[Q, \bar{Q} \to (Q\bar{Q}Q\bar{Q}) + Q, \bar{Q}]$, which correspond to the left and right diagrams of Fig.~\ref{fig:FF_diagrams}, respectively, and are encoded in the relevant SDCs.

By kinematic constraints, the minimum invariant mass required for the gluon-induced splitting is $\mu_{F,0}(g \to \TQQ) = 4 m_Q$, which we take as the evolution threshold for gluon fragmentation.
Similarly, the lowest allowed invariant mass for the heavy-quark-induced splitting is $\mu_{F,0}(Q \to \TQQ) = 5 m_Q$, which defines the corresponding threshold for the (anti)quark channel.

To properly account for the presence of both quark and gluon thresholds in the DGLAP evolution, we adopt a dedicated strategy rooted in the recently developed {\HFNRevo} scheme~\cite{Celiberto:2025euy,Celiberto:2024mex,Celiberto:2024bxu,Celiberto:2024rxa}.

This methodology is specifically designed to govern the DGLAP evolution of heavy-hadron FFs built from nonrelativistic inputs, and it is structured around three foundational elements: physical interpretation, evolution dynamics, and uncertainty quantification.
The interpretation step connects the low-transverse-momentum production mechanism to a two-parton fragmentation picture, as detailed in Sec.~\ref{ssec:FFs_intro}, thus enabling a consistent matching between FFNS and VFNS schemes.
The uncertainty quantification component provides a systematic way to assess missing higher-order uncertainties (MHOUs) through controlled variations of the evolution thresholds.

Originally designed to connect precision QCD predictions with a hadron-structure-driven perspective on quarkonium fragmentation, the {\HFNRevo} framework has recently been extended to accommodate rare and exotic matter production, yielding promising results.
Notable applications include initial studies of scalar ($0^{++}$) and tensor ($2^{++}$) tetraquarks using the {\tt TQ4Q1.0} functions~\cite{Celiberto:2024mab,Celiberto:2024beg}---precursors to the {\tt TQ4Q1.1} sets---as well as rare triply heavy baryons such as ${\rm \Omega}_{3Q}$~\cite{Celiberto:2025ogy}, investigated through the {\tt OMG3Q1.0} determinations. 
There {\HFNRevo} has proven to be a versatile scheme for evolving FFs that feature both heavy-quark and gluon initial-scale inputs.
This dual-channel nature requires a dedicated treatment of evolution thresholds, individually tailored to each partonic channel.

In the present study, which explores the {\HFNRevo} fragmentation of fully heavy tetraquarks, we defer the implementation of matching techniques. 
Our primary focus is instead placed on developing a consistent and accurate strategy for performing DGLAP evolution under the presence of both quark and gluon thresholds, as well as on the systematic treatment of MHOUs, whose role will be discussed in detail below.

According to {\HFNRevo}, the DGLAP evolution of a set of heavy-hadron FFs can be organized into two sequential steps.
Focusing on the $\TQQ$ case, the evolution begins from the partonic channel with the lowest threshold, namely the gluon.
We take gluon-FF input calculated within NRQCD, evaluated at the initial scale $\mu_{F,0}(g \to \TQc) = 4 m_Q$, and evolve it using DGLAP equations governed solely by the LO $P_{gg}$ kernel.
This evolution proceeds up to the threshold for the constituent heavy-(anti)quark channel, $\mu_{F,0}(Q \to \TQQ) = 5 m_Q$.

Within this range, the evolution exclusively generates collinear gluons.
Since it involves a single channel and is expanded perturbatively in powers of $\alpha_s$, this step can be performed analytically using the {\symJethad} plugin.

The second step starts at the heavy-quark threshold.
Here, the gluon FF evolved up to $Q_0 \equiv \mu_{F,0}(Q \to \TQQ) = 5 m_Q$ is combined with the (anti)quark FF NRQCD input.
Starting from this common scale, we carry out a full DGLAP evolution that includes all active channels, thereby generating the NLO {\tt TQ4Q1.1} functions released in {\tt LHAPDF} format~\cite{Buckley:2014ana}.

We refer to $Q_0$ as the \emph{evolution-ready} scale.
It corresponds to the maximum threshold among the partonic species involved and marks the point where the numerical DGLAP evolution is initialized.
For this second step, we employ the {\tt APFEL++} library~\cite{Bertone:2013vaa,Carrazza:2014gfa,Bertone:2017gds}, while future implementations will consider interfacing with {\tt EKO}~\cite{Candido:2022tld,Hekhorn:2023gul} as well.

An additional key ingredient of our framework is the treatment of perturbative-fragmentation scale uncertainties (F-MHOUs).  
These arise from the arbitrariness in the choice of energy scale, such as the initial scale $Q_0$, where FFs are matched to their NRQCD input.  
As mentioned, in our default setup, $Q_0$ is fixed at $5 m_Q$, which represents the ``natural'' value, suggested by LO kinematics.  
To estimate the associated uncertainty, we vary $Q_0$ by a factor $1/2$ to $2$ around this central value.  
For each variation, a corresponding replica of the initial conditions is generated, and the subsequent DGLAP evolution is carried out independently.  

The resulting set of replicas provides an envelope that quantifies the impact of F-MHOUs on the evolved FFs.  
Although these replicas are not Monte Carlo members in the statistical sense, they are organized in an analogous fashion, so that their spread can be directly propagated into collider-level observables.  
As detailed in Sec.~\ref{ssec:uncertainty}, this procedure enables us to disentangle the relative weight of F-MHOUs with respect to other sources of uncertainty, such as hard-scale variations (H-MHOUs) and LDMEs.

While our two-step evolution framework focuses on gluon and constituent heavy-quark channels, one may question the absence of light- and nonconstituent-heavy-quark contributions.
We have not included these channels at this stage.
Although light-quark initiated fragmentation into fully heavy tetraquarks has recently been computed~\cite{Bai:2024flh}, their implementation is beyond the scope of the present study and is left for future developments aimed at completing the set of NRQCD inputs.
As such, nonconstituent quarks are not assigned any initial-scale FFs and instead enter the evolution dynamically, being radiatively generated from the gluon and heavy-quark channels.

Nonetheless, insights from NRQCD studies on color-singlet pseudoscalar~\cite{Braaten:1993rw,Braaten:1993mp,Artoisenet:2014lpa,Zhang:2018mlo,Zheng:2021mqr,Zheng:2021ylc} and vector~\cite{Braaten:1993rw,Braaten:1993mp,Zheng:2019dfk} charmonia suggest that such suppressed fragmentation modes are phenomenologically negligible compared to the dominant gluon and constituent heavy-quark ones.

A quantitative assessment of their suppression can be inferred from recent NRQCD-based studies of light-quark initiated FFs into heavy quarkonia and tetraquarks.
For instance, Ref.~\cite{Bai:2024flh} shows that the light-quark FF into a fully charmed scalar or tensor tetraquark is suppressed by one to two orders of magnitude with respect to the gluon channel across a broad range of the momentum fraction $z$.
This is consistent with the aforementioned findings for quarkonia.

Moreover, in our phenomenological studies of tetraquark production at the LHC, the relevant range of longitudinal momentum fraction $x$ at which parton distribution functions (PDFs) are probed is approximately $10^{-4} \lesssim x \lesssim 10^{-2}$ (see, for instance, Refs.~\cite{Celiberto:2021dzy,Celiberto:2021fdp,Celiberto:2022dyf}).
In this regime, the gluon PDF dominates over light-quark PDFs by more than one order of magnitude.
Since our cross-section computations rely on collinear factorization at LO, where the partonic hard scattering convolves diagonally the incoming PDF and the outgoing FF, this further amplifies the relative weight of gluon-initiated channels.

\begin{figure*}[!t]
\centering

   \hspace{-0.00cm}
   \includegraphics[scale=0.430,clip]{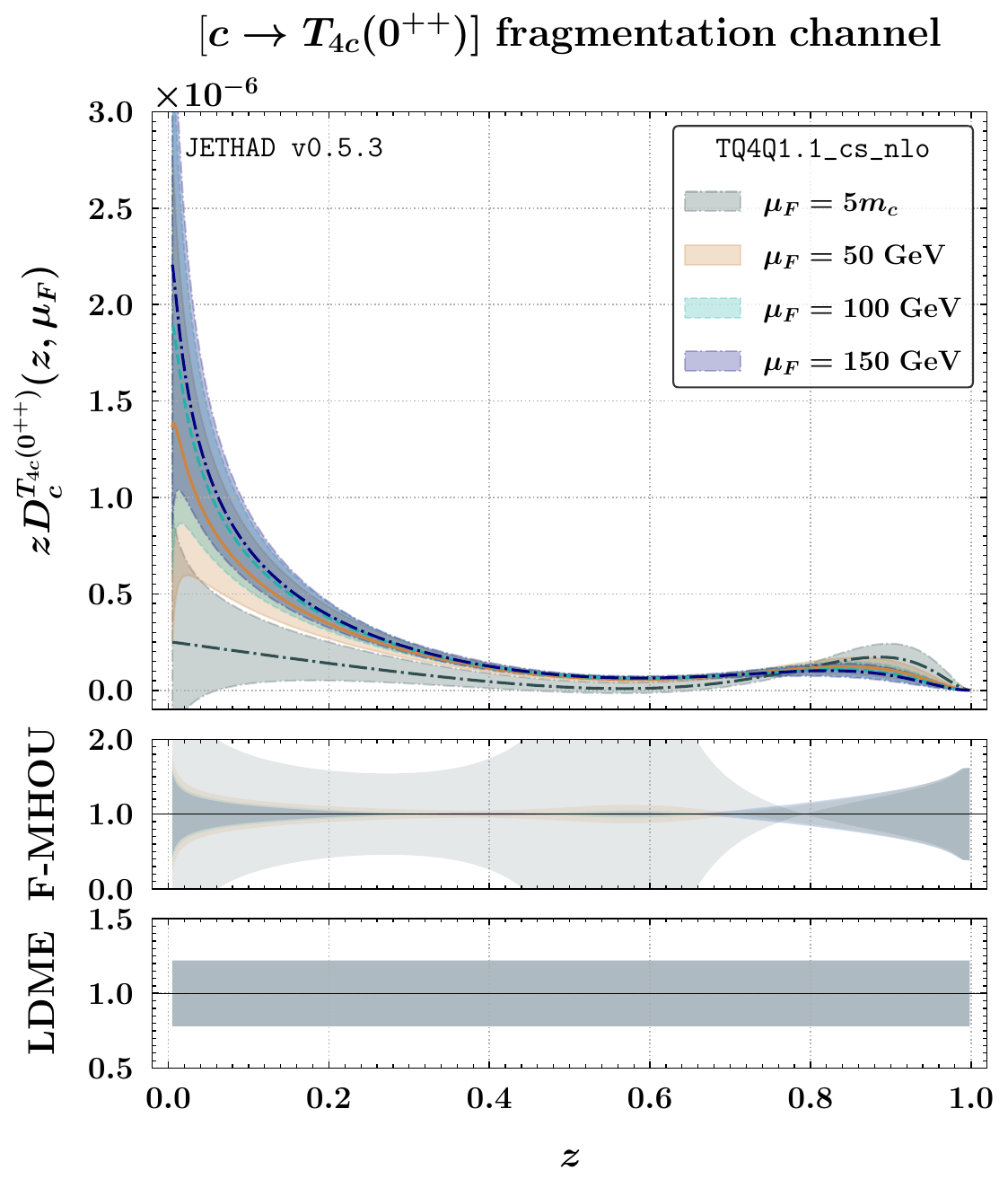}
   \hspace{0.90cm}
   \includegraphics[scale=0.430,clip]{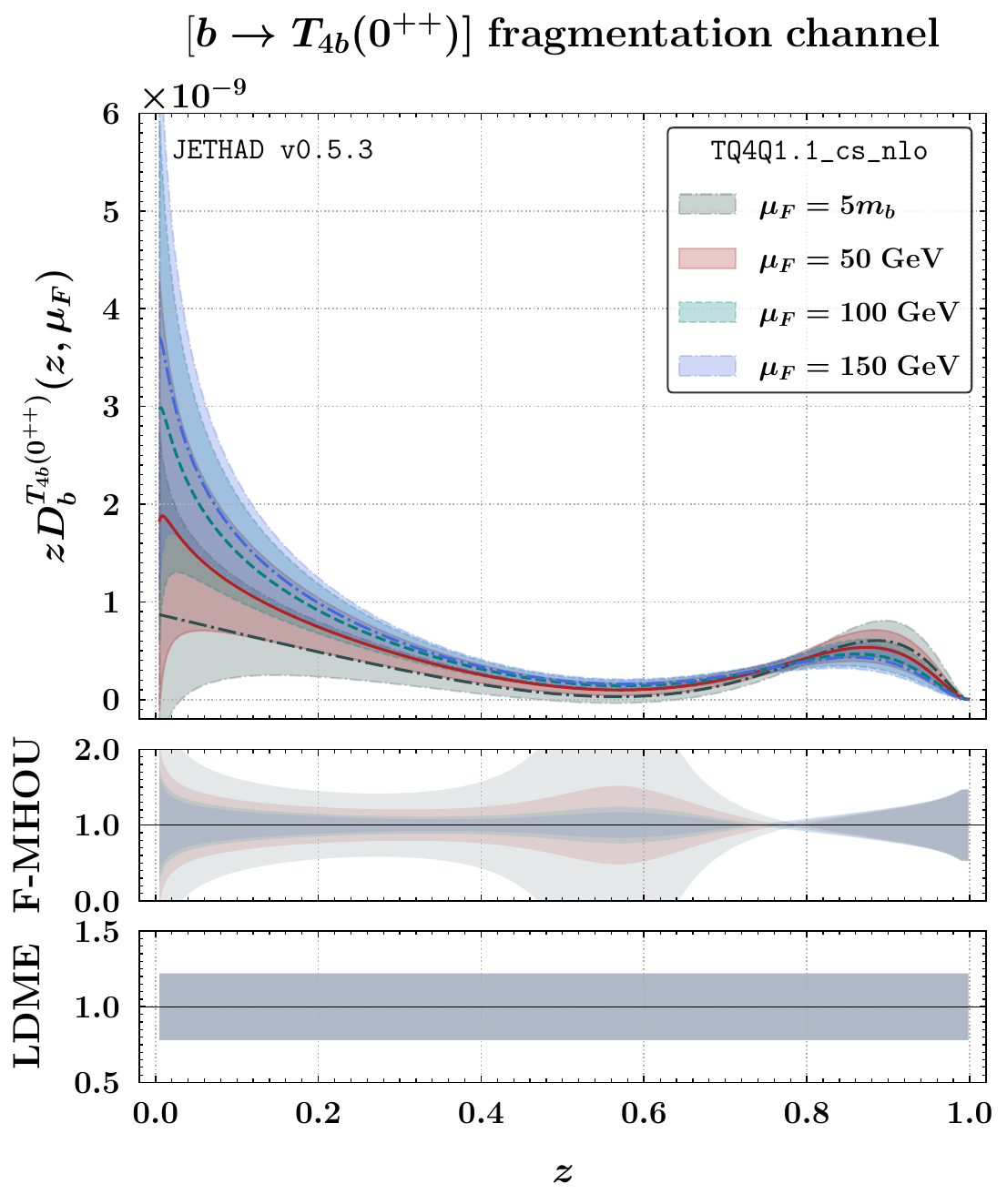}

   \vspace{0.35cm}

   \hspace{-0.00cm}
   \includegraphics[scale=0.430,clip]{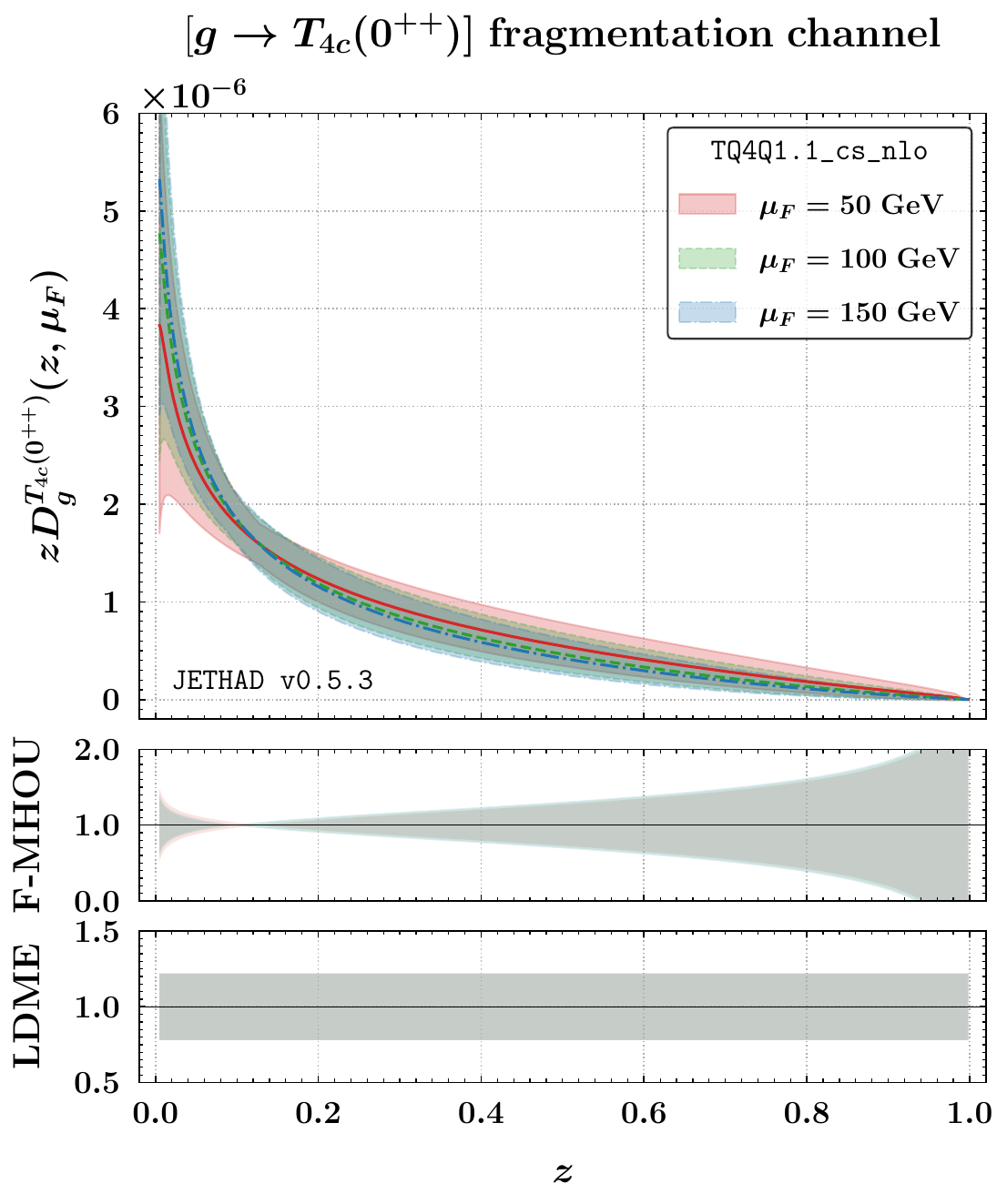}
   \hspace{0.90cm}
   \includegraphics[scale=0.430,clip]{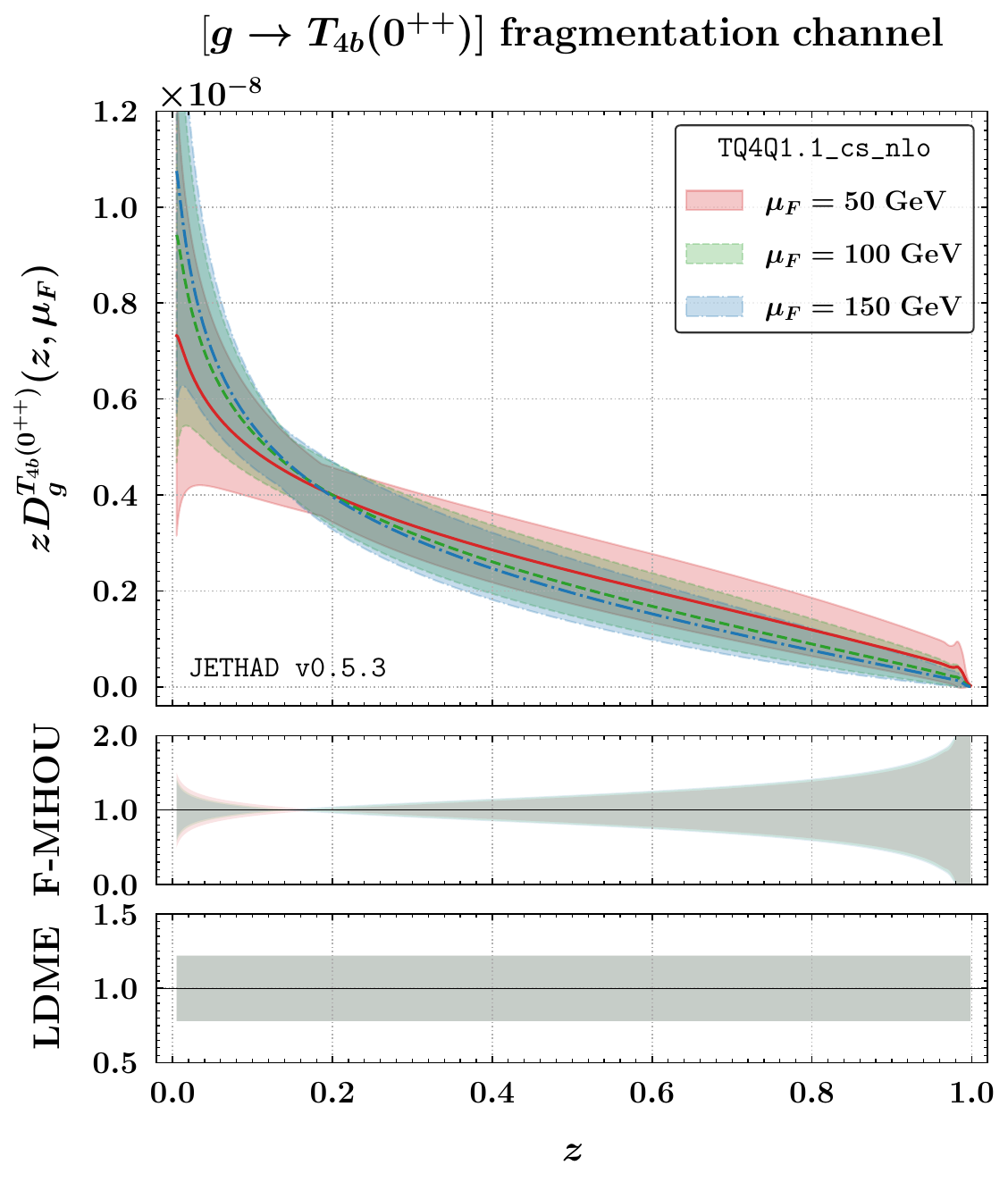}

\caption{
\justifying
\noindent
$z$-shape of the {\tt TQ4Q1.1} FFs for scalar tetraquarks $\TQcZpp$ (left) and $\TQbZpp$ (right), evaluated at various energy scales.
Upper (lower) plots refer to the heavy-quark (gluon) initiated channels.
Filled bands in the main panels represent the combined effect of uncertainties from perturbative F-MHOUs and nonperturbative LDME variations.
The first lower panel shows the impact of perturbative F-MHOUs (replica envelope normalized to the central one), while the second displays LDME variations as ratios to the central value.}
\label{fig:FFs-z_TQ0}
\end{figure*}

\begin{figure*}[!t]
\centering

   \hspace{-0.00cm}
   \includegraphics[scale=0.430,clip]{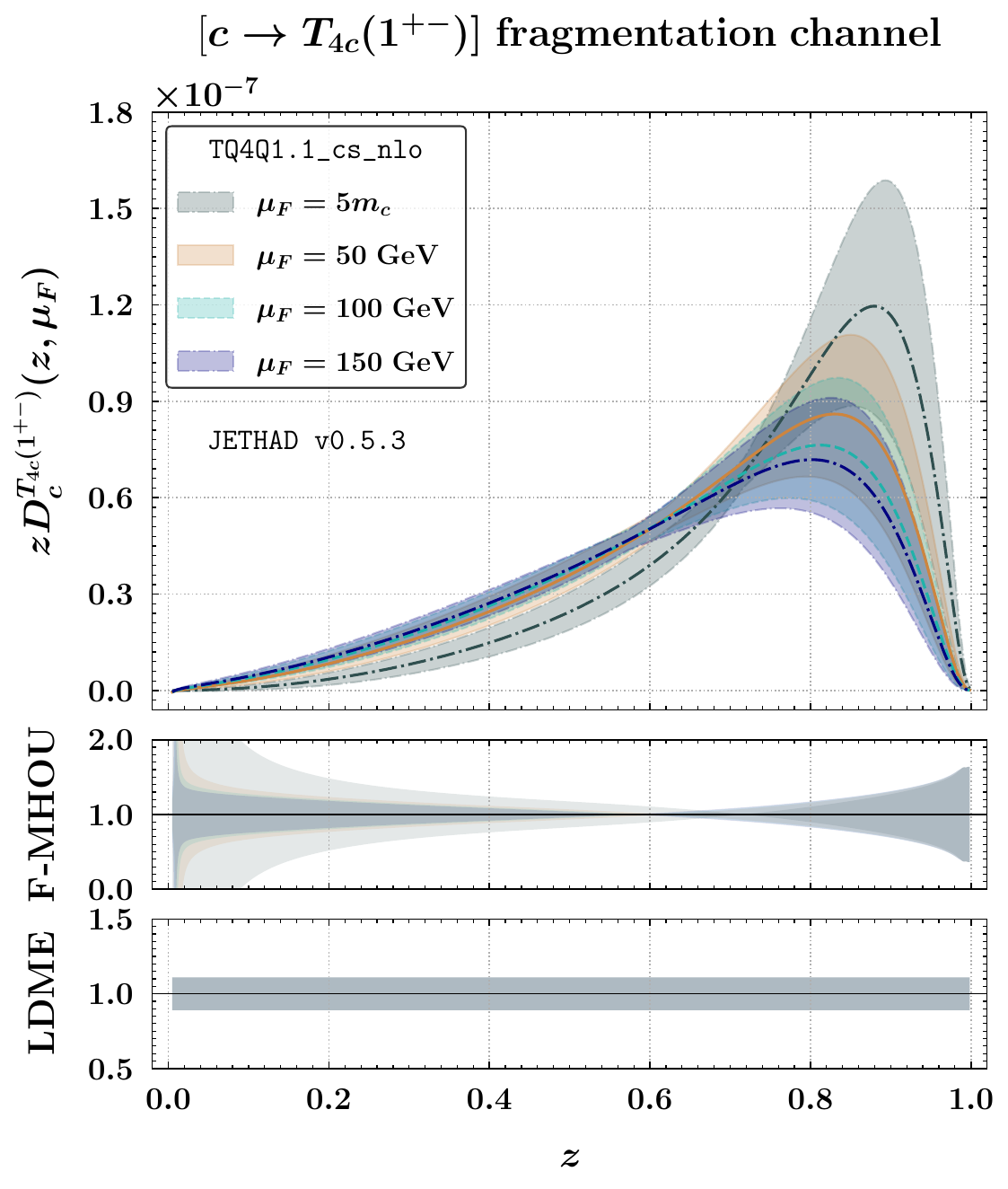}
   \hspace{0.90cm}
   \includegraphics[scale=0.430,clip]{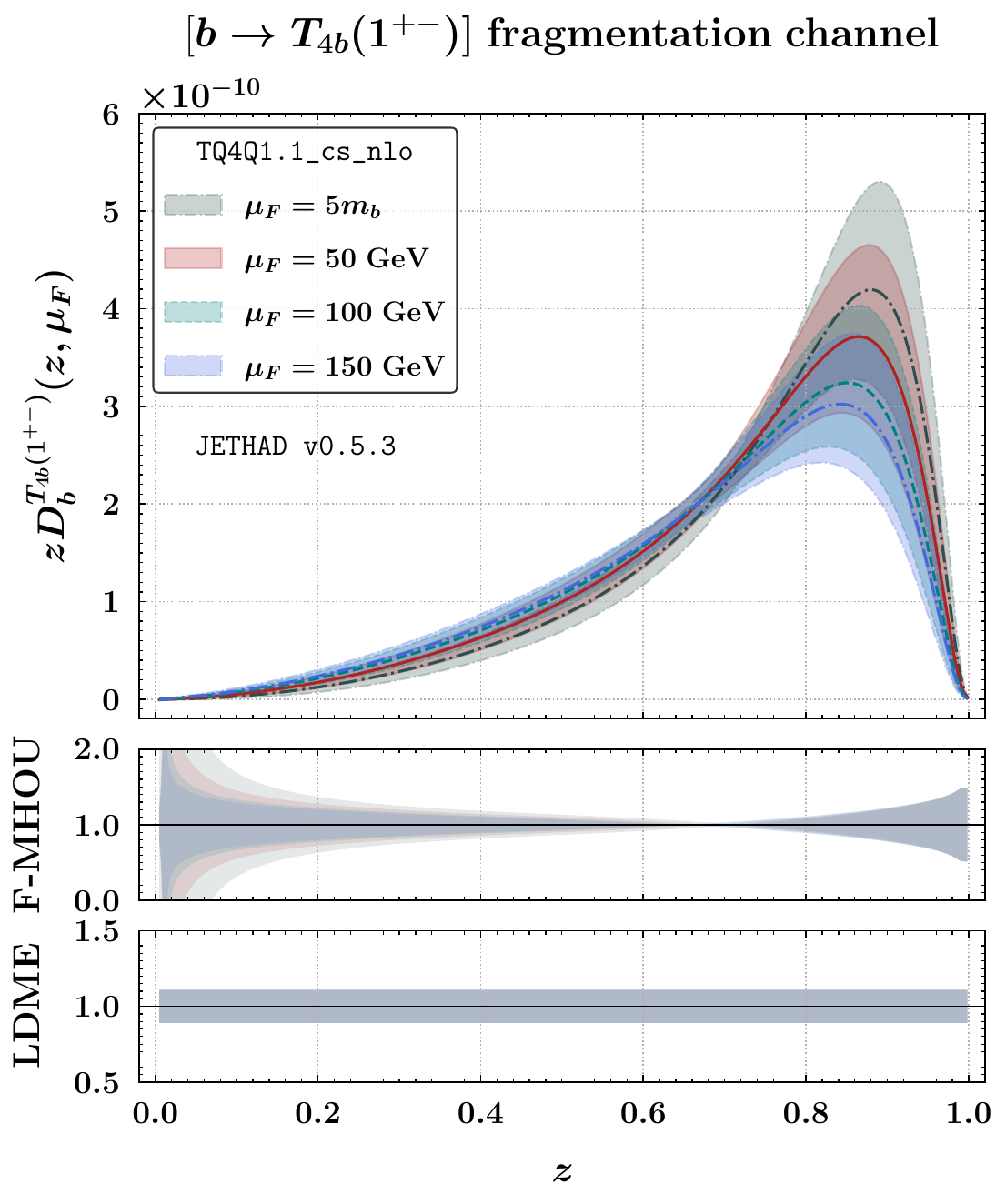}

   \vspace{0.35cm}

   \hspace{-0.00cm}
   \includegraphics[scale=0.430,clip]{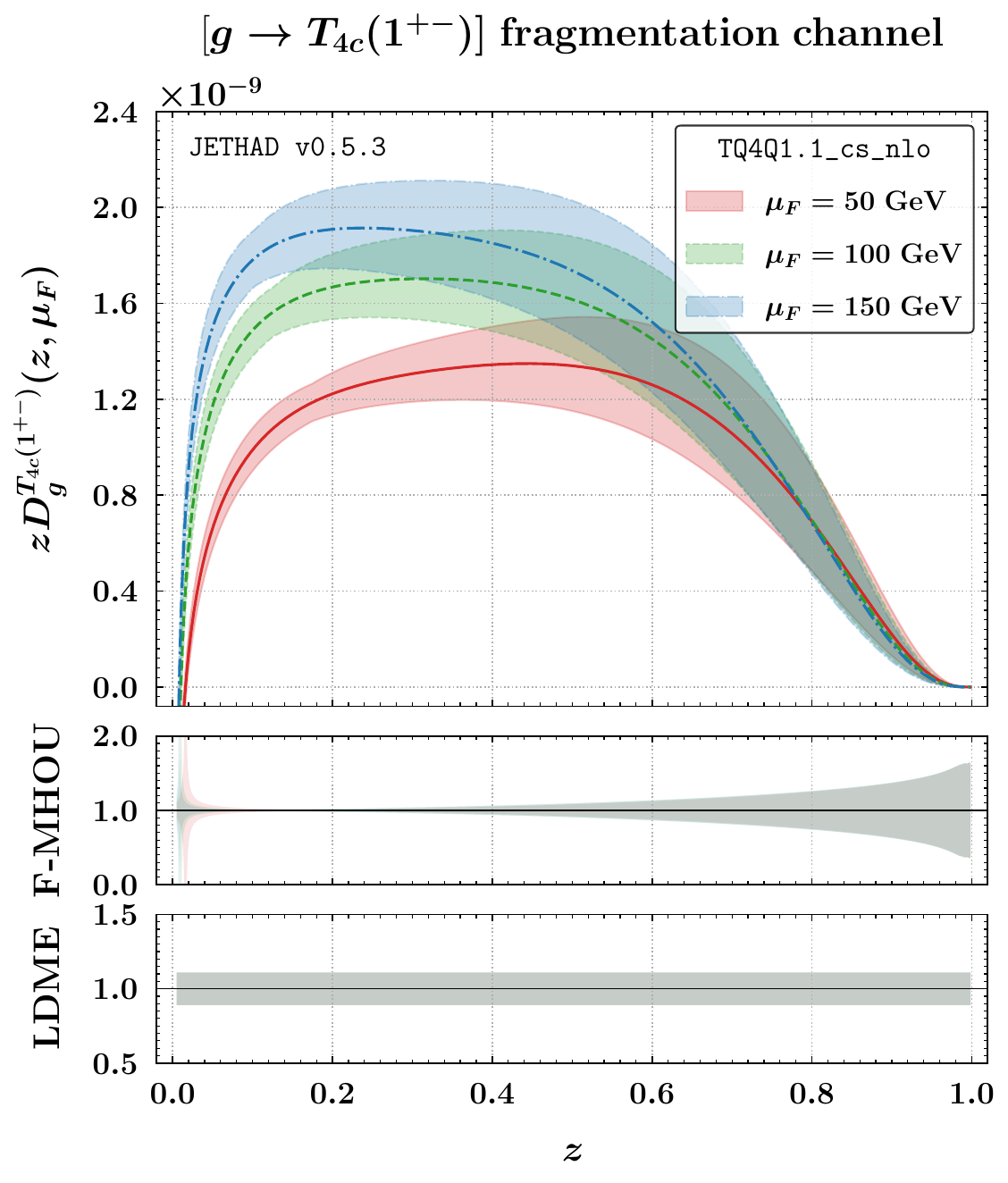}
   \hspace{0.90cm}
   \includegraphics[scale=0.430,clip]{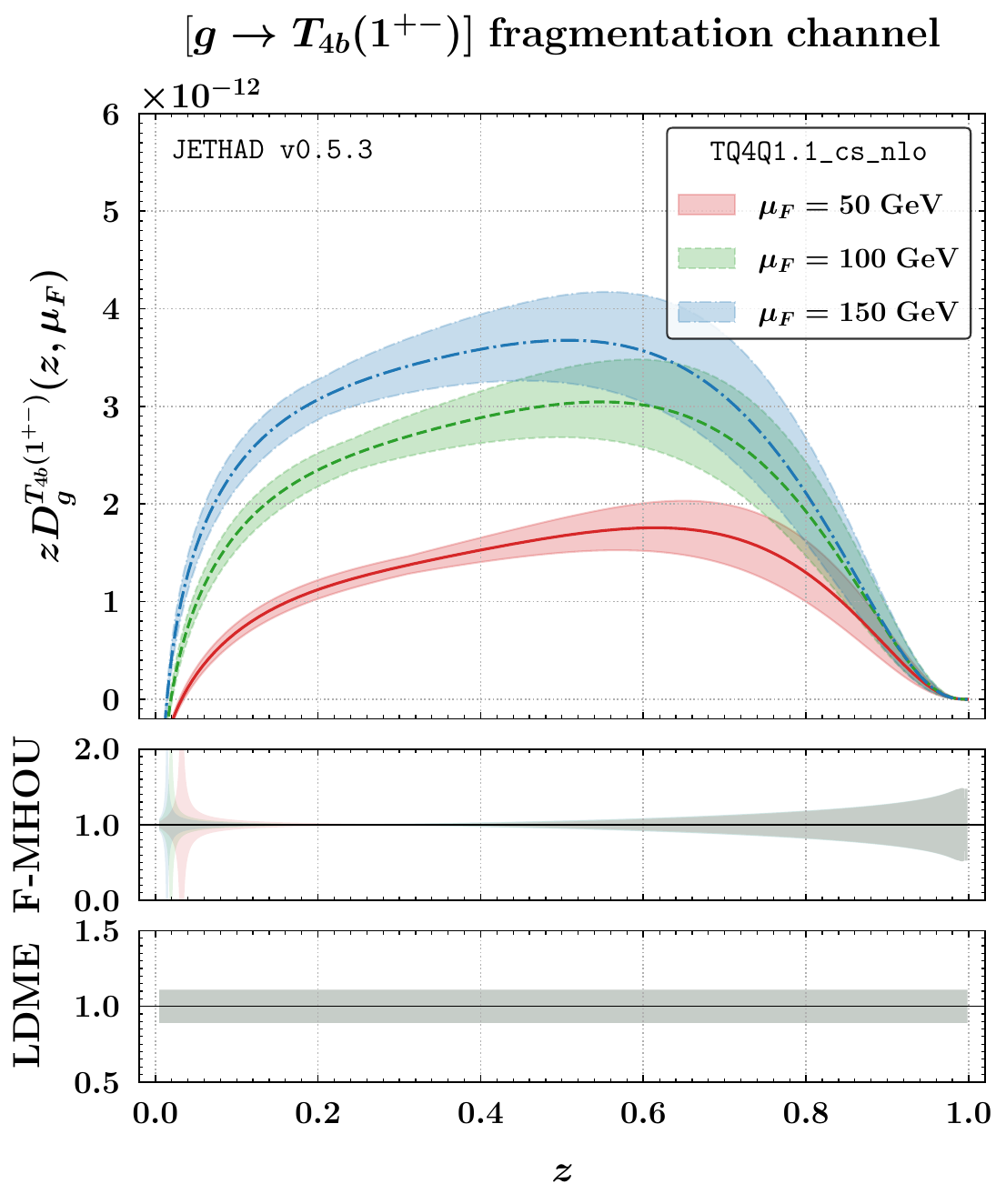}

\caption{
\justifying
\noindent
$z$-shape of the {\tt TQ4Q1.1} FFs for axial-vector tetraquarks $\TQcOpm$ (left) and $\TQbOpm$ (right), evaluated at various energy scales.
Upper (lower) plots refer to the heavy-quark (gluon) initiated channels.
Filled bands in the main panels represent the combined effect of uncertainties from perturbative F-MHOUs and nonperturbative LDME variations.
The first lower panel shows the impact of perturbative F-MHOUs (replica envelope normalized to the central one), while the second displays LDME variations as ratios to the central value.}
\label{fig:FFs-z_TQ1}
\end{figure*}

\begin{figure*}[!t]
\centering

   \hspace{-0.00cm}
   \includegraphics[scale=0.430,clip]{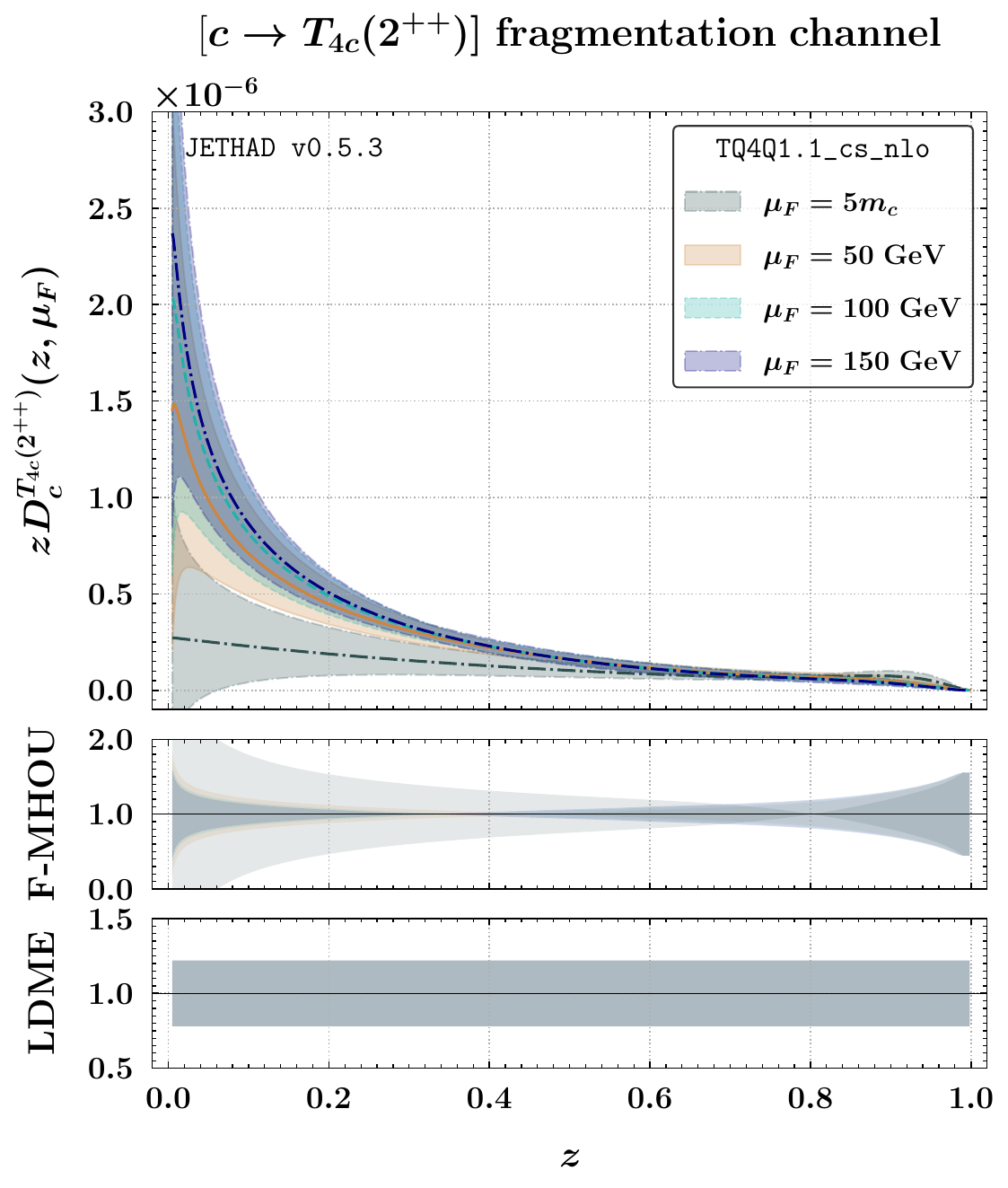}
   \hspace{0.90cm}
   \includegraphics[scale=0.430,clip]{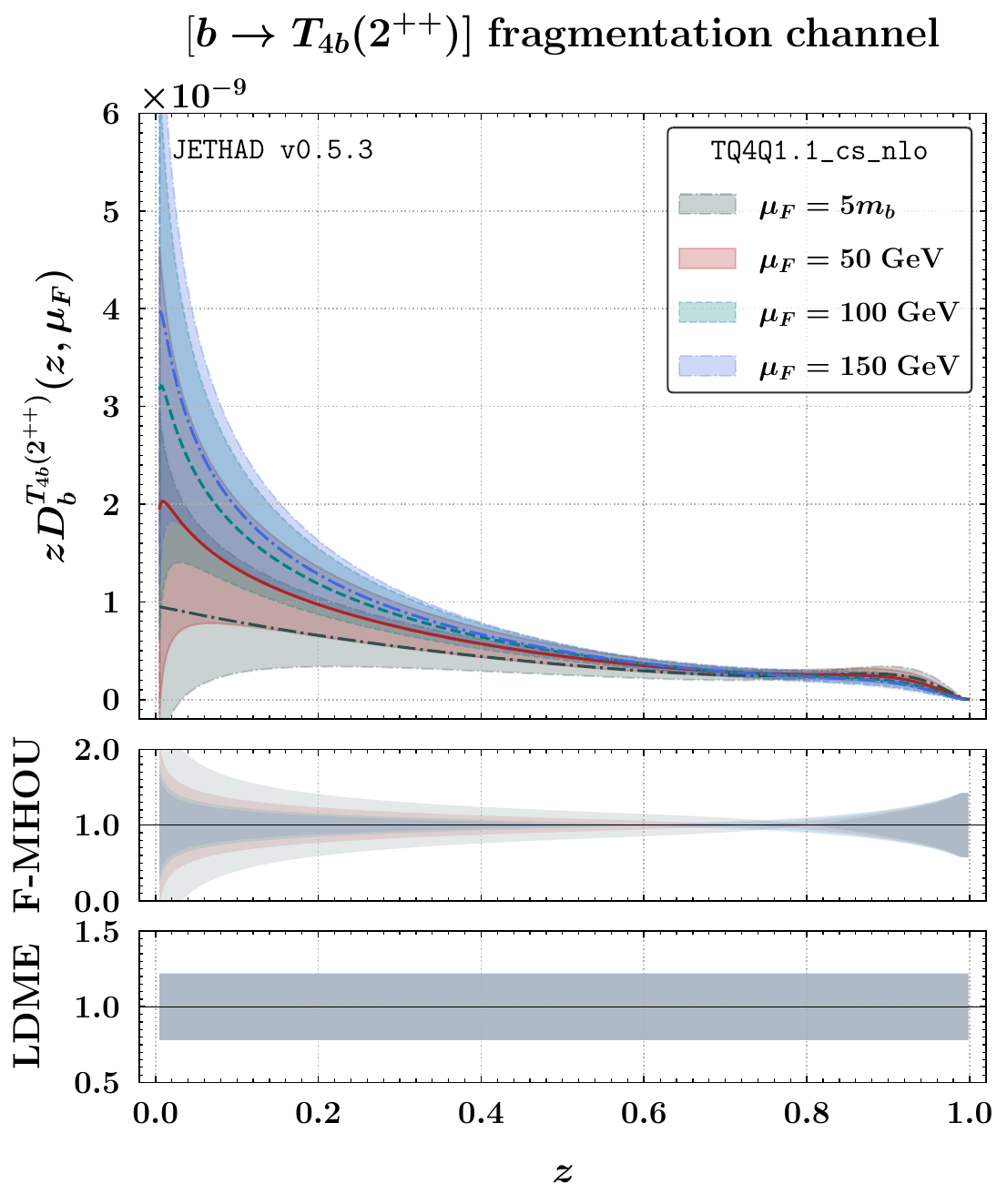}

   \vspace{0.35cm}

   \hspace{-0.00cm}
   \includegraphics[scale=0.430,clip]{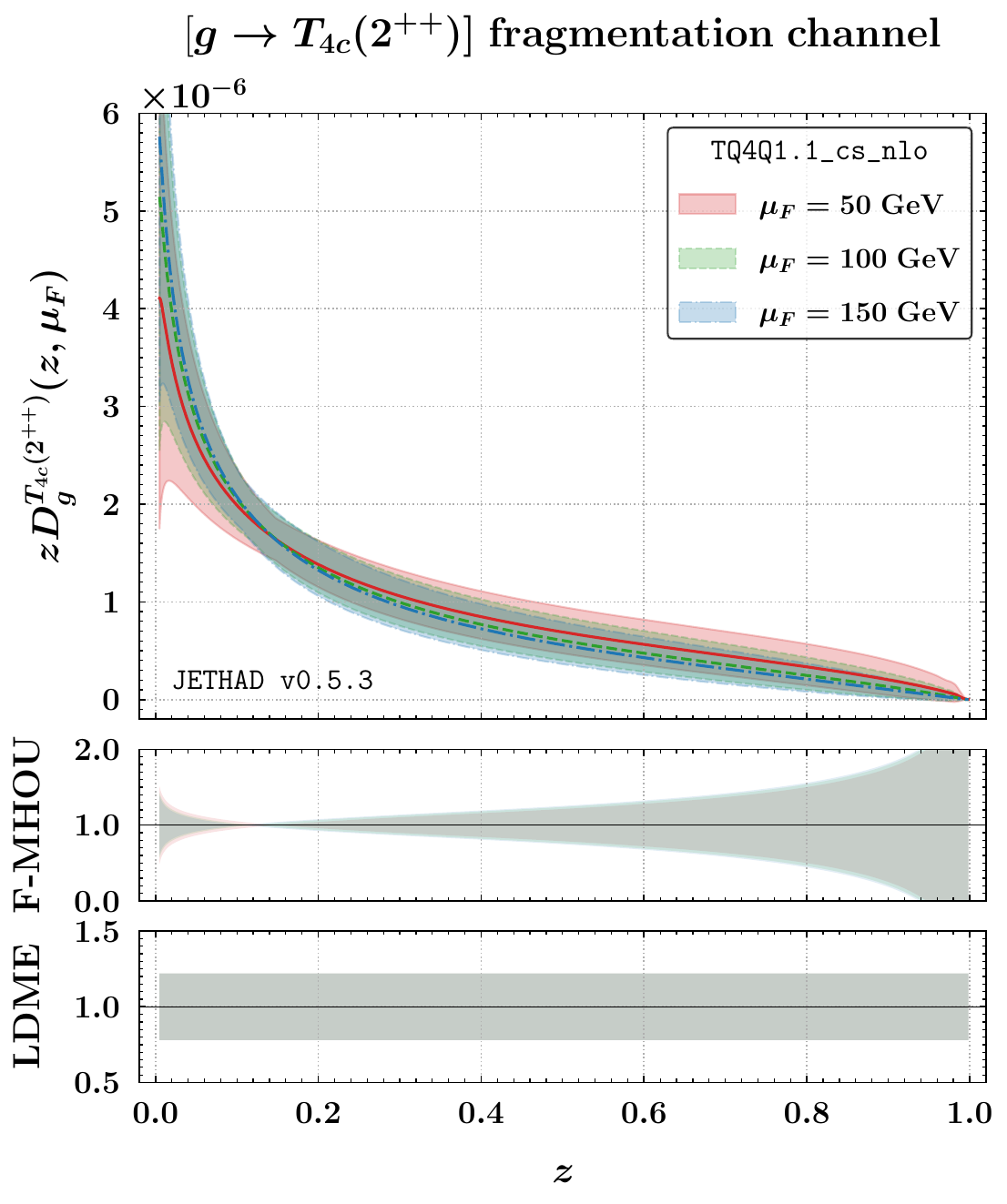}
   \hspace{0.90cm}
   \includegraphics[scale=0.430,clip]{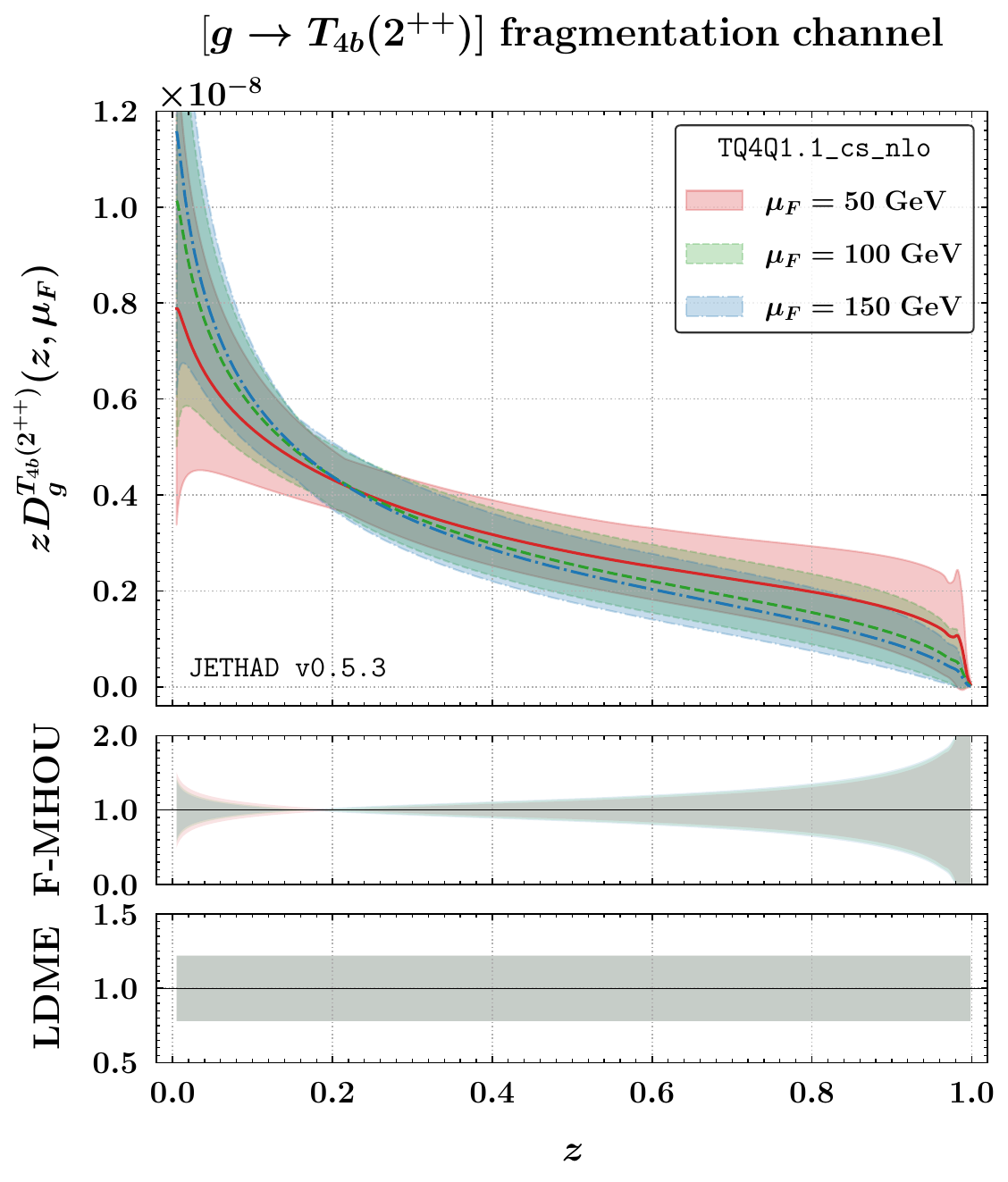}

\caption{
\justifying
\noindent
$z$-shape of the {\tt TQ4Q1.1} FFs for tensor tetraquarks $\TQcTpp$ (left) and $\TQbTpp$ (right), evaluated at various energy scales.
Upper (lower) plots refer to the heavy-quark (gluon) initiated channels.
Filled bands in the main panels represent the combined effect of uncertainties from perturbative F-MHOUs and nonperturbative LDME variations.
The first lower panel shows the impact of perturbative F-MHOUs (replica envelope normalized to the central one), while the second displays LDME variations as ratios to the central value.}
\label{fig:FFs-z_TQ2}
\end{figure*}

\vspace{1em}
\noindent
\textbf{Momentum dependence.}
In Figs.~\ref{fig:FFs-z_TQ0}, \ref{fig:FFs-z_TQ1}  and~\ref{fig:FFs-z_TQ2} we present the dependence on the momentum fraction $z$ of the {\tt TQ4Q1.1} FFs, multiplied by $z$, for the $0^{++}$, $1^{+-}$, and $2^{++}$ tetraquarks.  
Each figure presents a double panel structure, where the left (right) columns refer to the charm (bottom) sector, and the upper (lower) rows correspond to the constituent heavy-quark (gluon) fragmentation channels.
In all figures, the main panels display the combined effect of F-MHOU and LDME uncertainties, added in quadrature. 
The ancillary panels below disentangle these contributions: the first shows the impact of F-MHOUs, displayed as the envelope of replicas normalized to the central one, while the second illustrates the effect of LDME variations, again shown as ratios with respect to the central value.

To highlight the role of the perturbative QCD evolution, we show FFs at four representative values of the factorization scale $\mu_F$: the central value of the evolution-ready scale $Q_0 = 5m_Q$, and the evolved scales $\mu_F = 50$, $100$, and $150$~GeV.
At $Q_0$, the $[g \to \TQQ(1^{+-})]$ FF vanishes by construction, as it is generated radiatively through DGLAP evolution for $\mu_F > Q_0$.
For the scalar and tensor cases, a nonzero gluon FF is already present at the input scale.
However, for consistency across the plots, all gluon FFs are omitted at $Q_0$ and shown only for $\mu_F = 50$, $100$, and $150$~GeV.

The momentum-fraction dependence of the $[Q \to \TQQ]$ FFs displays distinct features for each quantum number.
In the scalar and tensor channels, the heavy-quark FFs show a characteristic two-phase structure: they start from a finite, nonzero value at $z = 0$, then decrease steeply with increasing $z$, before developing a moderate peak in the region around $z \simeq 0.9$ for the charm case, slightly shifted toward lower $z$ for the bottom case.
The descending tail from $z=0$ becomes increasingly steep with growing $\mu_F$, reflecting the enhancement of radiation at small momentum fractions induced by DGLAP evolution.

Despite this suppression, a small hump survives at large $z$, signaling a residual hard-fragmentation component.
This behavior is consistent with expectations from the heavy-quark limit and reflects the typical hard fragmentation pattern observed in other heavy-flavor systems~\cite{Suzuki:1977km,Bjorken:1977md}, where the heavy constituent quark tends to carry a sizable fraction of the hadron momentum.
It illustrates the soft-hard interplay characteristic of heavy-flavor fragmentation into multiquark bound states, where the fragmenting quark can retain either a small or a large share of the total momentum, depending on the configuration.

In our previous analysis of {\tt TQ4Q1.0} functions~\cite{Celiberto:2024mab}, precursors of the {\tt 1.1} determinations, we modeled the initial condition of the FF using a simplified kinematic approach inspired by the Suzuki model~\cite{Suzuki:1977km}, originally devised for heavy-light systems.
There, we emphasized that for singly heavy-flavored hadrons, such as $|Q\bar{q}\rangle$ mesons, the heavy quark and light antiquark must share the same velocity.
This requirement leads to a momentum-fraction peak at $\langle z \rangle \simeq 1 - \Lambda_q/m_Q$, where $\Lambda_q$ is a hadronic scale, thus enforcing a hard-fragmentation pattern directly from kinematics.
However, we also pointed out that such an argument does not hold in general for multiply heavy-flavored states, like quarkonia and tetraquarks, where no light constituent is present in the lowest Fock component and no natural soft scale exists.
In that context, we cautioned against using purely kinematic reasoning to predict the FF peak location.

Nevertheless, our present results show that, even in the absence of light quarks, a peak at large $z$ still emerges dynamically in the scalar and tensor channels, especially for the charm case.
This observation is in line with what is known for quarkonia~\cite{Braaten:1993mp}, where the FF of a heavy quark into $\eta_Q$ or $\Jpsi$ also displays a dominant peak at $z \gtrsim 0.7$, despite the absence of kinematic constraints.
Hence, the presence of a high-$z$ structure in the FFs for $T_{4Q}$ states can be interpreted as a genuine dynamical feature, reflecting the compactness of the bound state and the dominance of configurations where the fragmenting quark retains most of the momentum.
In this sense, our result confirms and extends the expected behavior for heavy-quark fragmentation into systems made exclusively of heavy constituents.

In contrast, the axial-vector heavy-quark FFs exhibit a sharply localized peak in the intermediate-to-large $z$ region, typically within $0.75 \lesssim z \lesssim 0.9$, and vanish in the limit $z \to 0$, for all values of $\mu_F$.
These functions are steeper and more collimated than those of the scalar and tensor cases, highlighting a strong preference for hard-fragmentation configurations and the reduced phase space available for soft emission.
The overall shape remains stable under scale evolution, especially in the peak region.

Finally, we observe that the FFs for the fully bottomed tetraquarks (right panels) are uniformly suppressed by approximately three orders of magnitude across the entire $z$ spectrum, when compared to their fully charmed counterparts (left panels).
This difference arises from the larger heavy-quark mass, which reduces the overall probability for fragmentation and shifts the dynamics toward even harder configurations.
Such a suppression overcomes the enhancement expected from the corresponding LDMEs, which are roughly 400 times larger for bottomed states (see Sec.~\ref{ssec:FFs_initial_scale}).
This confirms that the dominant control over the FF normalization is exerted by the perturbative coefficient, rather than by the nonperturbative matrix element.

Conversely, the $[g \to \TQQ]$ FFs exhibit a broader distribution, with significant support at lower values of $z$.
For scalar and tensor tetraquarks, the gluon FFs display a behavior that partially mirrors that of the corresponding heavy-quark FFs: they decrease monotonically with increasing $z$, starting from a finite value at $z = 0$, and do not develop any peak structure at large $z$.
In contrast, the axial-vector gluon FFs exhibit a distinct profile, characterized by a broad hump in the intermediate region, approximately within $0.15 \lesssim z \lesssim 0.7$, with a peak that shifts slightly toward higher $z$ as $\mu_F$ increases.
This structure becomes more pronounced under scale evolution, especially in the charm sector, and is accompanied by a strong suppression at both low and high $z$.

The qualitative difference between the axial-vector channel and the scalar or tensor ones is expected, since the gluon FF for $1^{+-}$ states is entirely generated via DGLAP evolution.
While this prevents us from probing the nonperturbative gluon input at the initial scale, it allows us to isolate and study the genuine effects of DGLAP dynamics.
The emergent hump structure thus provides insight into the intrinsic shape generated radiatively by the evolution itself, unaltered by model-dependent boundary conditions.

Finally, a similar suppression is observed for gluon-initiated FFs: those associated with fully bottomed tetraquarks (right panels) are approximately three orders of magnitude smaller across the entire $z$ spectrum compared to their fully charmed counterparts (left panels), consistently with what is found in the heavy-quark channel.

By comparing the relative magnitudes across the quantum numbers, we observe that the overall normalization of the axial-vector FFs is systematically lower than that of the scalar and tensor ones.
This suppression affects both $Q$ and $g$ channels , where the fragmentation probabilities toward the $1^{+-}$ state are reduced by more than one order of magnitude with respect to the scalar and tensor counterparts.
This hierarchy is consistent with the NRQCD picture and the features discussed in Sec.~\ref{ssec:FFs_initial_scale}.
In particular, the antisymmetric spin-color configuration of the $1^{+-}$ state leads to reduced overlap with the LO gluon production mechanisms and suppressed LDME values~\cite{Weng:2020jao,Feng:2023agq}.
Furthermore, the lack of orbital excitation in the scalar and tensor channels enhances their compatibility with collinear NRQCD selection rules, unlike the axial-vector case~\cite{Bodwin:2002cfe,Ma:2015yka,Xu:2021mju}.

As the factorization scale increases, all FFs experience a distortion of their $z$-shape.
For $Q$-initiated channels, the peak is preserved but slightly reduced in magnitude, while the tail at low $z$ becomes more populated.
This behavior reflects the typical softening induced by DGLAP evolution.
In the gluon channel, the growth of the low-$z$ tail with $\mu_F$ is evident, especially in the axial-vector cases.
Nevertheless, the shape distortion remains moderate even at the highest scale $\mu_F = 150$~GeV, confirming that the leading features of the fragmentation pattern are set at the input and preserved under evolution.

Finally, we remark that our extended analysis includes uncertainty bands for all quantum states, accounting for the combined effect of F-MHOUs and LDME uncertainties (added in quadrature) in the main panels, and displaying them separately in the corresponding ancillary panels below.
This completes and extends the partial picture offered in Ref.~\cite{Celiberto:2025dfe}, where error estimates were related to LDME variations only and limited to the $1^{+-}$ case.
Uncertainty bands for all the considered channels are moderate, reflecting the conservative strategy described in Sec.~\ref{ssec:FFs_initial_scale}.

\begin{figure*}[!t]
\centering

\begin{subfigure}[t]{0.475\textwidth}
\centering
   \hspace{-0.00cm} 
   \includegraphics[scale=0.400,clip]{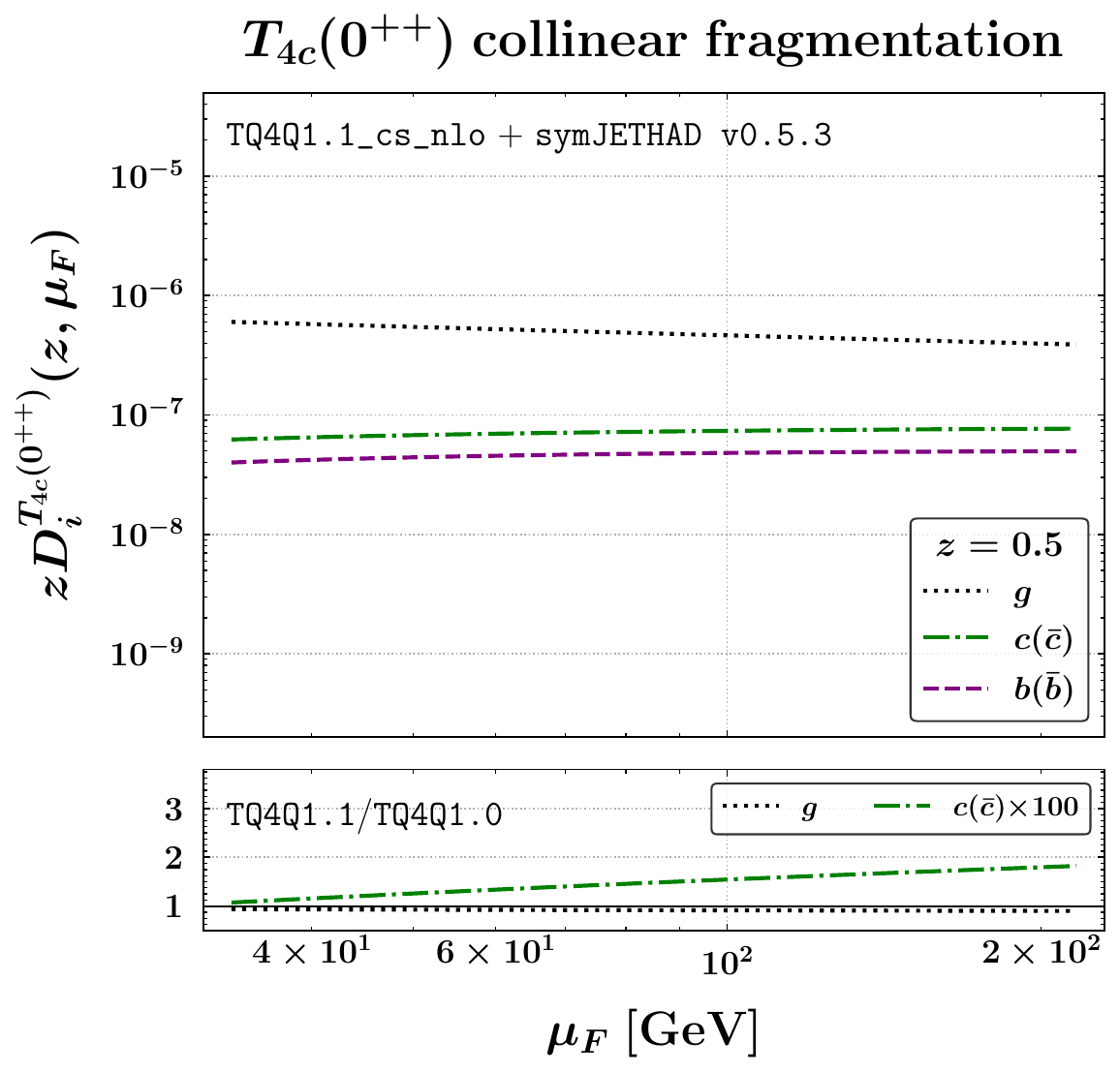}
\end{subfigure}
   \hspace{0.35cm}
\begin{subfigure}[t]{0.475\textwidth}
\centering
\raisebox{1.35cm}{
\includegraphics[scale=0.400,clip]{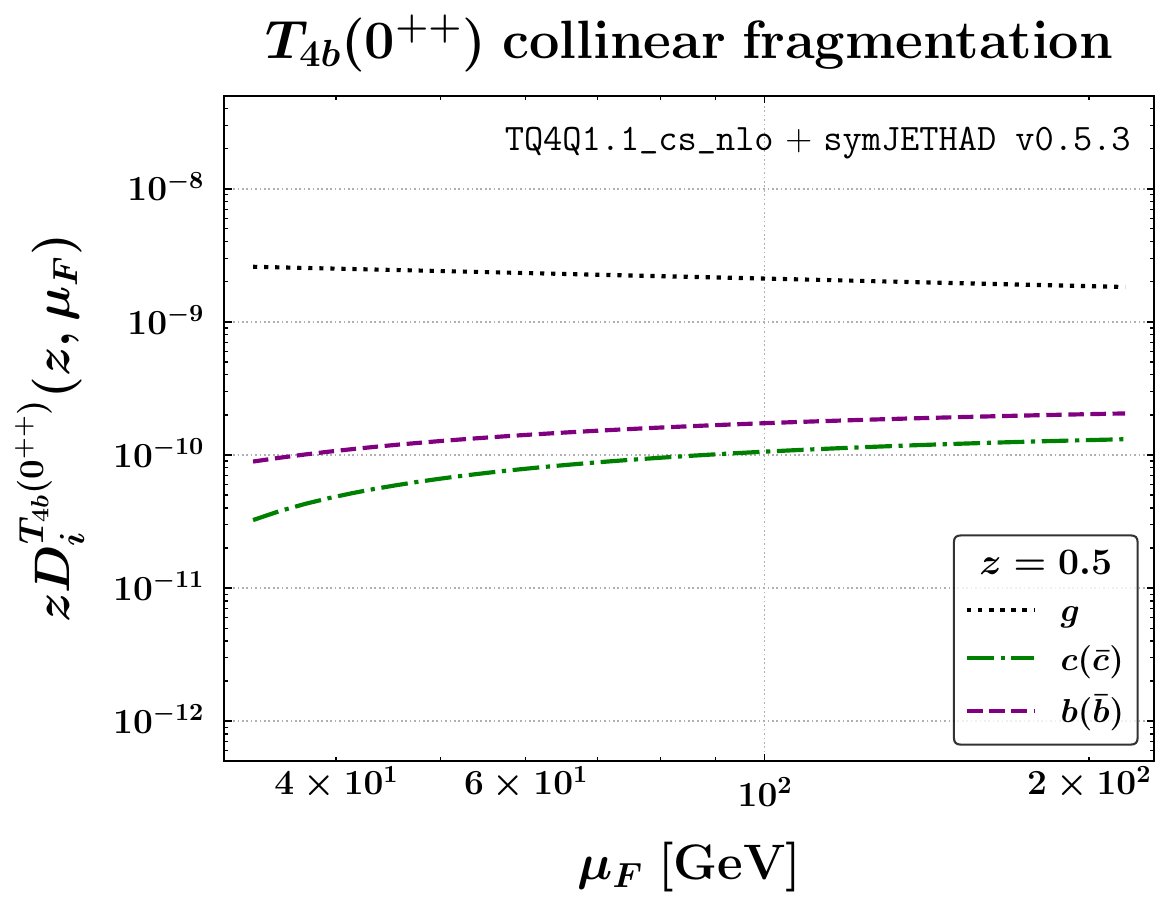}
}
\end{subfigure}

   \vspace{0.35cm}

\begin{subfigure}[t]{0.475\textwidth}
\centering
   \hspace{-0.00cm} 
   \includegraphics[scale=0.400,clip]{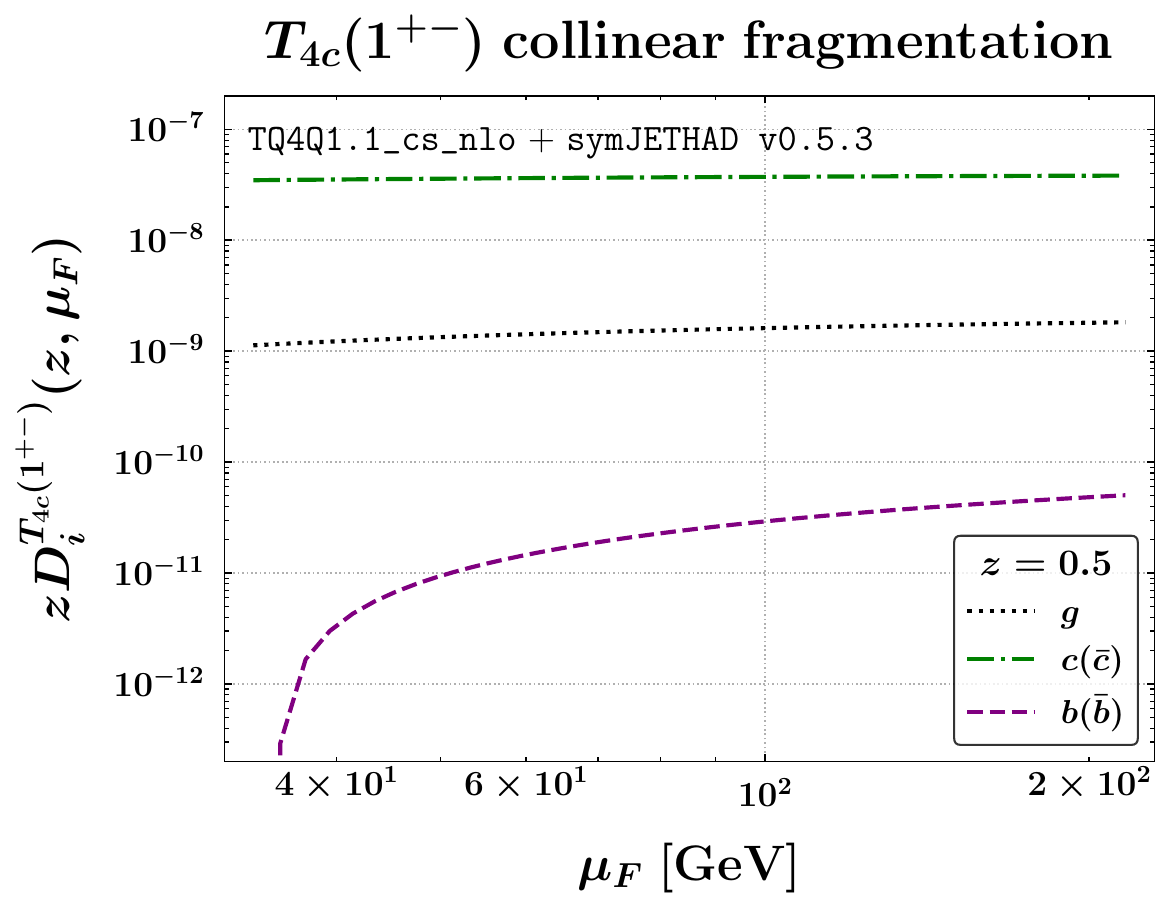}
\end{subfigure}
   \hspace{0.35cm}
\begin{subfigure}[t]{0.475\textwidth}
\centering
\raisebox{0.00cm}{
\includegraphics[scale=0.400,clip]{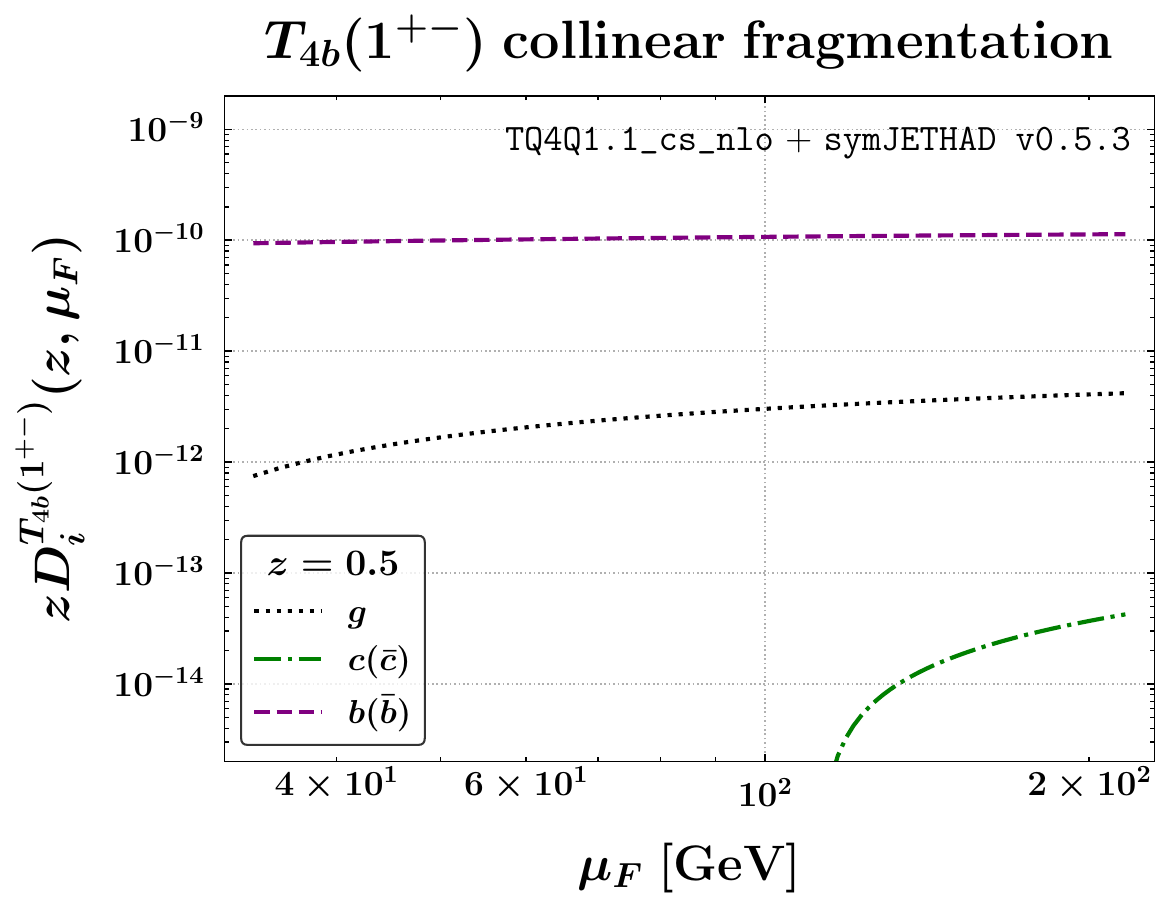}
}
\end{subfigure}

   \vspace{0.35cm}

\begin{subfigure}[t]{0.475\textwidth}
\centering
   \hspace{-0.00cm} 
   \includegraphics[scale=0.400,clip]{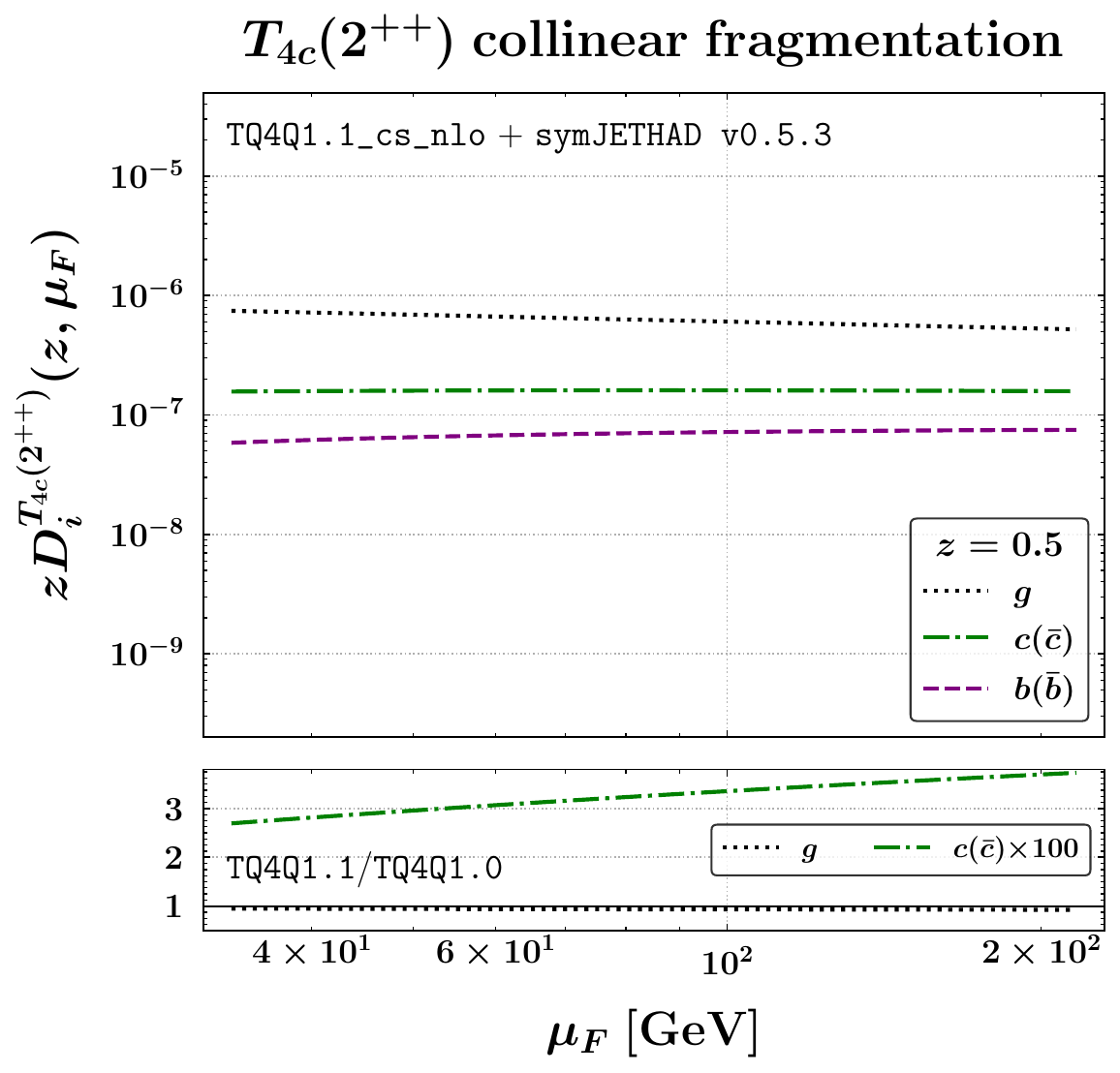}
\end{subfigure}
   \hspace{0.35cm}
\begin{subfigure}[t]{0.475\textwidth}
\centering
\raisebox{1.35cm}{
\includegraphics[scale=0.400,clip]{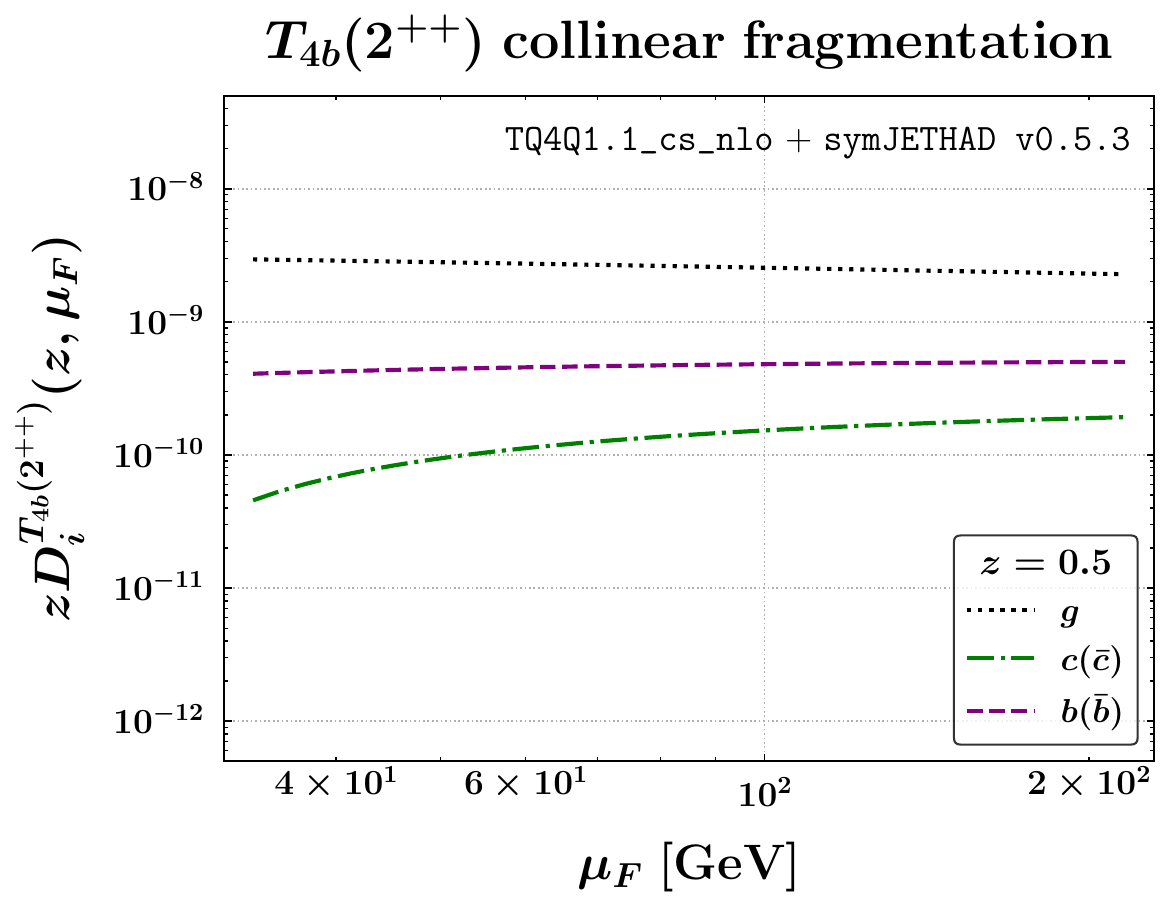}
}
\end{subfigure}

\caption{
\justifying
\noindent
Energy shape of the {\tt TQ4Q1.1} FFs for scalar (upper), axial-vector (central), and tensor tetraquarks (lower), evaluated at $z = \langle z \rangle \simeq 0.5$.
Left (right) panels are for fully charmed (bottomed) states.
Ratios between the new {\tt TQ4Q1.1} functions (this work) and the previous {\tt TQ4Q1.0} sets~\protect\cite{Celiberto:2024mab} are shown in the ancillary panels for scalar states.}
\label{fig:FFs-muF_TQQ}
\end{figure*}

\vspace{1em}
\noindent
\textbf{Energy dependence.}
For completeness, Fig.~\ref{fig:FFs-muF_TQQ} displays the energy dependence of the gluon, charm, and bottom {\tt TQ4Q1.1} FFs, multiplied by $z$, as a function of the factorization scale $\mu_F$.
The layout of the figure is organized by spin configuration, with scalar states shown in the upper panels, axial-vector in the middle, and tensor in the lower panels; the left columns correspond to fully charmed tetraquarks, while the right columns refer to fully bottomed ones.
The FFs are evaluated at a representative value of the momentum fraction, $z = \langle z \rangle \simeq 0.5$, which is typical in high-energy hadroproduction processes~\cite{Celiberto:2020wpk,Celiberto:2021dzy,Celiberto:2021fdp,Celiberto:2022dyf,Celiberto:2022keu,Celiberto:2024omj}.
For simplicity, only the central value of the FFs is shown, without propagating uncertainties.

We first examine the scalar channel ($0^{++}$, upper row), where both the gluon and the constituent heavy quarks are active at the input scale $Q_0$.
For the fully charmed case (left panel), the gluon FF dominates the spectrum across the full $\mu_F$ range, remaining nearly constant with a value of order $10^{-6}$.
This reflects the significant short-distance enhancement of the $[g \to \TQcZpp]$ SDC, and is consistent with the observed radiative stability under evolution.
The charm FF lies about one order of magnitude below, with a slightly increasing trend in $\mu_F$, as expected from timelike DGLAP evolution with positive splitting contributions from gluon radiation.
The evolution-generated bottom channel, although present, is further suppressed.
The pattern is qualitatively preserved in the fully bottomed case (right panel), although all curves are rescaled downward by approximately two orders of magnitude due to the heavier quark mass, which reduces both phase space and coupling strength at the input.
Here, the gluon FF retains its dominance, with the bottom FF as the subdominant contribution, and the evolution-generated charm FF being further suppressed, consistent with a mass-threshold origin.

Moving to the axial-vector channel ($1^{+-}$, middle row), a crucial structural difference emerges from the absence of the gluon initial-scale input, since no $[g \to \TQQ(1^{+-})]$ operator can be constructed in the color-singlet NRQCD framework at LO~\cite{Feng:2020riv,Yu:2022lak}.
This feature manifests clearly in the $\TQcOpm$ plot (left panel), where the gluon FF grows slowly through evolution, yet remains subdominant over the entire scale range.
The dominant contribution here is from the charm quark, with a flat profile driven by the relatively large input LDME and timelike DGLAP evolution at fixed $z$.
The $[b(\bar{b}) \to \TQcOpm]$ component is strongly suppressed and exhibits a logarithmiclike growth with $\mu_F$.
In the bottom sector (right panel), the pattern reverses, as expected; the bottom FF dominates, while the charm-induced one starts from zero and becomes visible only at $\mu_F \gtrsim 100$ GeV.
Analogously to the $\TQcOpm$ case, the gluon FF remains two to one orders of magnitude subdominant across the full scale range, further confirming that gluon-initiated fragmentation into axial tetraquarks is a radiatively generated and genuinely subleading channel.
This structural hierarchy, supported by the $[3,3]$ dominance in the color-spin basis~\cite{Bai:2024ezn}, is a distinctive signature of the $1^{+-}$ state and highlights its model-intrinsic stability.

Finally, the tensor channel ($2^{++}$, bottom row) exhibits a configuration broadly similar to the scalar one.
In both the fully charmed (left panel) and fully bottomed (right panel) cases, the gluon FF is the leading function across $\mu_F$, starting from a nonzero value at $Q_0$ and evolving very mildly.
As emphasized in Ref.~\cite{Feng:2022inv}, this behavior stems from the large number of contributing diagrams in the $[g \to \TQQ(2^{++})]$ SDC, due to the rich Lorentz structure of the final state.
The quark-induced FFs follow the same hierarchy observed in the scalar channel; charm dominates over bottom in the $\TQcTpp$ case, and the order is reversed in $\TQbTpp$, with cross-flavor contributions being significantly suppressed.
The overall $\mu_F$ dependence is again mild, with slightly steeper trends in the subdominant components, reflecting DGLAP-driven flavor mixing and radiative broadening at moderate $z$.

For comparison, the ancillary panels below the main plots in Fig.~\ref{fig:FFs-muF_TQQ} show the ratio between the {\tt TQ4Q1.1} (this work) and the previously released {\tt TQ4Q1.0} gluon and charm FFs (see Ref.~\cite{Celiberto:2024mab}).
These panels are provided only for the charmed scalar and tensor states, which were originally included in the {\tt TQ4Q1.0} release.
The charm FF is instead significantly suppressed. 
For $\TQcZpp$ states, the relative weight of the charm contribution remains below the percent level across scales (see the $\times 100$ rescaling shown in the panel), while for $\TQcTpp$ states it amounts to only a few percent, slowly increasing with $\mu_F$.
This pattern reflects the expected difference between NRQCD- and Suzuki-inspired FFs, with gluon fragmentation remaining dominant and charm contributions consistently subleading.
A dedicated comparison of the phenomenological consequences is provided in Sec.~\ref{ssec:I}.

An overarching feature that emerges from Fig.~\ref{fig:FFs-muF_TQQ} is the smooth behavior of the gluon FFs as the factorization scale $\mu_F$ increases.
In all spin configurations and for both charm and bottom sectors, the gluon FFs either show a mild growth with $\mu_F$ or display, at most, a very soft suppression, especially in the scalar and tensor cases.
This trend is not only theoretically expected from the structure of timelike DGLAP evolution, but it also plays a critical role in phenomenological stability.
Indeed, gluon FFs with such a regular $\mu_F$ pattern act as \emph{natural stabilizers}~\cite{Celiberto:2022grc} of high-energy distributions, particularly those sensitive to semi-inclusive emissions of heavy-flavored hadrons.

This concept of natural stability was first introduced in the context of singly charmed~\cite{Celiberto:2021dzy} and singly bottomed~\cite{Celiberto:2021fdp} mesons, and later confirmed in studies of vector quarkonia~\cite{Celiberto:2022dyf}, $B_c$-like states~\cite{Celiberto:2022keu,Celiberto:2024omj}, and early precision-QCD analyses of exotic hadrons~\cite{Celiberto:2023rzw,Celiberto:2024mab,Celiberto:2025dfe,Celiberto:2025ipt}.
Our results for the {\tt TQ4Q1.1} family reinforce this picture: the gluon channel, either initialized at the evolution-ready scale $Q_0$ (as in the scalar and tensor cases) or radiatively generated above it (as in the axial-vector case), contributes in a smooth and controlled fashion across the full $\mu_F$ range, ensuring robust theoretical predictions in the presence of NLO corrections and MHOUs (see Sec.~\ref{sec:phenomenology} for our phenomenological applications).

To conclude, the energy profiles of the {\tt TQ4Q1.1} FFs provide insight into the relative strength and scale sensitivity of different fragmentation channels.
The axial-vector configuration emerges as the cleanest and most theoretically stable scenario, due to its well-defined color-spin structure and radiative hierarchy.
The scalar and tensor cases, while quantitatively richer, are more model-dependent and sensitive to the chosen LDME set.
Altogether, this systematic characterization as a function of $\mu_F$ complements the $z$-dependent analysis presented previously, and sets the stage for phenomenological applications exotic hadron production at large transverse momenta.

\section{Hadron-collider phenomenology}
\label{sec:phenomenology}

To support phenomenology, we present predictions for rapidity and azimuthal-angle differential distributions sensitive to the inclusive production of tetraquark-jet systems at HL-LHC and FCC energies.
As mentioned, our reference formalism is the $\NLLp$ HyF, where the standard collinear factorization at NLO is consistently enhanced by including the resummation of high-energy logarithms within the next-to-leading accuracy and going partially beyond.

This hybrid approach is particularly suited for semihard observables that probe QCD at next-generation hadron colliders.
A comprehensive overview of the theoretical setup of the $\NLLp$ resummation formalism implemented in our analysis is provided in Appendix~\hyperlink{app:B}{B}.

The production of identified hadrons and jets provides a key window into the high-energy, semi-hard regime of QCD, where large logarithms of energy can disrupt standard perturbative expansions.\footnote{The semi-hard regime of QCD is defined by the scale hierarchy $\Lambda_{\rm QCD}^2 \ll Q^2 \ll s$, where $Q^2$ denotes a process typical hard scale and $s$ is the squared center-of-mass energy. In this regime, transverse momenta are large enough to allow for a perturbative treatment, but not asymptotically large. As a consequence, both collinear and high-energy logarithms can become sizable, and their joint resummation is required for reliable theoretical predictions.} 
The Balitsky-Fadin-Kuraev-Lipatov (BFKL) framework~\cite{Fadin:1975cb,Kuraev:1977fs,Balitsky:1978ic} addresses this by resumming both leading ($\alpha_s^n \ln s^n$) and next-to-leading ($\alpha_s^{n+1} \ln s^n$) energy logarithms.

In this context, cross sections are expressed as transverse-momentum convolutions of a universal NLO BFKL Green’s function~\cite{Fadin:1998py,Ciafaloni:1998gs} with process-dependent emission functions (also known as forward impact factors) that embed collinear inputs, including PDFs and FFs. 
This straightforwardly leads to the aforementioned HyF scheme that unifies high-energy resummation and collinear dynamics~\cite{Colferai:2010wu,Celiberto:2015yba,Celiberto:2017ptm,Celiberto:2020wpk}.

Over time, BFKL resummation has been widely applied to processes such as Mueller-Navelet jets \cite{Mueller:1986ey,Ducloue:2013hia,Colferai:2015zfa,Celiberto:2015yba,Celiberto:2015mpa,Celiberto:2016ygs,Celiberto:2017ius,Caporale:2018qnm,deLeon:2021ecb,Celiberto:2022gji,Baldenegro:2024ndr}, dihadron systems \cite{Celiberto:2016hae,Celiberto:2017ptm,Celiberto:2017ius,Celiberto:2020rxb,Celiberto:2022rfj}, hadron-jet \cite{Bolognino:2018oth,Bolognino:2019cac,Bolognino:2019yqj,Celiberto:2020wpk,Celiberto:2020rxb,Celiberto:2022kxx} and multijet tags \cite{Caporale:2015int,Caporale:2016soq,Caporale:2016xku,Celiberto:2016vhn,Caporale:2016zkc,Celiberto:2017ius}, forward Higgs \cite{Hentschinski:2020tbi,Celiberto:2022fgx,Celiberto:2020tmb,Mohammed:2022gbk,Celiberto:2023rtu,Celiberto:2023uuk,Celiberto:2023eba,Celiberto:2023nym,Celiberto:2023rqp,Celiberto:2022zdg,Celiberto:2024bbv}, Drell-Yan \cite{Celiberto:2018muu,Golec-Biernat:2018kem}, and heavy-flavor emissions \cite{Celiberto:2017nyx,Boussarie:2017oae,Bolognino:2019ouc,Bolognino:2019yls,Bolognino:2021mrc,Celiberto:2021dzy,Celiberto:2021fdp,Celiberto:2022dyf,Celiberto:2023fzz,Celiberto:2022grc,Bolognino:2022paj,Celiberto:2022keu,Celiberto:2022kza,Celiberto:2024omj,Gatto:2025kfl}.

Studies of single-forward emissions have offered key insights into small-$x$ gluon dynamics through unintegrated gluon distributions (UGDs), with analyses performed at HERA~\cite{Anikin:2011sa,Besse:2013muy,Bolognino:2018rhb,Bolognino:2018mlw,Bolognino:2019bko,Bolognino:2019pba,Celiberto:2019slj,Bolognino:2021bjd,Luszczak:2022fkf} and the EIC~\cite{Bolognino:2021niq,Bolognino:2021gjm,Bolognino:2021bjd,Bolognino:2022uty,Bolognino:2022ndh}.
This has led to the development of resummed PDFs~\cite{Ball:2017otu,Abdolmaleki:2018jln,Bonvini:2019wxf,Silvetti:2022hyc,Silvetti:2023suu,Rinaudo:2024hdb} and improved small-$x$ TMDs~\cite{Bacchetta:2020vty,Bacchetta:2024fci,Celiberto:2021zww,Bacchetta:2021oht,Bacchetta:2021lvw,Bacchetta:2021twk,Bacchetta:2022esb,Bacchetta:2022crh,Bacchetta:2022nyv,Celiberto:2022omz,Bacchetta:2023zir}.

Heavy-flavor emissions, such as ${\rm \Lambda}_c$~\cite{Celiberto:2021dzy} and $b$-hadron production~\cite{Celiberto:2021fdp}, have revealed strategies to overcome difficulties in modeling semihard processes at natural scales. 
Unlike light-hadron emissions, which suffer from large NLL corrections and threshold effects~\cite{Bolognino:2018oth,Celiberto:2020wpk}, heavy-flavored hadrons show a trend of \emph{natural stabilization}~\cite{Celiberto:2022grc}, stemming from VFNS-based collinear fragmentation.

This behavior motivated the construction of VFNS DGLAP-evolved FFs using NRQCD inputs~\cite{Braaten:1993mp,Zheng:2019dfk,Braaten:1993rw,Chang:1992bb,Braaten:1993jn,Ma:1994zt,Zheng:2019gnb,Zheng:2021sdo,Feng:2021qjm,Feng:2018ulg}, extending from vector quarkonia~\cite{Celiberto:2022dyf,Celiberto:2023fzz} to charmed $B$ mesons~\cite{Celiberto:2022keu,Celiberto:2024omj}.
Stability at high energies also opened new directions in exotic-hadron phenomenology, enabling fragmentation studies of doubly~\cite{Celiberto:2023rzw,Celiberto:2024beg} and fully heavy tetraquarks~\cite{Celiberto:2024mab,Celiberto:2024beg,Celiberto:2025dfe}, as well as charmed pentaquarks~\cite{Celiberto:2025ipt} and triply heavy baryons~\cite{Celiberto:2025ogy}.

In Sec.~\ref{ssec:uncertainty}, we outline our strategy for a systematic study of the uncertainties affecting the observables of interest.
In Secs.~\ref{ssec:I} and~\ref{ssec:I_phi} we present results for rapidity distributions and angular multiplicities, respectively.
All numerical calculations were performed by making use of {\Jethad}~\cite{Celiberto:2020wpk,Celiberto:2022rfj,Celiberto:2023fzz,Celiberto:2024mrq,Celiberto:2024swu}, a hybrid \textsc{Python}/\textsc{Fortran} multimodular interface designed for the computation, handling, and postprocessing of physical observables across multiple theoretical formalisms.

While our predictions are \emph{inclusive} in the final state (\emph{i.e.}, not restricted to specific tetraquark decay channels),  they remain fully compatible with existing strategies based on $[\Jpsi + \Jpsi]$ reconstruction. 
In this context, observables such as rapidity separation and jet angular multiplicity can provide valuable handles for signal isolation when used in conjunction with double‑quarkonium triggers. Similar approaches have been explored in past experimental studies: for instance, CMS investigated prompt double‑$\Jpsi$ production at 7~TeV~\cite{CMS:2014cmt}, and D0 at the Tevatron explored single from DPS contributions in double‑$\Jpsi$ production~\cite{D0:2014vql}. 
We believe that extending these analyses to encompass fully heavy tetraquark candidates is both feasible and timely.

\subsection{Uncertainty estimation}
\label{ssec:uncertainty}

A reliable phenomenological analysis requires a systematic assessment of theoretical uncertainties.  
In the present study, we explicitly disentangle and quantify the impact of the main sources of error that enter our framework, both of perturbative and nonperturbative origin.  
This decomposition allows us to evaluate not only the individual role of each contribution but also their combined effect on collider-level observables.  
In particular, we consider:
\begin{itemize}

 \item[$(i)$]
 \textbf{Perturbative H-MHOUs}. 
 These arise from the freedom in the choice of factorization and renormalization scales entering the partonic-subprocess hard factor.  
 Their variation around the central values by a factor $1/2$ to $2$ serves to gauge the impact of unknown subleading contributions.

 \item[$(ii)$]
 \textbf{Perturbative F-MHOUs}.
 These reflect the uncertainty in the perturbative initial conditions of the FFs at the starting scale.  
 As highlighted in Sec.~\ref{ssec:FFs_TQ4Q11}, for instance, the \emph{evolution-ready} scale $Q_0$ is varied around its natural value, $5m_Q$, by a factor $1/2$ to $2$, and the resulting envelope is taken as the associated uncertainty.  
 This procedure captures the impact of unknown subleading effects in the FF evolution.

 \item[$(iii)$]
 \textbf{Nonperturbative LDMEs}.  
 These encode the hadron-specific, long-distance dynamics of the hadronization process.  
 Uncertainties are estimated by varying the relevant LDMEs within ranges consistent with potential-model calculations, as discussed in Sec.~\ref{ssec:FFs_initial_scale}. 
 The resulting bands quantify the impact of hadronization-model ambiguities on collider-level observables.

 \item[$(iv)$]
 \textbf{Proton PDFs}. 
 These represent an additional source of uncertainty, since collinear proton PDFs are genuinely nonperturbative objects extracted from global fits to experimental data, unlike our proxy-model FFs.  
 Nevertheless, our dedicated numerical tests for tetraquark-jet production confirmed that variations among different PDF parametrizations or replicas remain below the $1\%$ level.  
 Accordingly, we restrict our analysis to the central member of the {\tt NNPDF4.0} set~\cite{NNPDF:2021uiq,NNPDF:2021njg}, without propagating the broader PDF-fit uncertainty commonly considered in global analyses, since it is subleading compared to the dominant MHOUs and LDME effects.

 \item[$(v)$]
 \textbf{Phase-space numerical integration}.
 The dominant source of numerical uncertainty originates from the multidimensional integration over the final-state phase space (see Eq.~\eqref{DY_distribution}) and over the Mellin variable $\nu$ (see Eqs.~\eqref{CnNLL}), \eqref{CnLL}), and~\eqref{CnHENLO}) in the Appendix~\hyperlink{app:B}{B}).  
 These integrals are performed through the native routines of {\tt JETHAD}, with errors systematically controlled below the $1\%$ level.  
 Subleading contributions arise from the one-dimensional integration over the partonic longitudinal momentum fractions $x$, which define the convolution of PDFs and FFs in the LO and NLO tetraquark emission function (see Eq.~\eqref{LOHEF}).  
 Dedicated tests confirm that these uncertainties are negligible compared to the dominant multidimensional integration.

\end{itemize}

\subsection{Rapidity distributions}
\label{ssec:I}

The first class of observables investigated in our phenomenological study is the rapidity distribution, defined as the cross section differential with respect to the rapidity separation $\DY = y_1 - y_2$ between the two final-state particles.
One has
\begin{equation}
\label{DY_distribution}
 \frac{\drv \sigma(\DY, s)}{\drv \DY} \, \equiv \, C_{n=0} \;,
\end{equation}
where $C_{n=0}$ stands for the $\varphi$-summed azimuthal-angle coefficient, integrated over the rapidity and momentum final-state phase space, and taken at fixed $\DY$ (see Appendix~\hyperlink{app:B}{B}).
By isolating the $n=0$ conformal spin, we suppress all transverse-angle modulations and retain the leading energy-dependent component of the cross section, which is particularly sensitive to resummation effects.

The phenomenological analysis of rapidity-interval distributions offers a direct insight into the interplay between high-energy dynamics and the collinear structure of hadrons, especially in semi-inclusive topologies involving a fully heavy tetraquark and a recoiling jet.

The kinematic cuts adopted in our phenomenological analysis are designed to reflect the acceptance and reconstruction capabilities of the CMS detector at the LHC. 
In particular, the rapidity range $|y_1| < 2.5$ for the tetraquark is based on the barrel calorimeter coverage~\cite{Chatrchyan:2012xg}, while the jet rapidity cut $|y_2| < 4.7$ corresponds to the extended acceptance provided by the end cap calorimeters~\cite{Khachatryan:2016udy}.
The transverse momentum of the $\TQQ(J^{PC})$ tetraquark is varied between $30$ and $120$~GeV, while the accompanying light jet spans from $40$~GeV to~$120$~GeV.
These ranges are compatible with those employed in current and projected LHC analyses involving hadronic and jet final states~\cite{Khachatryan:2016udy,Khachatryan:2020mpd}.
The use of \emph{asymmetric} transverse momentum cuts magnifies high-energy resummation effects over the fixed-order baseline~\cite{Celiberto:2015yba,Celiberto:2015mpa,Celiberto:2020wpk}.

\begin{figure*}[!t]
\centering

   \hspace{0.00cm}
   \includegraphics[scale=0.415,clip]{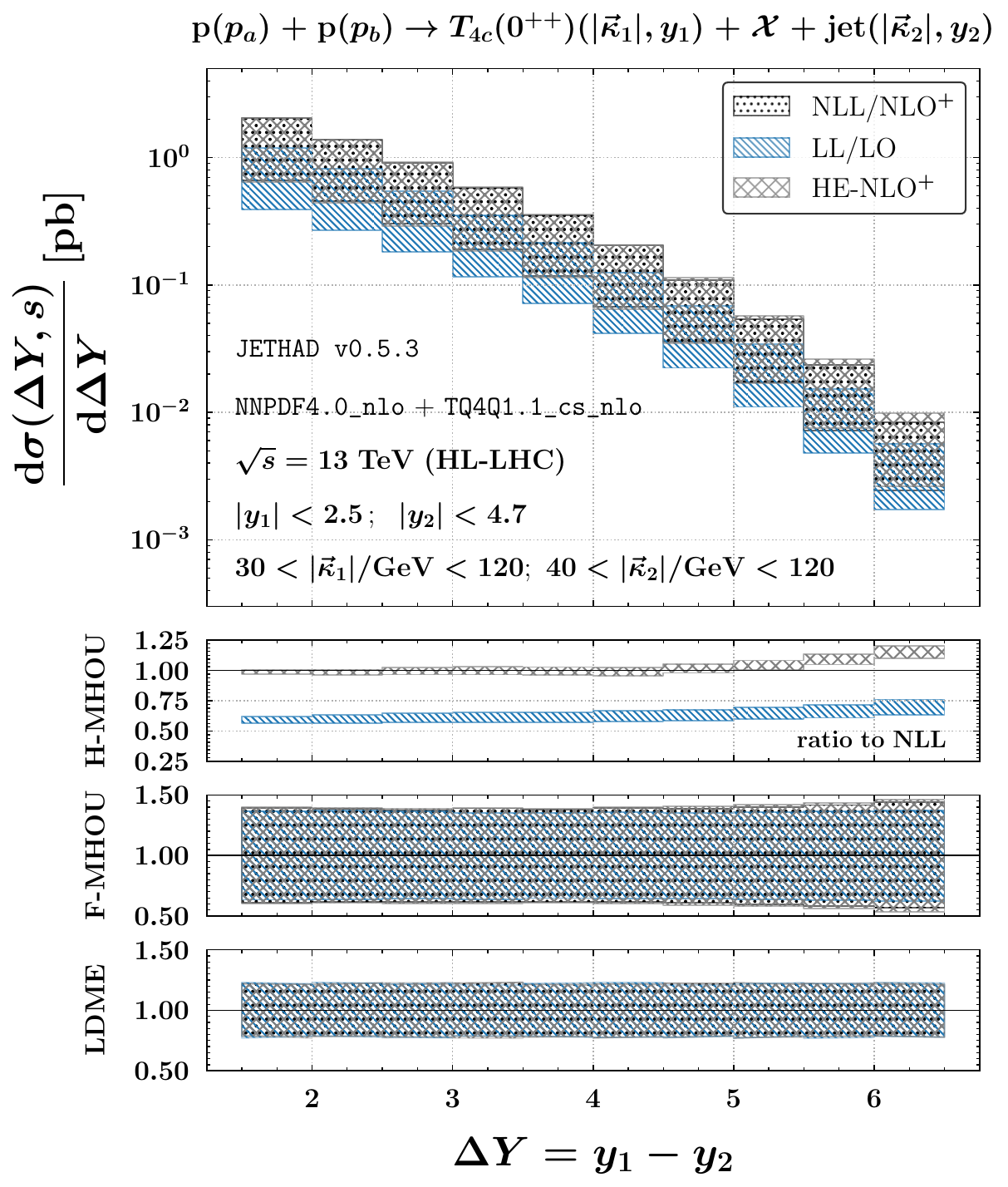}
   \hspace{-0.00cm}
   \includegraphics[scale=0.415,clip]{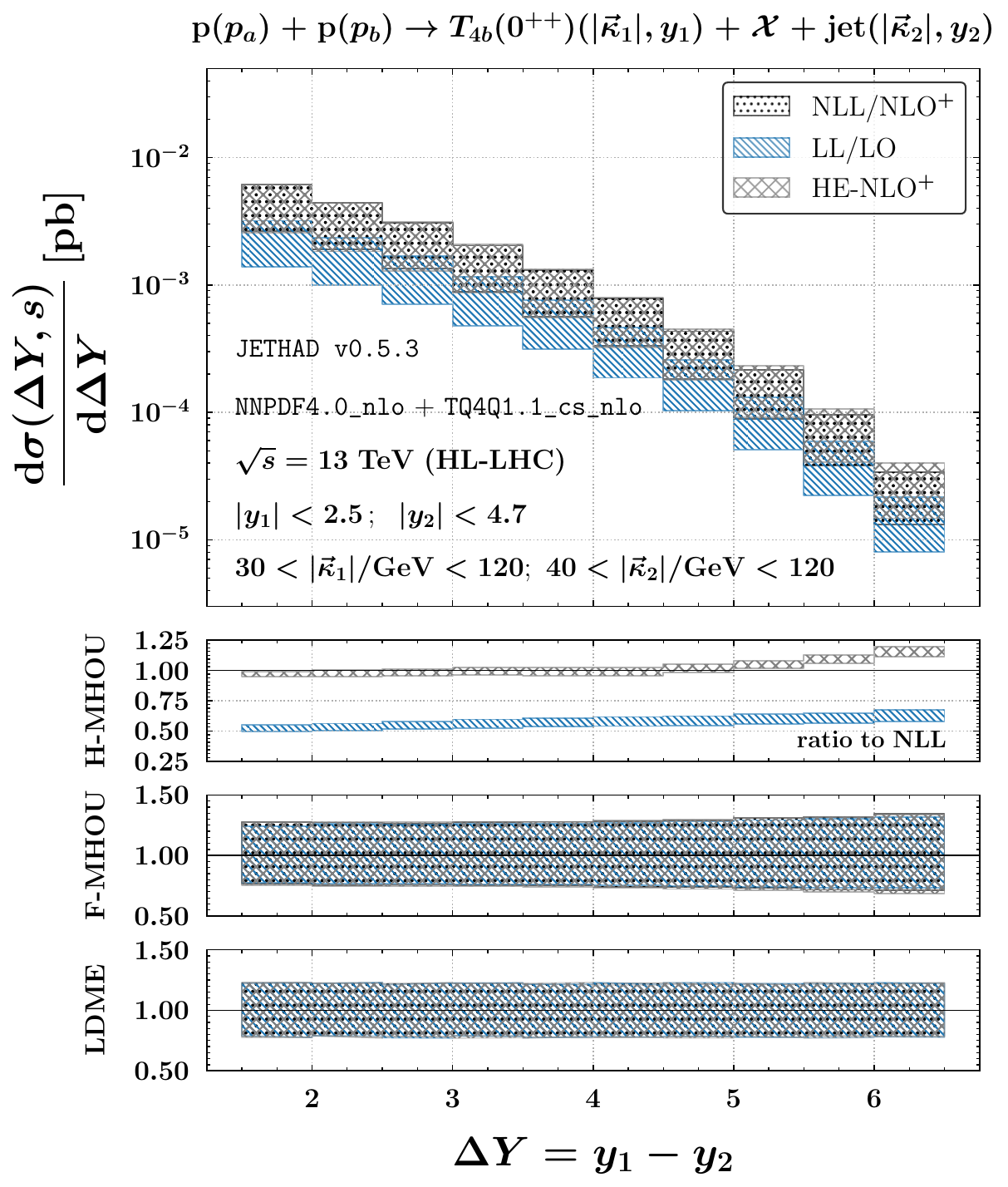}

   \vspace{0.35cm}

   \hspace{0.00cm}
   \includegraphics[scale=0.415,clip]{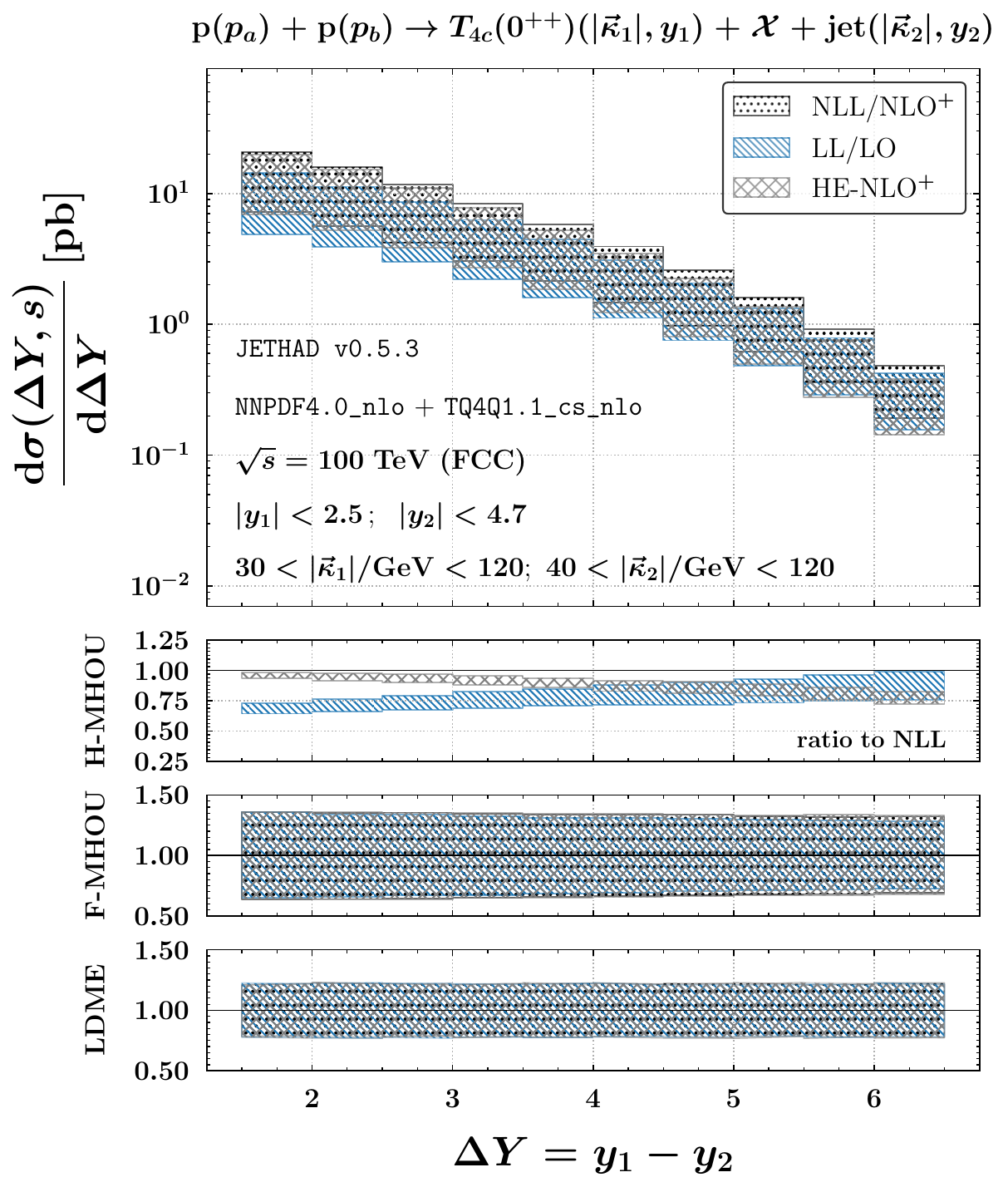}
   \hspace{-0.00cm}
   \includegraphics[scale=0.415,clip]{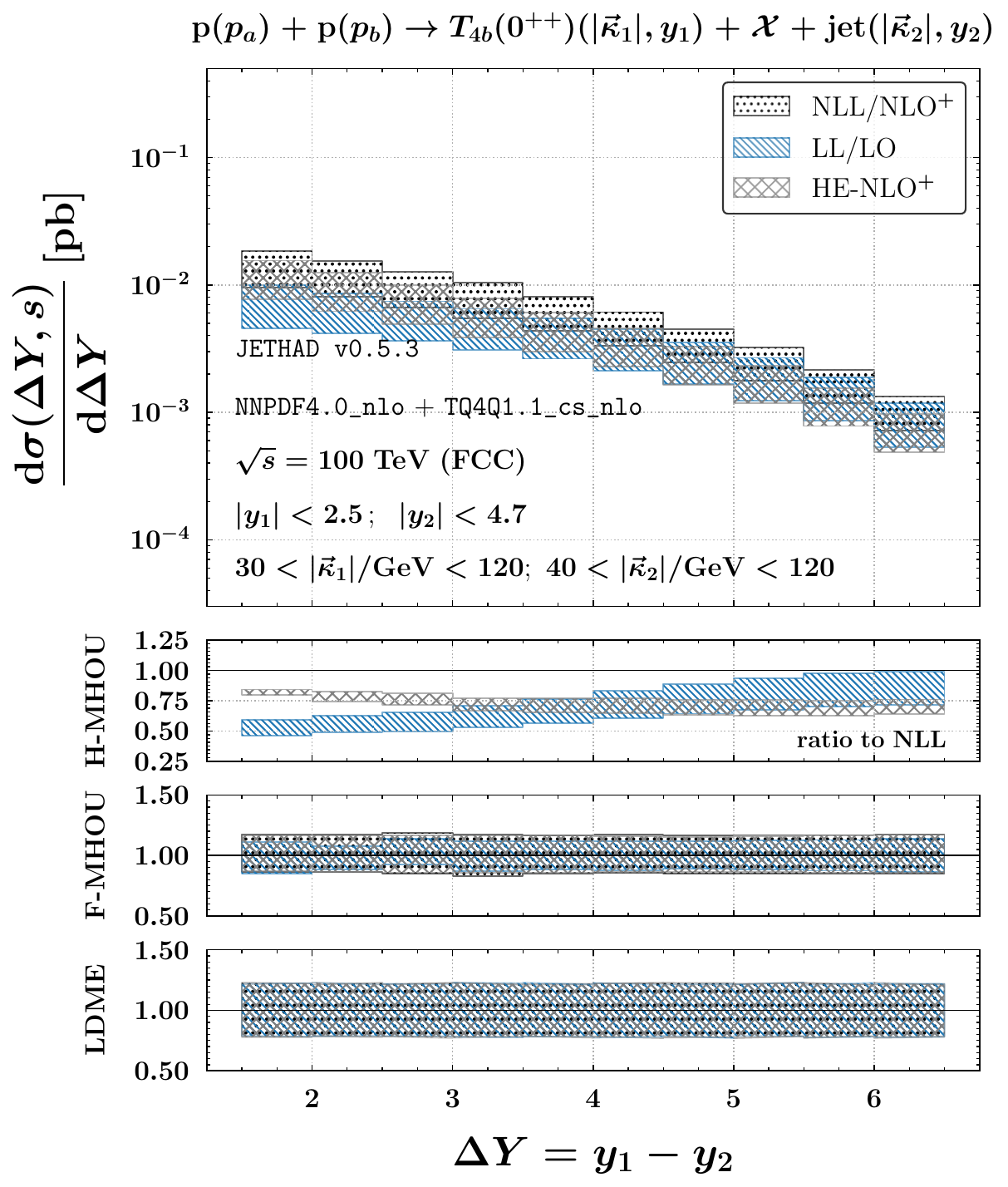}

\caption{
\justifying
\noindent
Rapidity distributions for scalar tetraquarks $\TQcZpp$ (left) and $\TQbZpp$ (right) produced in association with a jet at $\sqrt{s} = 13$ TeV (HL-LHC, top) and $100$ TeV (nominal FCC, bottom). 
Filled bands in the main panels indicate the total uncertainty, obtained by combining H-MHOUs, F-MHOUs, LDME variations, and phase-space integration.
Ancillary panels: $(i)$ ratios of $\LL$ and $\HENLOp$ to the $\NLLp$ baseline with H-MHOUs only; $(ii)$ F-MHOUs as the replica envelope normalized to the central curve; $(iii)$ LDME uncertainties as ratios to the central value.
}
\label{fig:I_TQ0}
\end{figure*}

\begin{figure*}[t]
\centering

   \hspace{0.00cm}
   \includegraphics[scale=0.415,clip]{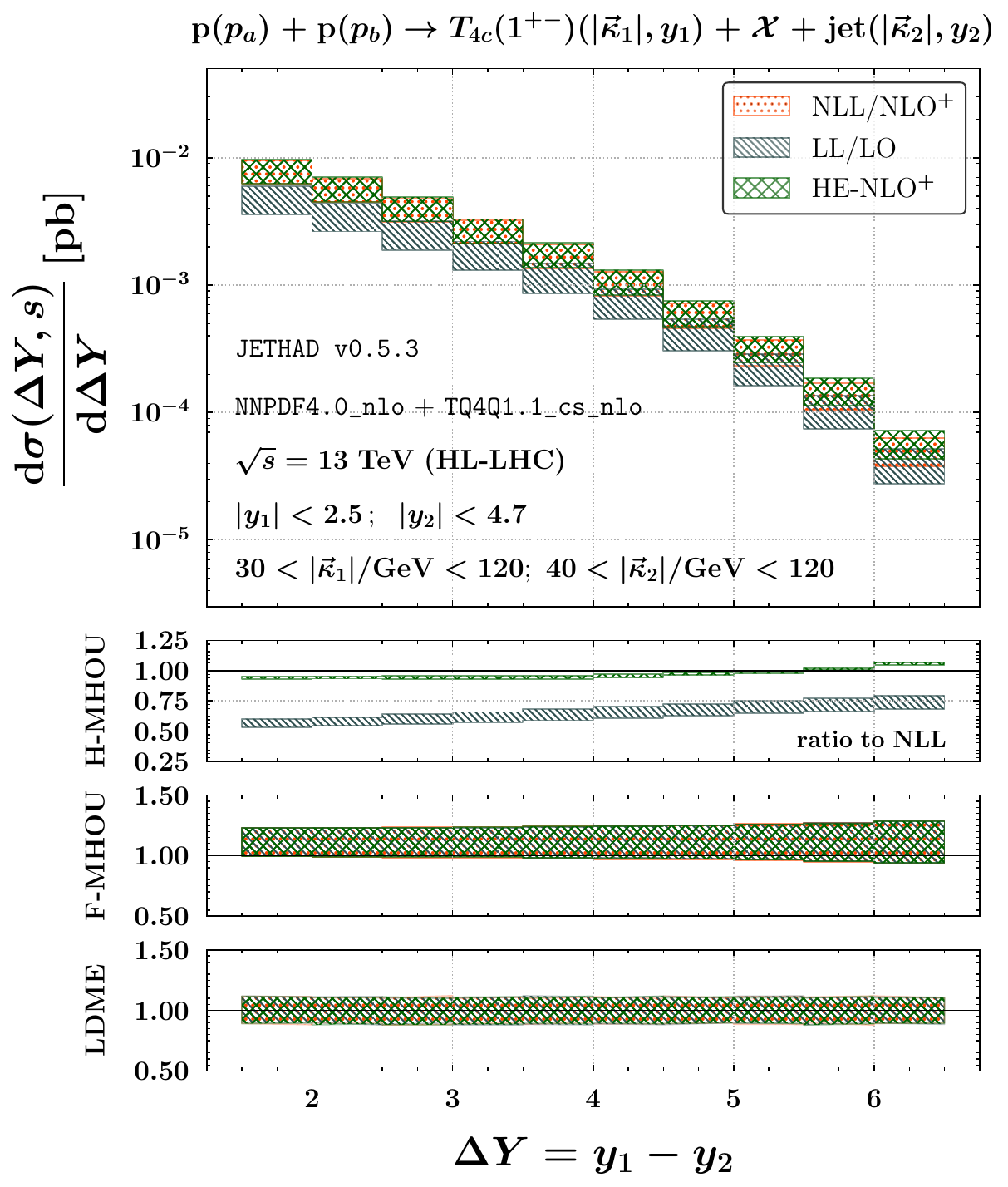}
   \hspace{-0.00cm}
   \includegraphics[scale=0.415,clip]{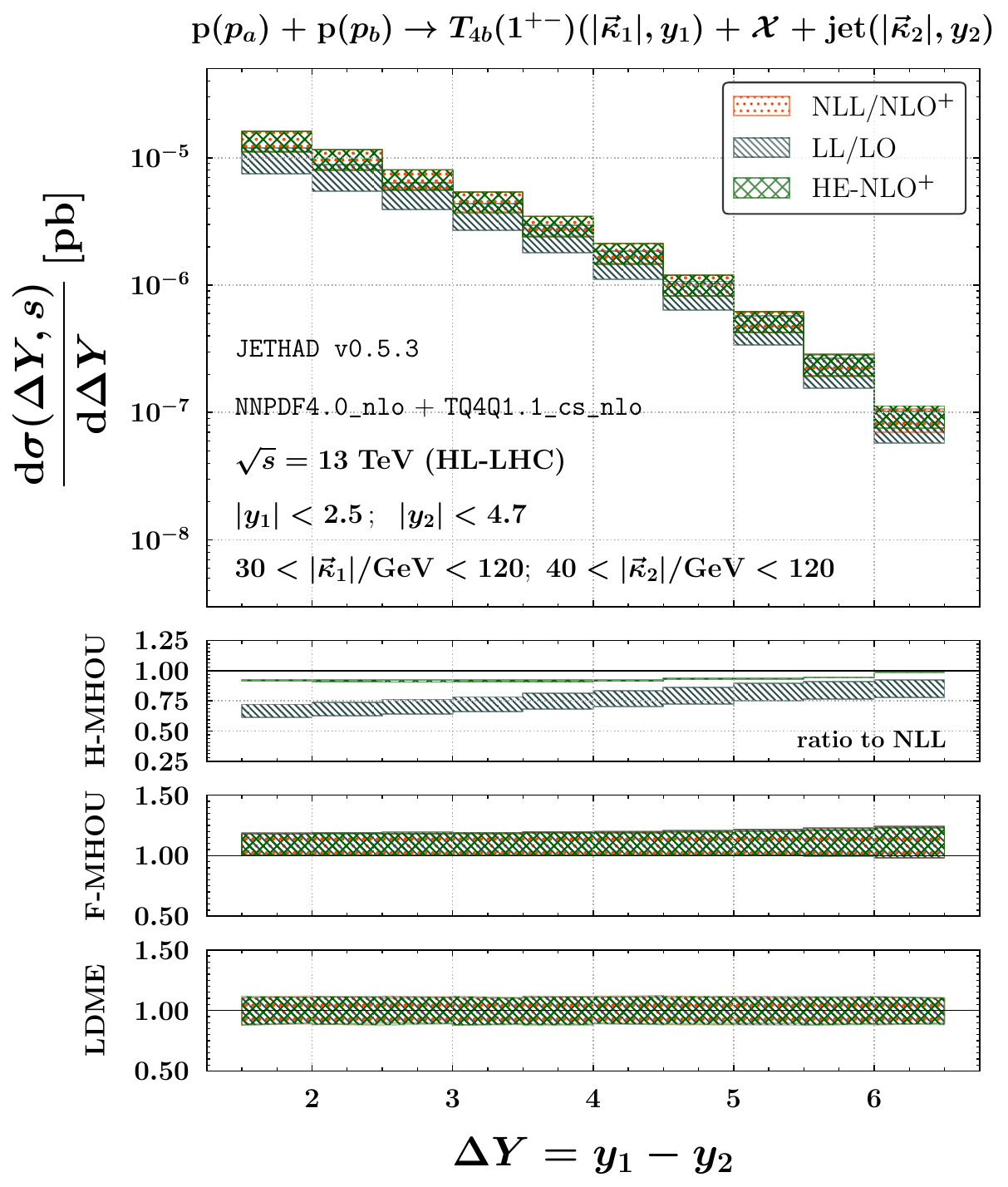}

   \vspace{0.35cm}

   \hspace{0.00cm}
   \includegraphics[scale=0.415,clip]{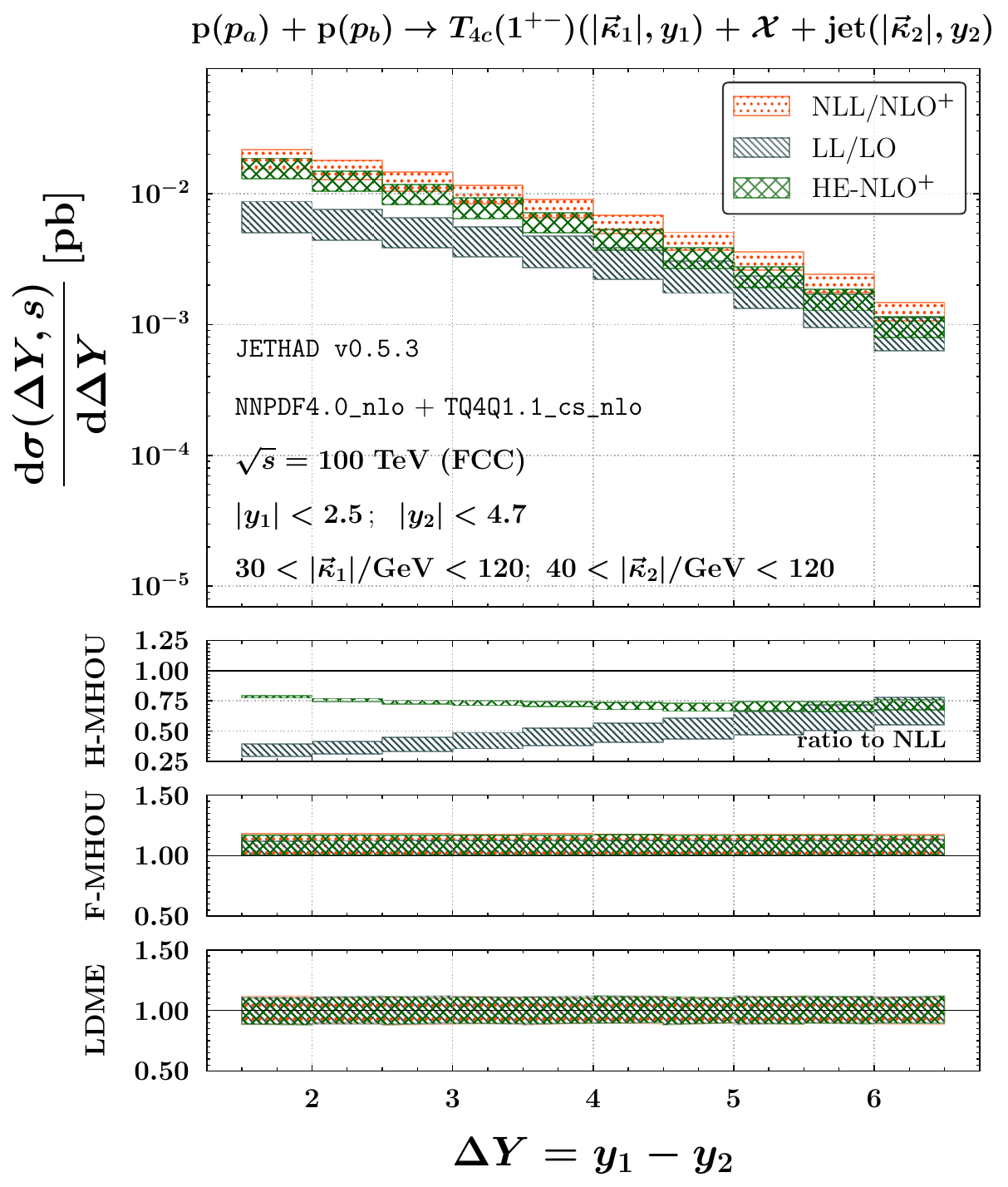}
   \hspace{-0.00cm}
   \includegraphics[scale=0.415,clip]{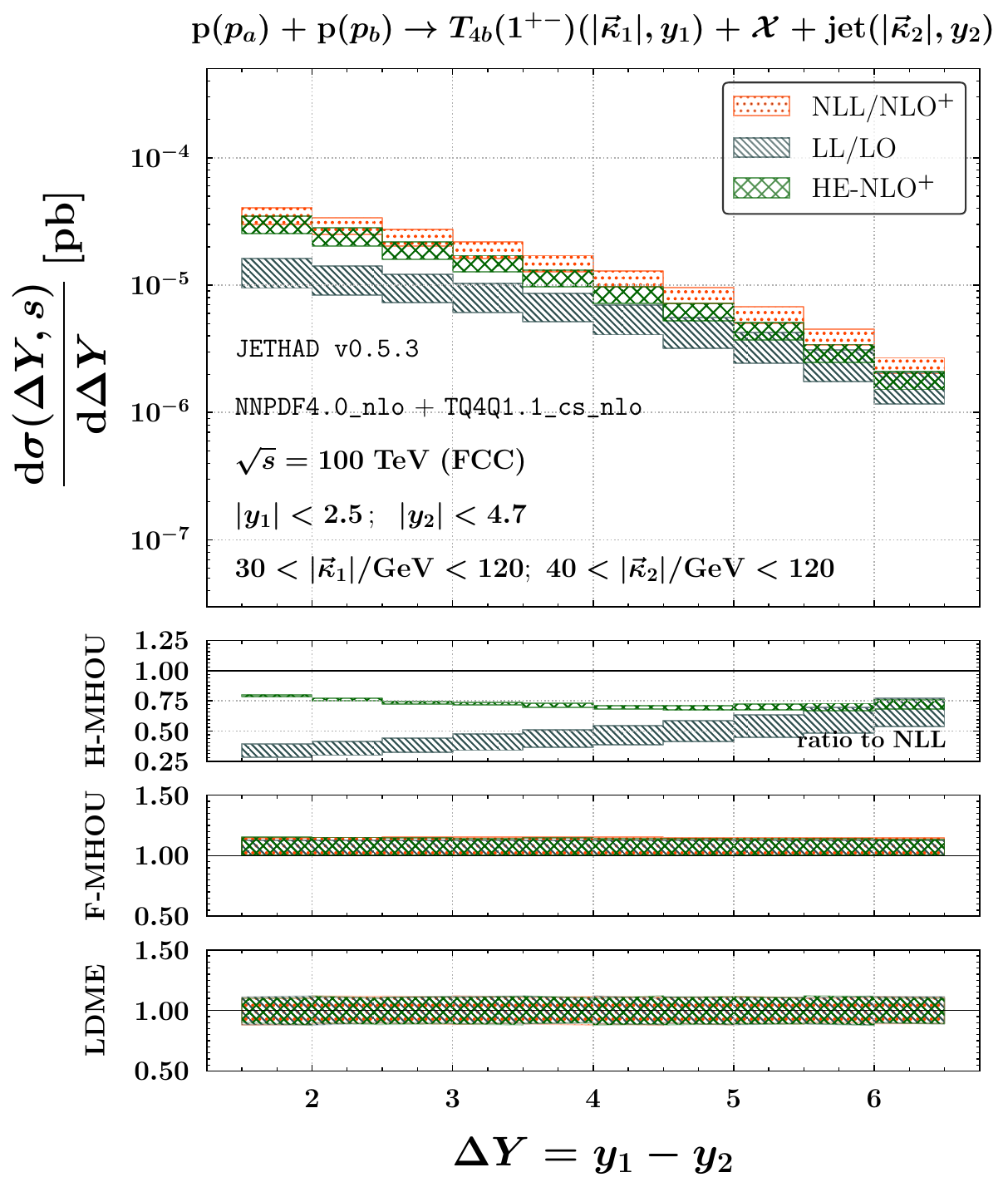}

\caption{
\justifying
\noindent
Rapidity distributions for axial-vector tetraquarks $\TQcOpm$ (left) and $\TQbOpm$ (right) produced in association with a jet at $\sqrt{s} = 13$ TeV (HL-LHC, top) and $100$ TeV (nominal FCC, bottom). 
Filled bands in the main panels indicate the total uncertainty, obtained by combining H-MHOUs, F-MHOUs, LDME variations, and phase-space integration.
Ancillary panels: $(i)$ ratios of $\LL$ and $\HENLOp$ to the $\NLLp$ baseline with H-MHOUs only; $(ii)$ F-MHOUs as the replica envelope normalized to the central curve; $(iii)$ LDME uncertainties as ratios to the central value.
}
\label{fig:I_TQ1}
\end{figure*}

\begin{figure*}[!t]
\centering

   \hspace{0.00cm}
   \includegraphics[scale=0.415,clip]{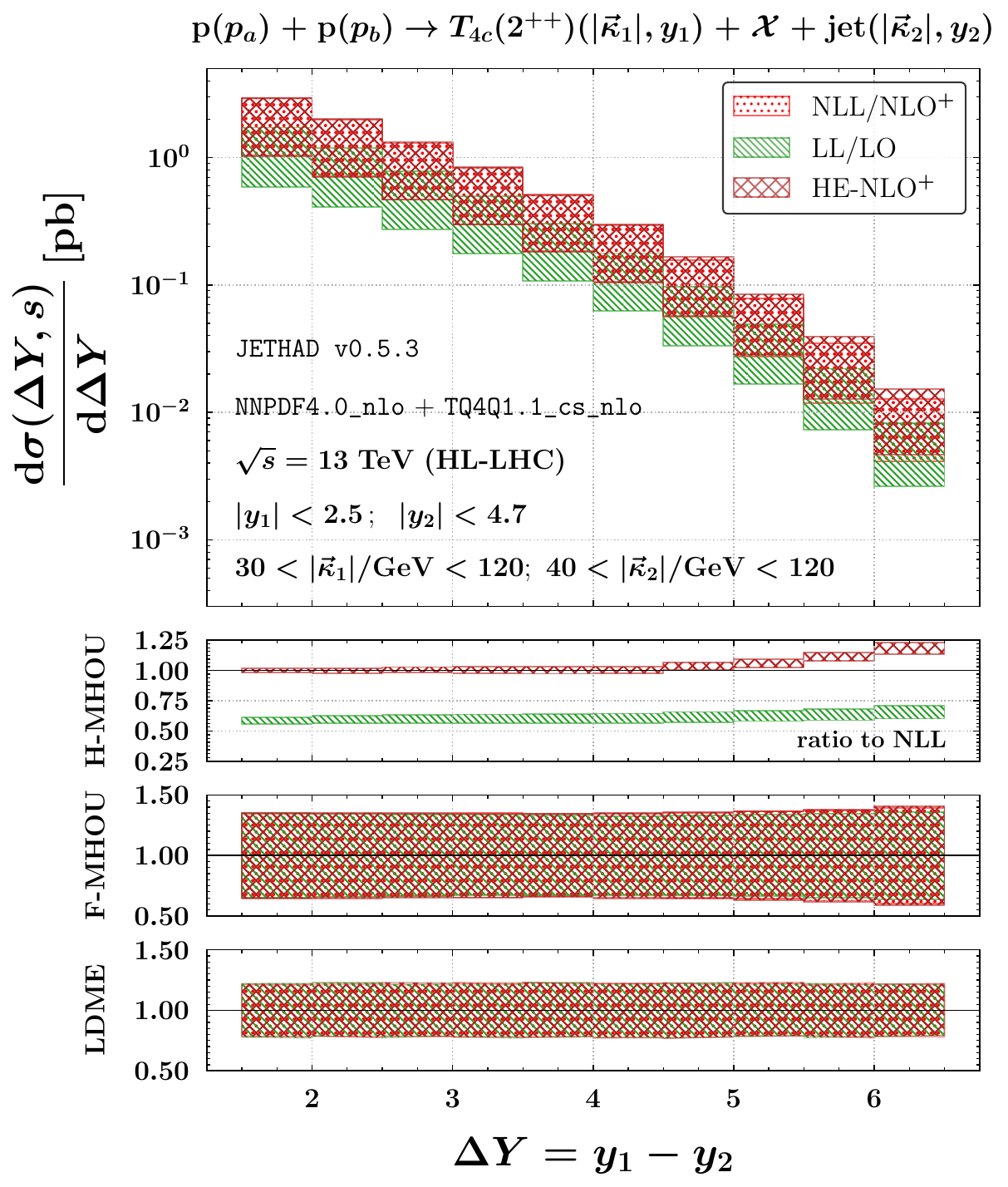}
   \hspace{-0.00cm}
   \includegraphics[scale=0.415,clip]{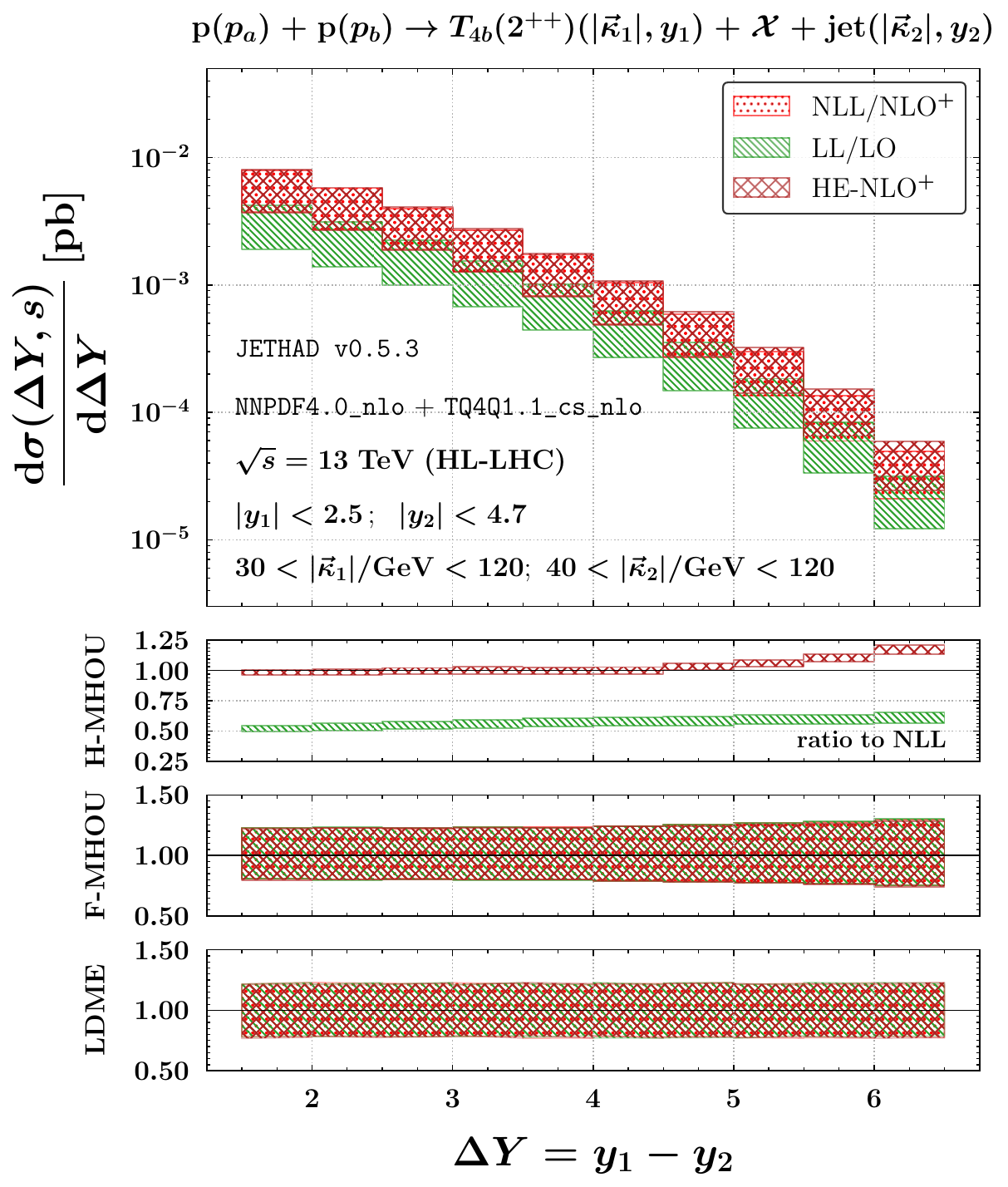}

   \vspace{0.35cm}

   \hspace{0.00cm}
   \includegraphics[scale=0.415,clip]{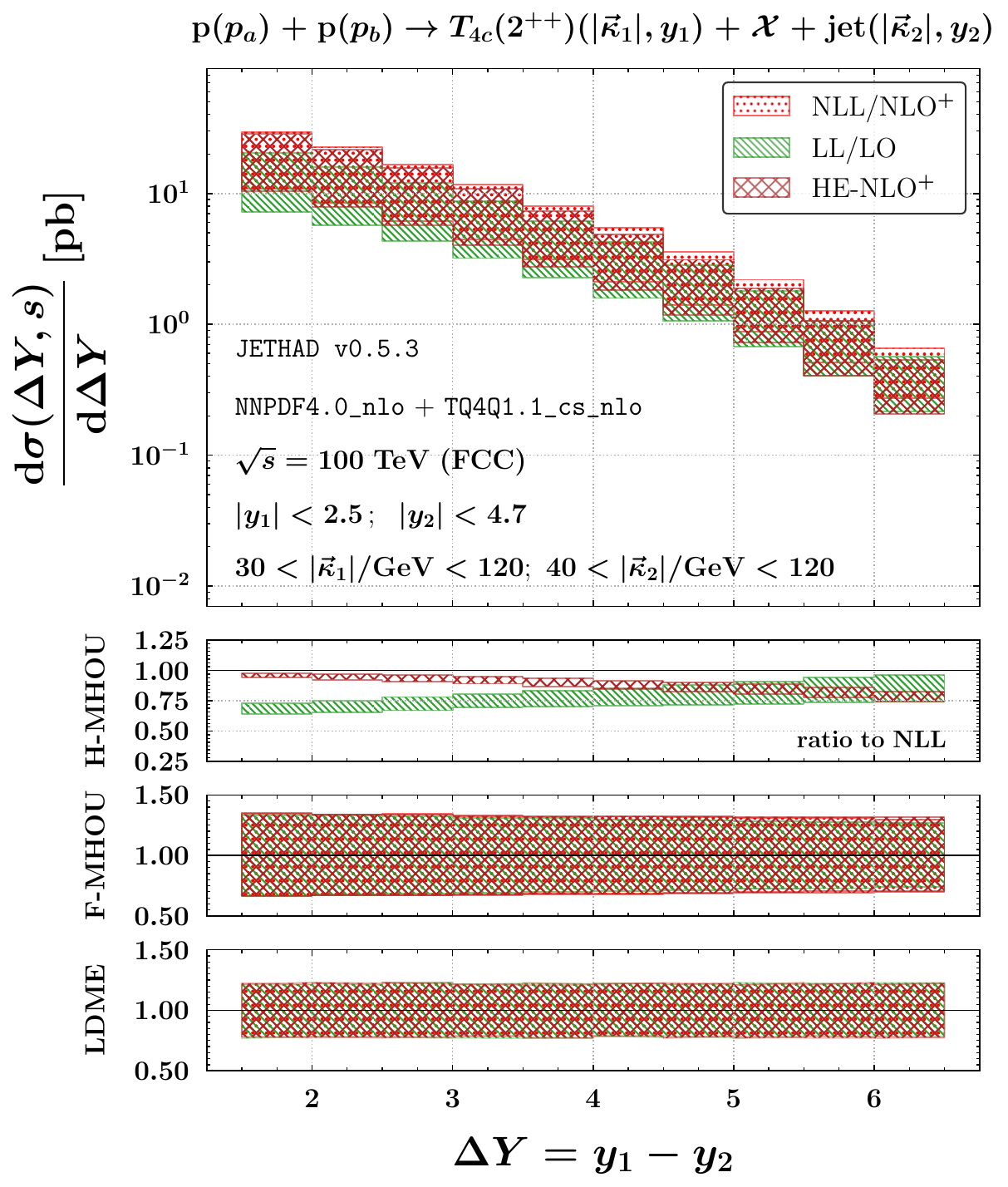}
   \hspace{-0.00cm}
   \includegraphics[scale=0.415,clip]{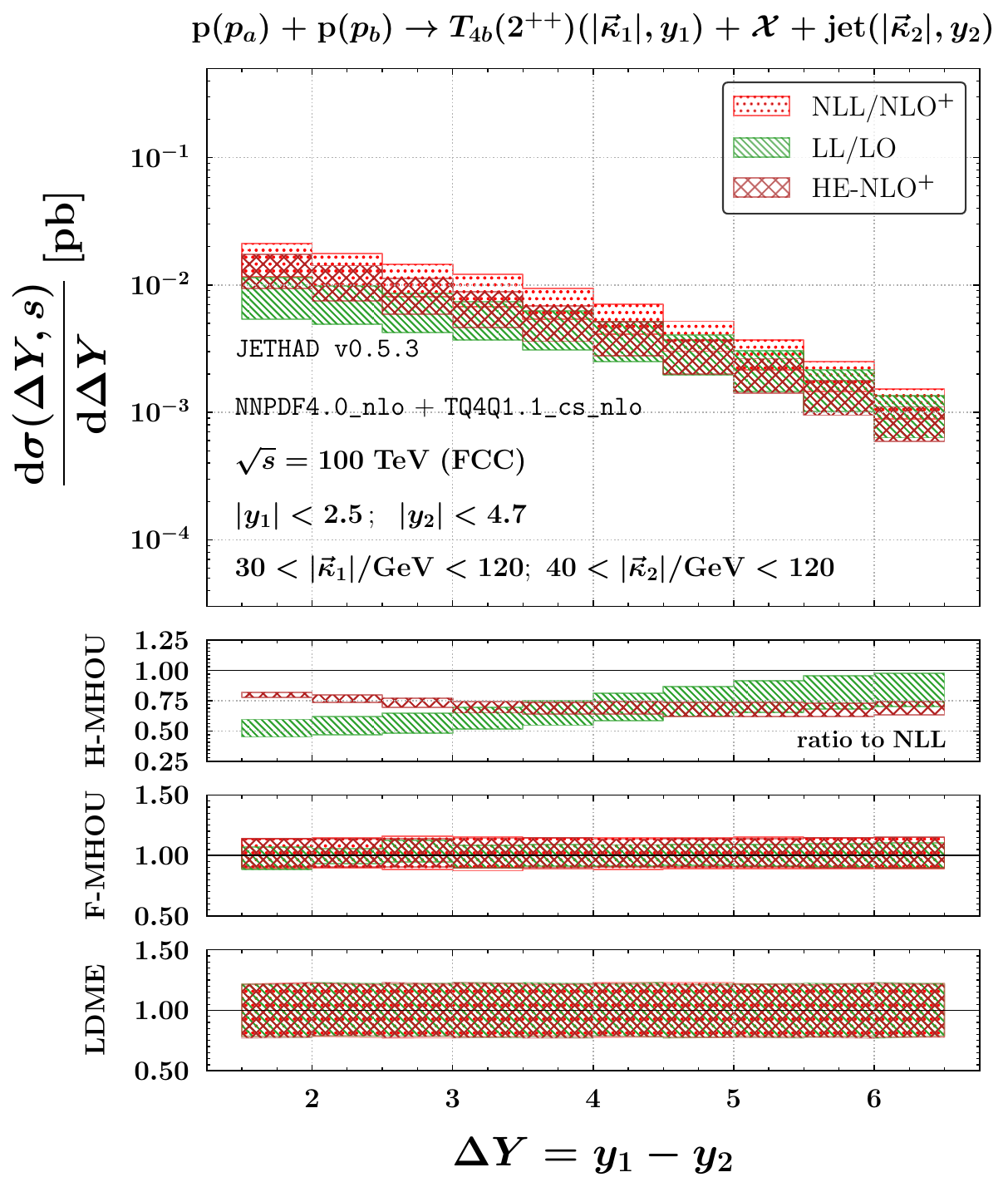}

\caption{
\justifying
\noindent
Rapidity distributions for tensor tetraquarks $\TQcTpp$ (left) and $\TQbTpp$ (right) produced in association with a jet at $\sqrt{s} = 13$ TeV (HL-LHC, top) and $100$ TeV (nominal FCC, bottom). 
Filled bands in the main panels indicate the total uncertainty, obtained by combining H-MHOUs, F-MHOUs, LDME variations, and phase-space integration.
Ancillary panels: $(i)$ ratios of $\LL$ and $\HENLOp$ to the $\NLLp$ baseline with H-MHOUs only; $(ii)$ F-MHOUs as the replica envelope normalized to the central curve; $(iii)$ LDME uncertainties as ratios to the central value.
}
\label{fig:I_TQ2}
\end{figure*}

In Figs.~\ref{fig:I_TQ0} to~\ref{fig:I_TQ2} we show our theoretical predictions for $\drv \sigma / \drv \DY$ in the production of scalar ($0^{++}$), axial-vector ($1^{+-}$), and tensor ($2^{++}$) tetraquarks with charm (left) or bottom content (right), in association with a light jet.
Upper and lower panels refer to HL-LHC ($\sqrt{s} = 13$~TeV) and FCC ($\sqrt{s} = 100$~TeV) configurations, respectively.
Each plot displays the absolute differential cross section in the main panel and the ratio of LL and HE-NLO$^+$ to $\NLLp$ results in the ancillary panel below.
Uncertainty bands in the main plots combine H-MHOUs, F-MHOUs, LDMEs, and multidimensional phase-space integration errors, added in quadrature. 
The three ancillary panels below each main plot separately show: 
$(i)$ the ratios of $\LL$ and $\HENLOp$ predictions to the $\NLLp$ baseline, with bands reflecting H-MHOUs only; 
$(ii)$ the effect of F-MHOUs, displayed as the envelope of replicas normalized to the central one; 
$(iii)$ the impact of LDME variations, again shown as ratios with respect to the central value.

The bin width in $\DY$ is set to $0.5$ throughout.

We note a universal trend across all final states: the cross section decreases with increasing $\DY$.
This behavior stems from a balance between two competing mechanisms.
On the one hand, the resummed partonic coefficient grows with energy (\emph{i.e.}, with $\DY$), as predicted by the BFKL formalism.
On the other hand, this growth is counteracted by the convolution with collinear PDFs and FFs, which are strongly suppressed at large momentum fractions.
As a result, the observable exhibits a relatively moderate decrease at small $\DY$ values, say $\DY \sim 2.5 \div 3$, and a steady decline thereafter.

Despite this shared trend, some differences emerge when comparing the three channels.

\begin{figure*}[!t]
\centering

   \hspace{0.00cm}
   \includegraphics[scale=0.415,clip]{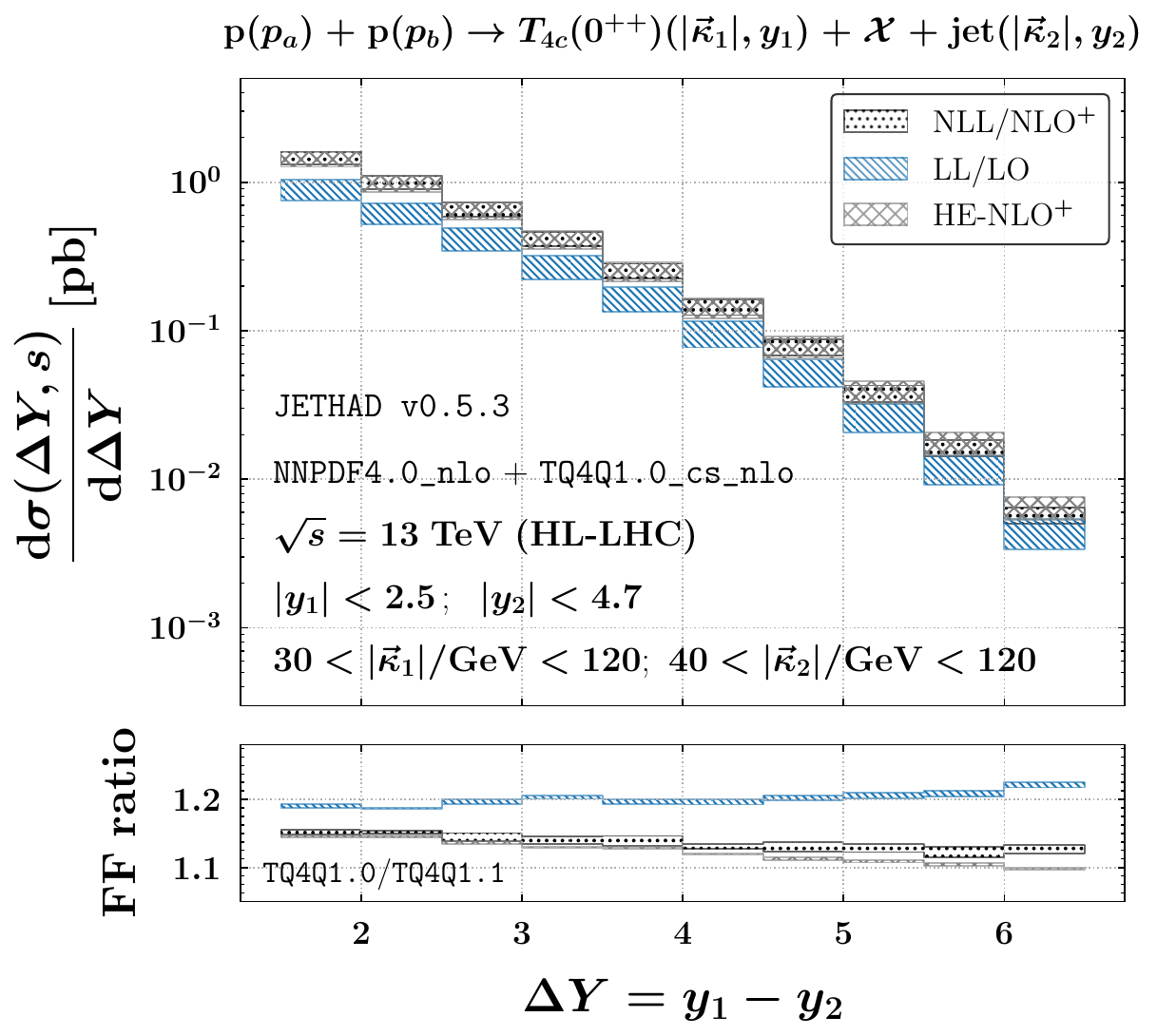}
   \hspace{-0.00cm}
   \includegraphics[scale=0.415,clip]{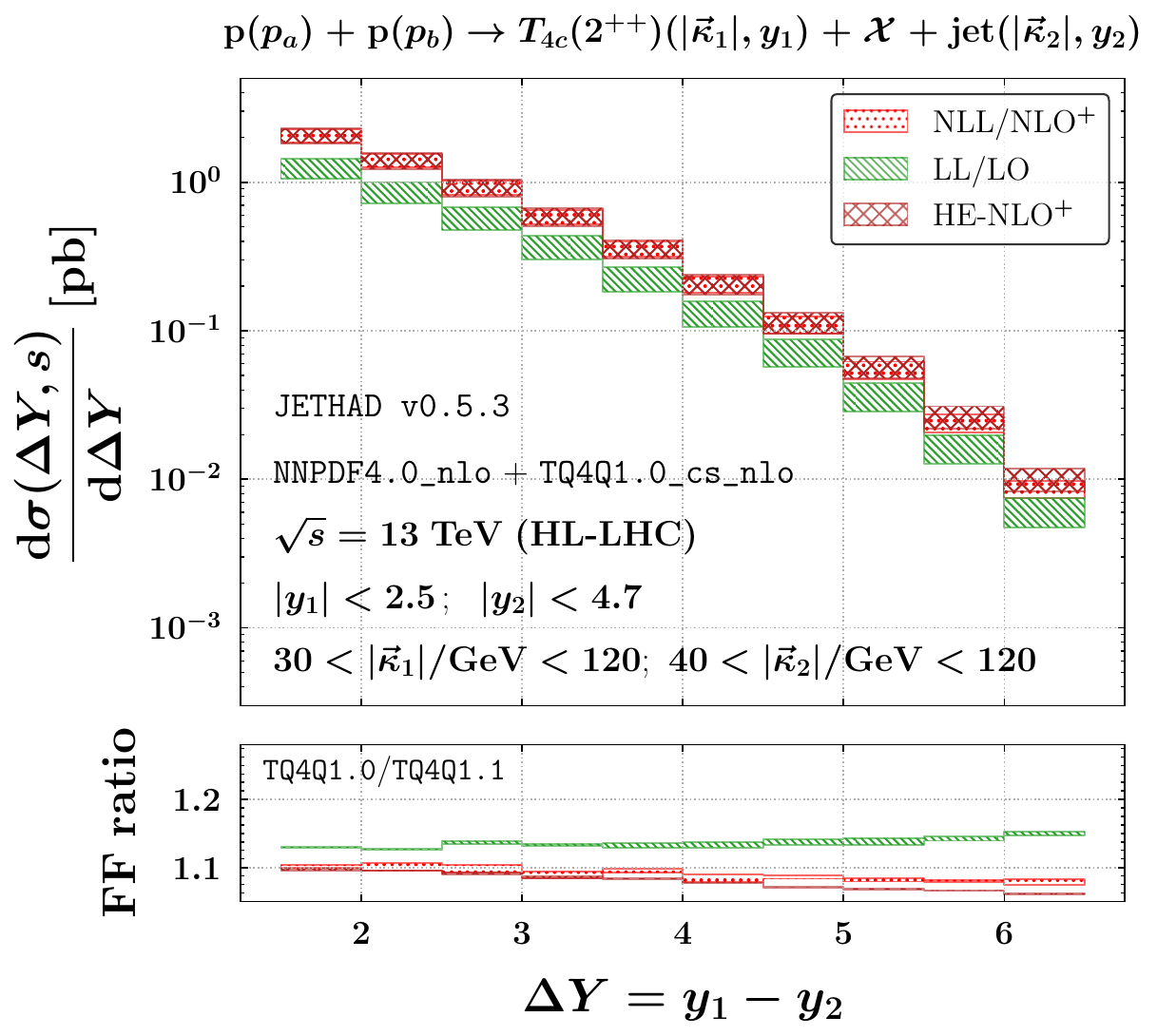}

   \vspace{0.35cm}

   \hspace{0.00cm}
   \includegraphics[scale=0.415,clip]{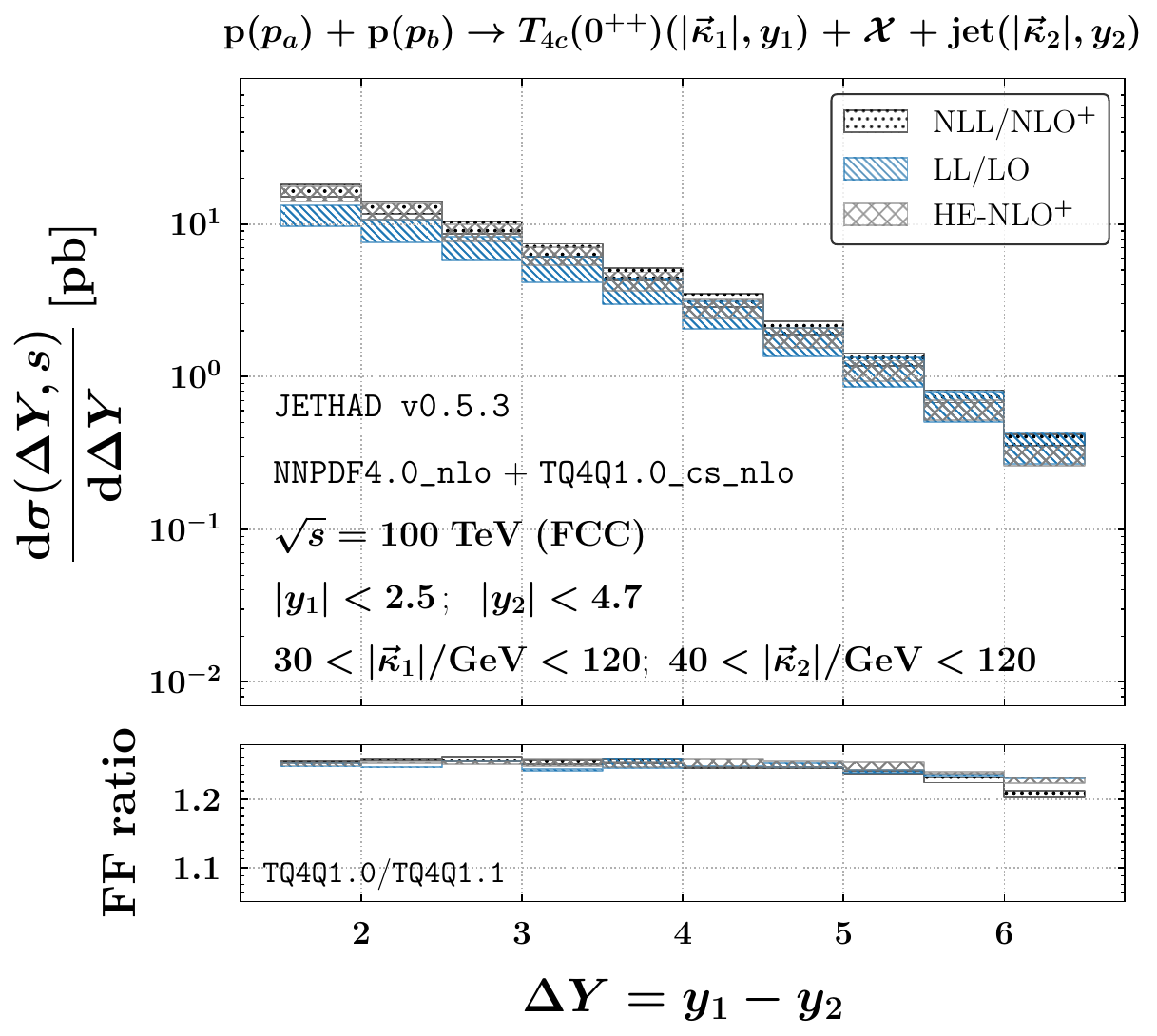}
   \hspace{-0.00cm}
   \includegraphics[scale=0.415,clip]{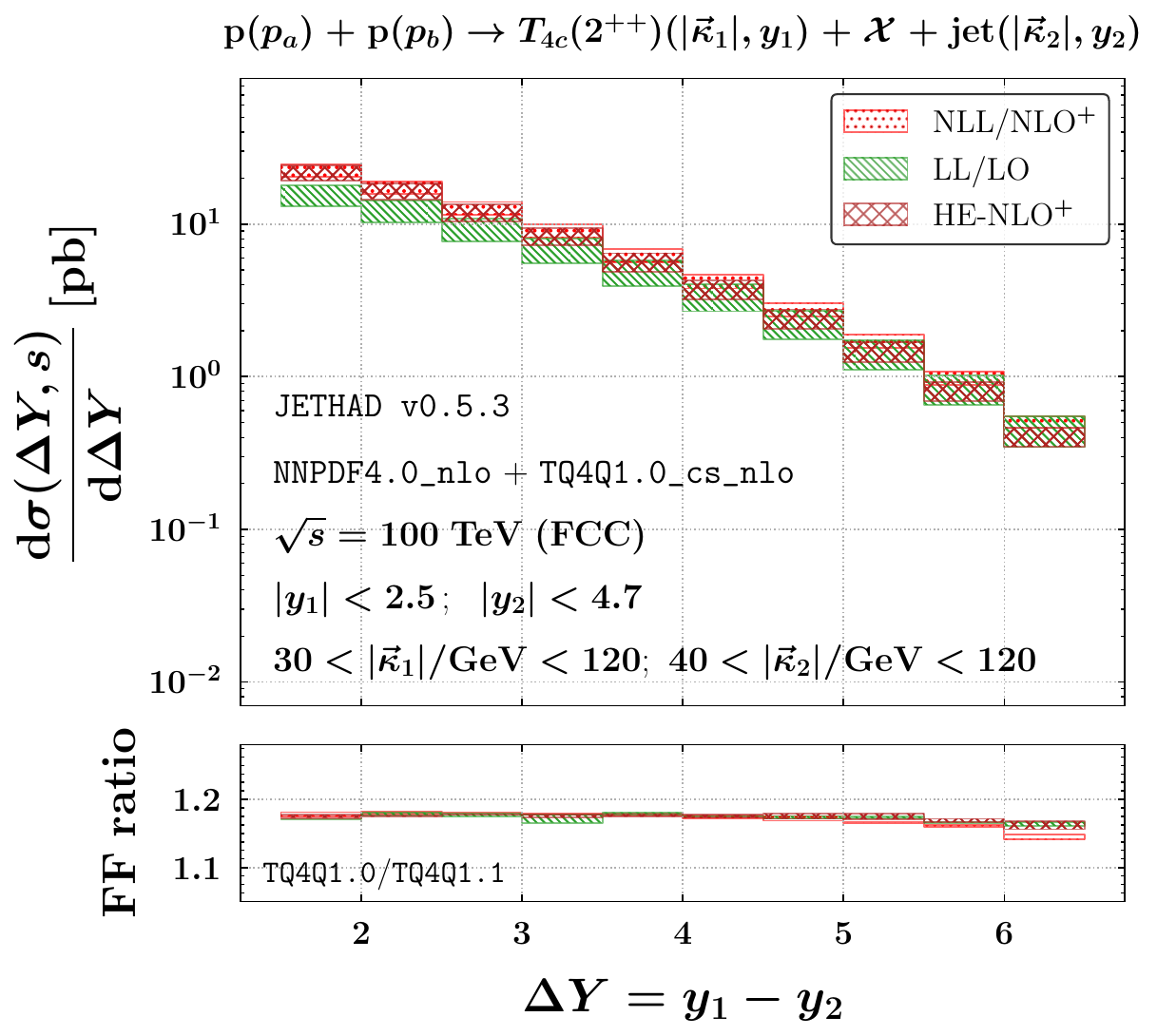}

\caption{
\justifying
\noindent
Rapidity distributions for scalar and tensor tetraquarks, $\TQcZpp$ (left) and $\TQcTpp$ (right), produced in association with a jet at $\sqrt{s} = 13$~TeV (HL-LHC, top) and $100$~TeV (nominal FCC, bottom). 
Filled bands in the main panels represent the combined effect of H-MHOUs and phase-space integration.
Ancillary panels display ratios between predictions from the first-generation {\tt TQ4Q1.0} FFs~\protect\cite{Celiberto:2024mab} and the updated {\tt 1.1} set.
}
\label{fig:I_FF-comp}
\end{figure*}

\vspace{1em}
\noindent
\textbf{Scalar channel ($0^{++}$).}
Predictions for the $\TQcZpp$ and $\TQbZpp$ states are shown in Fig.~\ref{fig:I_TQ0}.
Here, we observe the highest overall cross sections among all considered cases, with values ranging from around $10^{-2}$~pb to $10$~pb for charm and from $10^{-5}$~pb to $10^{-2}$~pb for bottom, depending on $\DY$ and $\sqrt{s}$.
This is in line with expectations, given the absence of spin suppression and the compactness of the wave function for scalar $S$-wave configurations.

The uncertainty bands due to H-MHOUs are modest, remaining well below a $1.5$ relative size.
The $\LL$ curves tend to overestimate the $\NLLp$ result in the low-$\DY$ region, while the two predictions becomes closer at large $\DY$.
This convergence indicates a good resummation control and validates the use of NLL-improved hybrid factorization in this context.
The rise in the cross section from HL-LHC to FCC is significant---roughly an order of magnitude---suggesting excellent prospects for observing such states at future colliders.

A closer inspection of the lower panels in Fig.~\ref{fig:I_TQ0} reveals a nontrivial pattern in the ratio between fixed-order $\HENLOp$ predictions and the resummed $\NLLp$ result.
At small values of $\DY$, the $\HENLOp$ versus $\NLLp$ ratio is close to unity for both HL-LHC and FCC setups, confirming that high-energy effects are mild and under control in this domain.
However, as $\DY$ increases, we observe opposite behaviors at the two energies:
at $13$~TeV~HL-LHC, the $\HENLOp$ result tends to overestimate the $\NLLp$ one, while at $100$~TeV~FCC, the opposite happens---the $\HENLOp$ curve lies systematically below the $\NLLp$ prediction.

This inversion strongly suggests that genuine BFKL resummation effects become increasingly relevant as the available rapidity interval enlarges and the energy grows.
We interpret this as a threefold indication of the potential impact of high-energy logarithms at future hadron colliders:
\begin{enumerate}
 \item The suppression of the $\HENLOp$ cross section relative to $\NLLp$ at large $\DY$ and 100~TeV points to an enhanced role of resummation terms, which dominate over fixed-order ones when $\log s$ becomes sizable.
 \item Under current fiducial cuts, such behavior implies that it will be feasible to isolate the genuinely resummed signal at FCC energies and large $\DY$, even without the need to tune additional observables or selection strategies.
 \item These features are consistent with prior findings on Higgs production at high energies~\cite{Bonvini:2018ixe,Bonvini:2018iwt}, where BFKL effects were shown to become essential to achieve accurate predictions beyond fixed-order.
\end{enumerate}

Consequently, our results provide yet another data point in support of the idea that the high-energy limit of QCD, when coupled to proper collinear evolution and heavy-flavor fragmentation, is not only theoretically consistent but also phenomenologically accessible, especially at next-generation hadron colliders like the FCC.

\vspace{1em}
\noindent
\textbf{Axial-vector channel ($1^{+-}$).}
The $\TQcOpm$ and $\TQbOpm$ cases, detailed in Fig.~\ref{fig:I_TQ1}, feature substantially lower rates, due to the suppression in the nonperturbative transition from a fragmenting parton to a spin-$1$ tetraquark, as discussed in Sec.~\ref{ssec:FFs_initial_scale}.
Event yields span from $10^{-4}$~pb to about $10^{-2}$~pb for charm and from $10^{-7}$~pb to about $10^{-5}$~pb for bottom.

Despite these low rates, the axial-vector channel stands out for its distinctive theoretical features, particularly concerning uncertainty control and the structure of radiative corrections.
Among all the spin configurations considered, the $1^{+-}$ exhibits the smallest uncertainty bands over the full $\DY$ range.
This stability is driven by two intertwined effects:
\begin{enumerate}
 \item As illustrated in Fig.~\ref{fig:FFs-muF_TQQ}, the gluon FF grows significantly with the factorization scale in the axial-vector case, unlike the scalar and tensor ones, where it tends to decrease slightly.
 This trend amplifies the \emph{natural stabilization} power of the HyF resummation; the faster the FF evolves, the more it compensates higher-order instabilities and H-MHOU-related uncertainties~\cite{Celiberto:2021dzy,Celiberto:2021fdp,Celiberto:2022grc};
 \item As discussed in Sec.~\ref{ssec:FFs_initial_scale}, the initial-scale FF for the $1^{+-}$ state is built upon a single heavy-quark fragmentation channel, and thus the LDME-induced uncertainties propagate from a unique, localized source, rather than combining multiple contributions as in the other spin channels.
\end{enumerate}

The comparison between the $\HENLOp$ and the $\NLLp$ curves reveals closer proximity in the axial-vector case, especially at $\sqrt{s} = 13$~TeV.
This is clearly visible in the ancillary panels, where the $\HENLOp$ versus $\NLLp$ ratio stays systematically nearer to one, with deviations remaining below $50\%$ across the entire $\DY$ spectrum.
This behavior implies that, in the $1^{+-}$ channel, the unresummed high-energy limit already captures a relevant portion of the NLL-enhanced corrections.
Still, the role of subleading logarithms is far from negligible, as evidenced by the $\LL$ versus $\NLLp$ separation.
At $100$~TeV, all channels display a similar gap between $\HENLOp$ and $\NLLp$, reinforcing the conclusion that BFKL-type resummation becomes increasingly relevant at higher energies, and the axial-vector case is no exception.

In summary, although the $1^{+-}$ signal is intrinsically weaker, it displays the cleanest radiative structure, and the most robust separation of resummation dynamics.
Thus, it offers a promising laboratory to validate the HyF formalism and probe subleading high-energy corrections.
The favorable perturbative convergence and reduced uncertainty bands observed at the FCC make the axial-vector channel a promising candidate for precision tests of high-energy resummation effects.

\vspace{1em}
\noindent
\textbf{Tensor channel ($2^{++}$).}
The tensor case, presented in Fig.~\ref{fig:I_TQ2}, lies between the scalar and axial-vector patterns, with cross sections slightly exceeding those in the $0^{++}$ channel.
As such, the $2^{++}$ state emerges as one of the most promising configurations in terms of experimental visibility.
The overall pattern of $\DY$ dependence follows that of the scalar state, with a mild decrease around $\DY \sim 2.5$ and more rapid falloff at large rapidity separations.

Uncertainty bands under $\NLLp$ remain controlled and competitive, and the gap between $\LL$ and $\NLLp$ is qualitatively similar to the scalar case.
This suggests that subleading logarithms have the same impact on both scalar and tensor productions, especially at lower $\DY$.
Interestingly, we observe that $\HENLOp$ curves tend to lie between $\LL$ and $\NLLp$ throughout the kinematic range, validating their interpretation as a fixed-order limit of the resummed approach.

Overall, the tensor channel confirms the broader narrative: heavy tetraquark production via fragmentation is \emph{naturally stable} under resummation, but sensitive enough to reveal slight differences between spin structures.

\vspace{1em}
\noindent
\textbf{Comparative summary.}
The detailed inspection of all rapidity distributions across spin states allows us to draw several key conclusions.

First, resummation effects are visible and under control across all channels, with $\NLLp$ offering a stable and physically motivated baseline.
Second, the impact of spin is twofold: it modulates the absolute normalization of the cross section through FF suppression and shapes the $\DY$ profile via the scale evolution of the underlying FFs.
Third, among all channels, the axial-vector one provides the clearest window to isolate subleading logarithmic corrections, due to the combination of low rate, strong scale evolution, and mild contamination from higher-spin or orbital-momentum mixing.

From an experimental point of view, these findings suggest that searches for heavy tetraquarks at high-energy colliders should prioritize inclusive observables such as $\DY$ distributions.
Although suppressed, axial-vector channels could act as clean indicators of genuine high-energy behavior, whereas scalar and tensor channels offer higher yields and complementarity.

We emphasize that these patterns hold across both $\TQc$ and $\TQb$ sectors and persist from 13~TeV to 100~TeV, confirming their universality and robustness.

Interestingly, we observe that the perturbative F-MHOU and the nonperturbative LDME uncertainties are of the same order, with the F-MHOUs being slightly larger in most kinematic configurations. 
However, the impact of the F-MHOUs is reduced when moving from the HL-LHC at $\sqrt{s}=13$~TeV to the FCC at $\sqrt{s}=100$~TeV. 

This behavior is expected and can be traced back to the scale dependence of the FF evolution. 
At higher center-of-mass energies, the typical factorization scales probed in the FFs are larger, leading to a smaller value of the strong coupling. 
As a consequence, the sensitivity of the FFs to variations in their evolution scale is reduced. Moreover, the longer DGLAP evolution path at the FCC acts as a smoothing mechanism, washing out differences induced at the input scale. Kinematically, the FCC also probes smaller momentum fractions, where the gluon-driven evolution is more stable, further suppressing the relative effect of scale variations in the FF sector.

\vspace{1em}
\noindent
\textbf{From {\tt TQ4Q1.0} to {\tt TQ4Q1.1}.}
For comparison, Fig.~\ref{fig:I_FF-comp} shows $\drv \sigma / \drv \DY$ distributions for scalar (left) and tensor (right) tetraquarks, produced in association with a jet at $\sqrt{s} = 13$~TeV (HL-LHC, top) and $100$~TeV (FCC, bottom), using predictions based on the first-generation {\tt TQ4Q1.0} functions~\cite{Celiberto:2024mab}, prior to the NRQCD-inspired update of the heavy-quark input implemented in this work.
Since the {\tt TQ4Q1.0} sets did not include systematic uncertainty studies related to fragmentation---neither from SDCs (F-MHOUs) nor from LDMEs---the error bands in the main panels reflect only phase-space integration and H-MHOUs as evaluated within the current setup.
Ancillary panels show the ratio between predictions based on the {\tt TQ4Q1.0} FFs and those obtained with the updated {\tt 1.1} set.
This comparison at the level of physical observables complements the direct analysis of FF evolution and normalization presented in Fig.~\ref{fig:FFs-muF_TQQ}.

The overall behavior of the main panels in Fig.~\ref{fig:I_FF-comp} remains consistent with the results shown in Figs.~\ref{fig:I_TQ0} and~\ref{fig:I_TQ2}, confirming that the updated {\tt TQ4Q1.1} FFs do not significantly alter the shape of the differential cross sections within uncertainties.
In contrast, a direct inspection of the ancillary panels reveals that the use of the new {\tt 1.1} set leads to a modest but systematic reduction in the $\DY$ spectra compared to the previous {\tt 1.0} functions.
This reduction is most pronounced for tensor tetraquarks and becomes more evident as $\DY$ increases, reflecting the different scaling behavior of the input FFs (see also Fig.~\ref{fig:FFs-muF_TQQ}).
Such an effect underlines the importance of using consistent and updated FF parametrizations when modeling observables at large rapidity intervals---especially in forward-plus-backward production setups---where differences between (HL-)LHC and FCC kinematics can amplify sensitivity to the fragmentation input and its evolution.

\vspace{1em}
\noindent
\textbf{Expected event yields.}
In order to assess the feasibility of future experimental searches, it is instructive to convert our predictions of the rapidity rate into expected event yields.
Assuming the full integrated luminosity collected by CMS at $\sqrt{s} = 13$ TeV during Run 2 (2015--2018), $\mathcal{L}^{\rm (CMS)} = 138.6~\text{fb}^{-1}$ (see Refs.~\cite{Giraldi:2022mwf,Radl:2024dvn} and references therein), we estimate the number of events for each $J^{PC}$ configuration via
\begin{equation}
\label{eq:event_yields}
 N_{\rm events}(s) \; = \; \mathcal{L}^{\rm (CMS)} \, \times \int_{}^{} \drv \DY \, \frac{\drv \sigma^{\rm NLL}}{\drv \DY} \;.
\end{equation}

The $\DY$ distribution is evaluated at $\NLLp$ accuracy and numerically integrated over the full rapidity range considered in our analysis. 
While our primary predictions focus on the fiducial interval $1.5 < |\DY| < 6.5$, chosen to ensure consistency with the high-energy behavior modeled by the HyF framework, we stress that the $\DY$ variable---defined as the rapidity separation between the tetraquark and the recoiling jet---is symmetric under forward-backward exchange. 
Although only the positive-$\DY$ region is explicitly shown in Figs.~\ref{fig:I_TQ0} to~\ref{fig:I_TQ2}, the underlying distributions implicitly account for both hemispheres. 
Since the CMS detector exhibits rapidity symmetry around the central region, it is natural and experimentally justified to include both forward and backward configurations in the estimation of total event yields. 
This effectively doubles the statistics with respect to the $\DY > 0$ case alone.

To provide a broader phenomenological perspective, it is instructive to also estimate event yields in the extended kinematic window $|\DY| < 6.5$. 
Although the region $|\DY| < 1.5$ lies outside the strict domain of applicability of our resummation approach, including it offers a more complete picture of the potential statistical reach. 
We therefore report yields for both ranges in our analysis.

The same luminosity is conservatively used also at $\sqrt{s} = 100$~TeV, to allow a baseline comparison of energy scaling, independently of specific FCC projections.
We note that employing $\mathcal{L}^{\rm (CMS)}$ at $13$~TeV is likewise conservative, given the significantly higher integrated luminosities expected at the HL-LHC.
Uncertainties are propagated from the $\DY$ distributions and account for the combined effect of scale variations, FF evolution, and LDME inputs.

Table~\ref{tab:event_yields_fiducial} summarizes the resulting yields for $\TQQ(J^{PC})$ states at both energies, within the range $1.5 < |\DY| < 6.5$, providing valuable insight into the potential observability of fully heavy tetraquarks at current and future hadron colliders.

For the scalar states ($J^{PC} = 0^{++}$), we observe some of the highest yields among all configurations.
In the $\TQc$ sector, the expected number of events exceeds $4 \times 10^5$ at 13~TeV and reaches over $6 \times 10^6$ at 100~TeV, suggesting excellent prospects for discovery or differential measurements, especially at HL-LHC and FCC.
Even for the $\TQb$ case, yields grow from several roughly 1700 at 13~TeV to a 8000 at 100~TeV, indicating that observation may be within reach, particularly if reconstruction through double-$\Jpsi$ decays is performed with high efficiency.

Axial-vector states ($J^{PC} = 1^{+-}$) exhibit the lowest yields, due to the suppressed nature of their FFs and more intricate spin structure.
For $\TQc$, we expect about 3300 events at 13~TeV and about 11000 at 100~TeV.
The $\TQb$ case is more challenging, with only a handful of events at 13~TeV and around 20 at 100~TeV.
Nonetheless, such rare signals may still be accessible if clean final states are available and background conditions are favorable, especially at ultra-high luminosities.

Tensor states ($J^{PC} = 2^{++}$) yield the largest number of events in our analysis.
In the $\TQc$ sector, expected yields exceed $7 \times 10^5$ at 13~TeV and approach $10^7$ at 100~TeV.
For $\TQb$, values range from just above 1000 to nearly 5000.
These figures confirm that scalar and tensor channels are the most promising for experimental searches, both in terms of production rates and reconstruction potential.

Overall, while the $\TQb$ sector remains more elusive, all six channels fall within the reach of future experimental programs, particularly when considering the luminosity upgrades of HL-LHC and the unprecedented dataset anticipated at FCC. 
The clear separation in yield hierarchies across spin channels also provides an opportunity to test the structure of heavy multiquark dynamics and validate the FF framework.

\begin{table}[t]
\centering
\begin{tabular}{c|c|r|r}
\toprule
\multicolumn{4}{c}{\small\textbf{HyF fiducial rapidity range:} \; $1.5 < |\Delta Y| < 6.5$} \\
\midrule
$\TQQ$ & $J^{PC}$ & Events [13 TeV Run 2] & Events [100 TeV FCC] \\
\midrule
$T_{4c}$ & $0^{++}$ & $489872 \pm 83529$ & $6706834 \pm 938912$ \\
$T_{4c}$ & $1^{+-}$ & $3354 \pm 471$ & $11424 \pm 348$ \\
$T_{4c}$ & $2^{++}$ & $718363 \pm 124716$ & $9606362 \pm 1299886$ \\
$T_{4b}$ & $0^{++}$ & $1750 \pm 296$ & $8291 \pm 384$ \\
$T_{4b}$ & $1^{+-}$ & $6 \pm 1$ & $22 \pm 3$ \\
$T_{4b}$ & $2^{++}$ & $2371 \pm 391$ & $9928 \pm 299$ \\
\bottomrule
\end{tabular}
\caption{
\justifying
\noindent
Expected event yields for fully heavy tetraquark production at $\sqrt{s} = 13$~TeV and $\sqrt{s} = 100$~TeV, obtained by integrating the NLL-resummed $\DY$ distributions over the HyF fiducial rapidity range $1.5 < |\DY| < 6.5$ considered in our analysis.
The 13~TeV yields assume an integrated luminosity of $\mathcal{L}^{\rm (CMS)} = 138.6~\text{fb}^{-1}$, corresponding to the total dataset collected by CMS during Run~2~\protect\cite{Giraldi:2022mwf,Radl:2024dvn}.
The same luminosity is conservatively adopted for 100~TeV to enable a baseline comparison of energy scaling, independently of specific FCC projections.
Uncertainties are propagated from the $\DY$ distributions and reflect the combined effect of scale variations, FF evolution, and LDME inputs.
}
\label{tab:event_yields_fiducial}
\end{table}

\begin{table}[t]
\centering
\begin{tabular}{c|c|r|r}
\toprule
\multicolumn{4}{c}{\small\textbf{Extended rapidity range:} \; $|\Delta Y| < 6.5$} \\
\midrule
$\TQQ$ & $J^{PC}$ & Events [13 TeV Run 2] & Events [100 TeV FCC] \\
\midrule
$T_{4c}$ & $0^{++}$ & $1522545 \pm 270309$ & $11888862 \pm 1801190$ \\
$T_{4c}$ & $1^{+-}$ & $9367 \pm 1323$ & $16976 \pm 462$ \\
$T_{4c}$ & $2^{++}$ & $2270257 \pm 417565$ & $17294873 \pm 2725859$ \\
$T_{4b}$ & $0^{++}$ & $4842 \pm 822$ & $12128 \pm 673$ \\
$T_{4b}$ & $1^{+-}$ & $16 \pm 2$ & $32 \pm 4$ \\
$T_{4b}$ & $2^{++}$ & $6531 \pm 1118$ & $14184 \pm 435$ \\
\bottomrule
\end{tabular}
\caption{
\justifying
\noindent
Expected event yields for fully heavy tetraquark production at $\sqrt{s} = 13$~TeV and $\sqrt{s} = 100$~TeV, obtained by integrating the NLL-resummed $\DY$ distributions over the extended rapidity range $|\DY| < 6.5$.
The 13~TeV yields assume an integrated luminosity of $\mathcal{L}^{\rm (CMS)} = 138.6~\text{fb}^{-1}$, corresponding to the total dataset collected by CMS during Run~2~\protect\cite{Giraldi:2022mwf,Radl:2024dvn}.
The same luminosity is conservatively adopted for 100~TeV to enable a baseline comparison of energy scaling, independently of specific FCC projections.
Uncertainties are propagated from the $\DY$ distributions and reflect the combined effect of scale variations, FF evolution, and LDME inputs.
}
\label{tab:event_yields_extended}
\end{table}

To quantify the impact of including the central rapidity region ($|\DY| < 1.5$), we compare the event yields obtained in the HyF fiducial and extended intervals.
As shown in Tables~\ref{tab:event_yields_fiducial} and~\ref{tab:event_yields_extended}, extending the integration range to $|\DY| < 6.5$ significantly enhances the total statistics across all spin configurations.
For instance, the $\TQc$ tensor state sees a rise from approximately $7.2 \times 10^5$ to $2.3 \times 10^6$ events at 13~TeV, and from $9.6 \times 10^6$ to nearly $1.7 \times 10^7$ at 100~TeV.
Similar enhancements, typically by a factor 2.5 to 3, are observed for the scalar and axial-vector states.

In the $\TQb$ sector, although absolute numbers remain smaller, the relative increase is comparably significant.
Yields for the tensor state grow from roughly 2400 to 6500 events at 13~TeV, and from 9900 to over 14000 at 100~TeV.
This consistent growth confirms that the central rapidity region, despite lying outside the strict domain of BFKL applicability, plays a nontrivial role to the total event count.
From an experimental perspective, this region is particularly relevant: central detectors such as CMS and ATLAS feature optimal acceptance and resolution at low rapidities, facilitating the reconstruction of final states such as double-$\Jpsi$ decays.
Thus, the extended range offers not only a statistical advantage but also practical benefits in terms of event detection and background control.

Finally, we note that the reconstruction of exotic hadron candidates decaying into double-quarkonium final states, such as $T_{4c} \to \Jpsi \Jpsi$, can be significantly affected by high pileup conditions, particularly at ATLAS and CMS during HL-LHC operations.
Multiple interactions per bunch crossing complicate the association of both $\Jpsi$ mesons to a common primary vertex, increasing the risk of combinatorial background. 
In addition, continuum double-quarkonium production via standard QCD processes (\emph{e.g.}, single or DPS mechanisms) constitutes a genuine physical background to exotic searches. 

Although the pileup itself is not the source of this background, it can degrade the experimental discrimination between signal and continuum. 
These effects may reduce the effective statistical reach of our yield estimates. 
Nevertheless, improvements in pileup mitigation strategies, vertex resolution, and forward detector capabilities (\emph{e.g.}, at LHCb) could help preserve sensitivity to fully heavy tetraquark production.

\subsection{Angular multiplicities}
\label{ssec:I_phi}

The second focal point of our phenomenological investigation is the angular distribution.
We perform, for the first time, a detailed analysis of this observable fully heavy tetraquark states.
In particular, we study the following normalized multiplicities
\begin{equation}
\label{angular_multiplicity}
\frac{1}{\sigma} \frac{\drv \sigma(\varphi, s)}{\drv \varphi} = \frac{1}{\pi} \left[ \frac{1}{2} + \sum_{n=1}^\infty
\langle \cos(n \varphi) \rangle \, \cos (n \varphi) \right] \;,
\end{equation}
with $\varphi = \phi_1 - \phi_2 - \pi$, $\phi_{1,2}$ the azimuthal angles of the two outgoing objects, $C_n$ the azimuthal coefficients (see Appendix~\hyperlink{app:B}{B}) and $\langle \cos(n \varphi) \rangle = C_n/C_0$ standing for the corresponding azimuthal correlation moments.
The coefficients $C_n$ are integrated in the full rapidity and transverse momentum windows as in Sec.~\ref{ssec:I}, and integrated over fixed bins of the rapidity interval $\DY$.

Originally developed to investigate the azimuthal decorrelation in light dijet systems~\cite{Marquet:2007xx,Ducloue:2013hia}, these angular multiplicities have proven to be powerful observables for probing the dynamics of QCD in the high-energy limit.
They encapsulate contributions from all azimuthal harmonics and offer a sensitive handle on the internal structure of the partonic interactions.

Their differential dependence on the azimuthal angle $\varphi$ makes them particularly suitable for experimental analyses, allowing for straightforward comparison with detector data even in the presence of nonuniform azimuthal acceptance over the full $2\pi$ range.
A recent investigation on dijet angular multiplicities highlighted two key advantages~\cite{Celiberto:2022gji}.
First, it helps alleviate persistent resummation instabilities observed in light-flavored final states at natural scales~\cite{Bolognino:2018oth,Celiberto:2020wpk}.
Second, it improves the agreement with CMS data at $\sqrt{s} = 7$~TeV~\cite{Khachatryan:2016udy}.

For the sake of simplicity, we consider only the fully charmed tensor resonance, $\TQcTpp$, which, as mentioned, represents the most favorable candidate for the $X(6900)$ state~\cite{LHCb:2020bwg}.
However, numerical tests have not shown significant discrepancies when considering the fully bottom counterpart, $\TQbTpp$.

\begin{figure*}[!t]
\centering

 \hspace{0.00cm}
 \includegraphics[scale=0.395,clip]{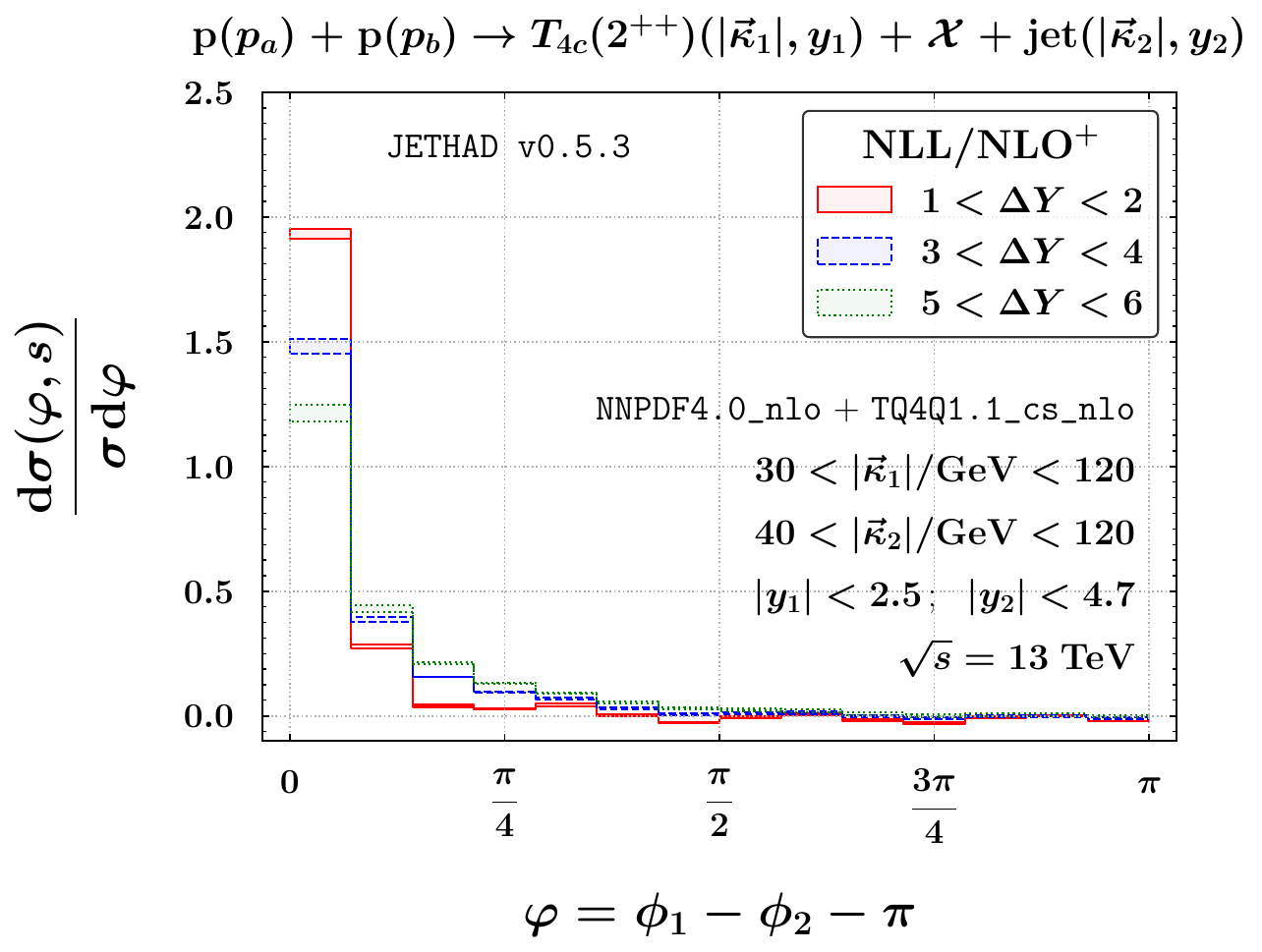}
 \hspace{-0.00cm}
 \includegraphics[scale=0.395,clip]{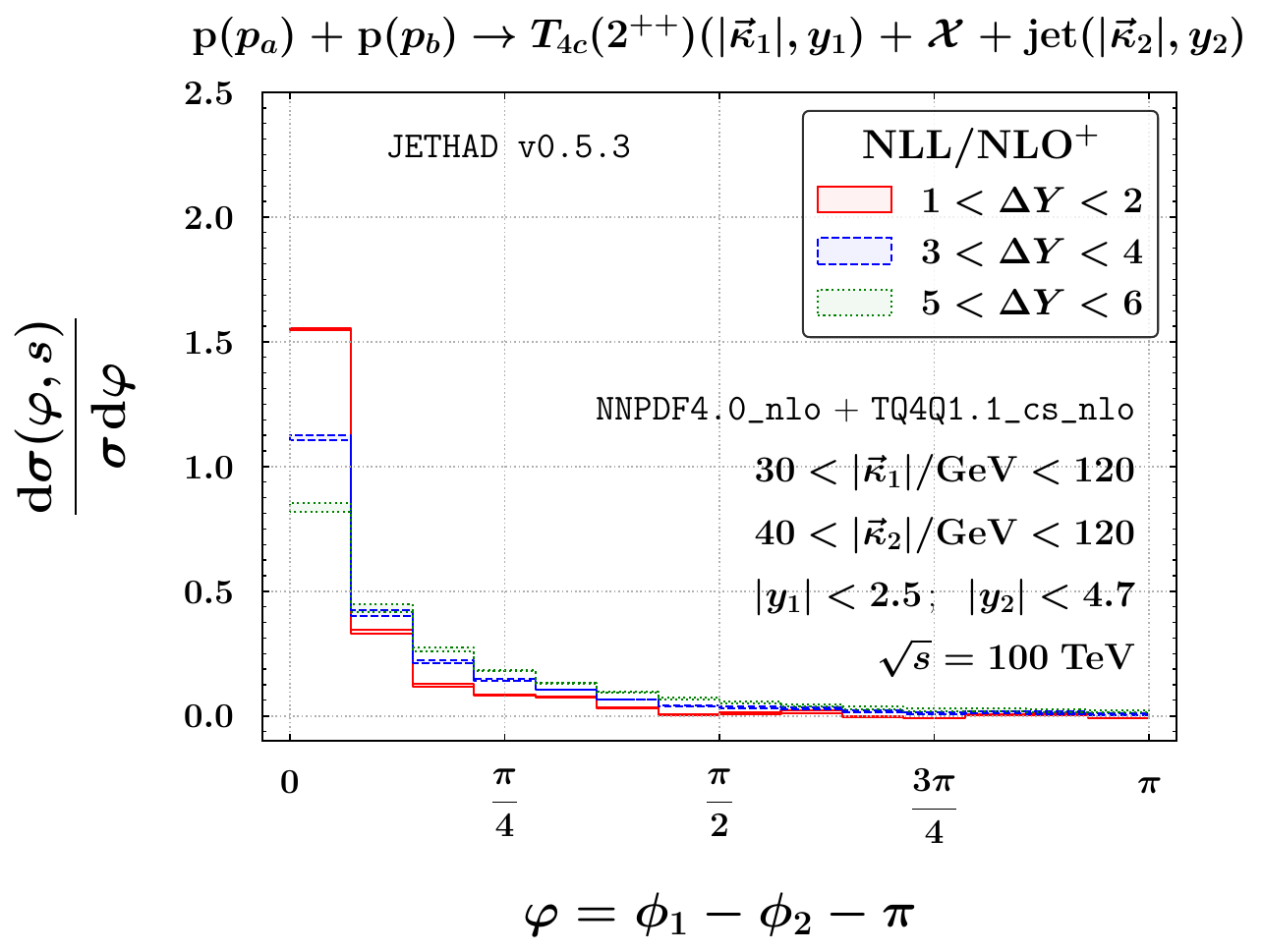}
 
 \vspace{0.35cm}
 
 \hspace{0.00cm}
 \includegraphics[scale=0.395,clip]{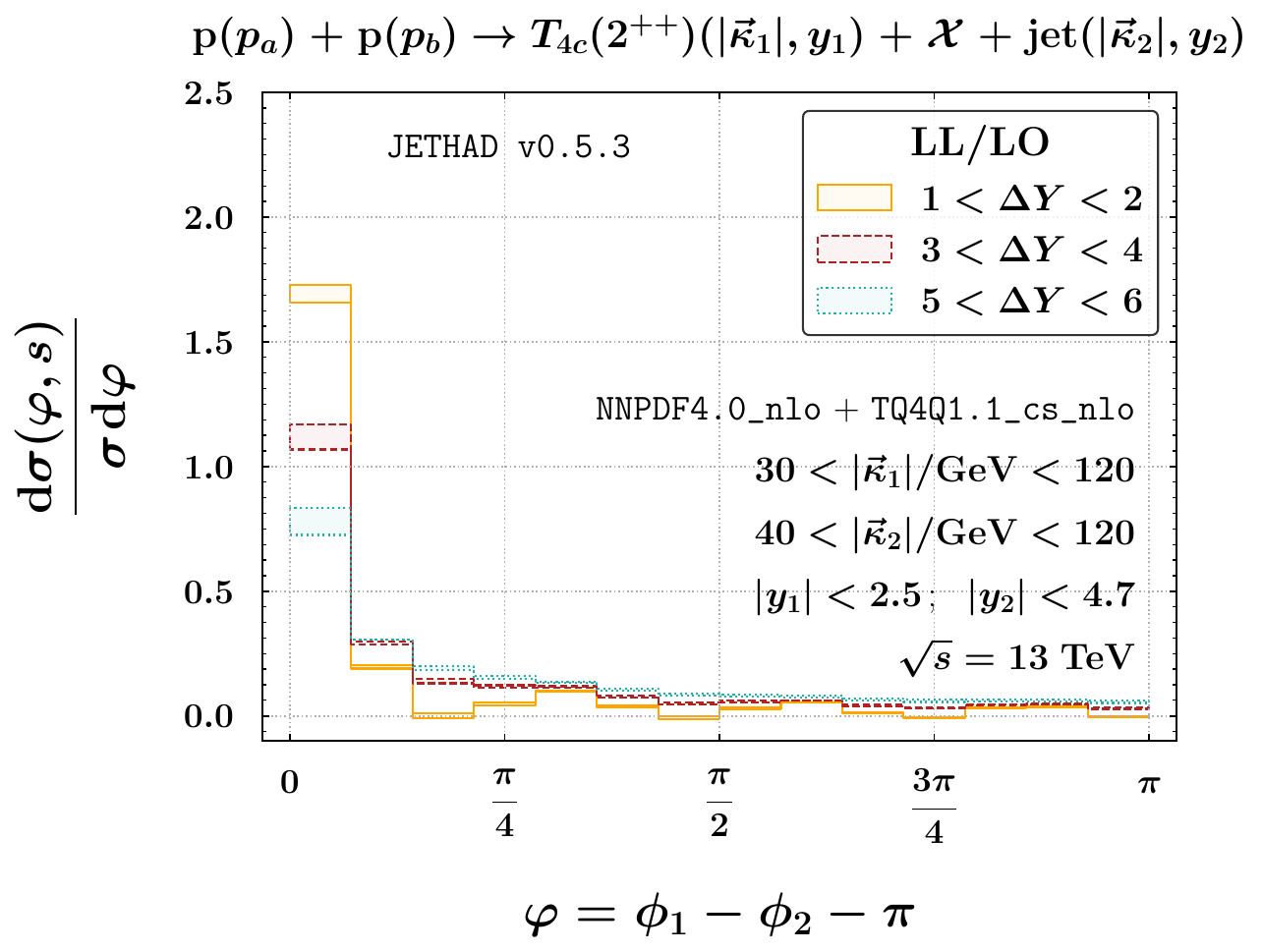}
 \hspace{-0.00cm}
 \includegraphics[scale=0.395,clip]{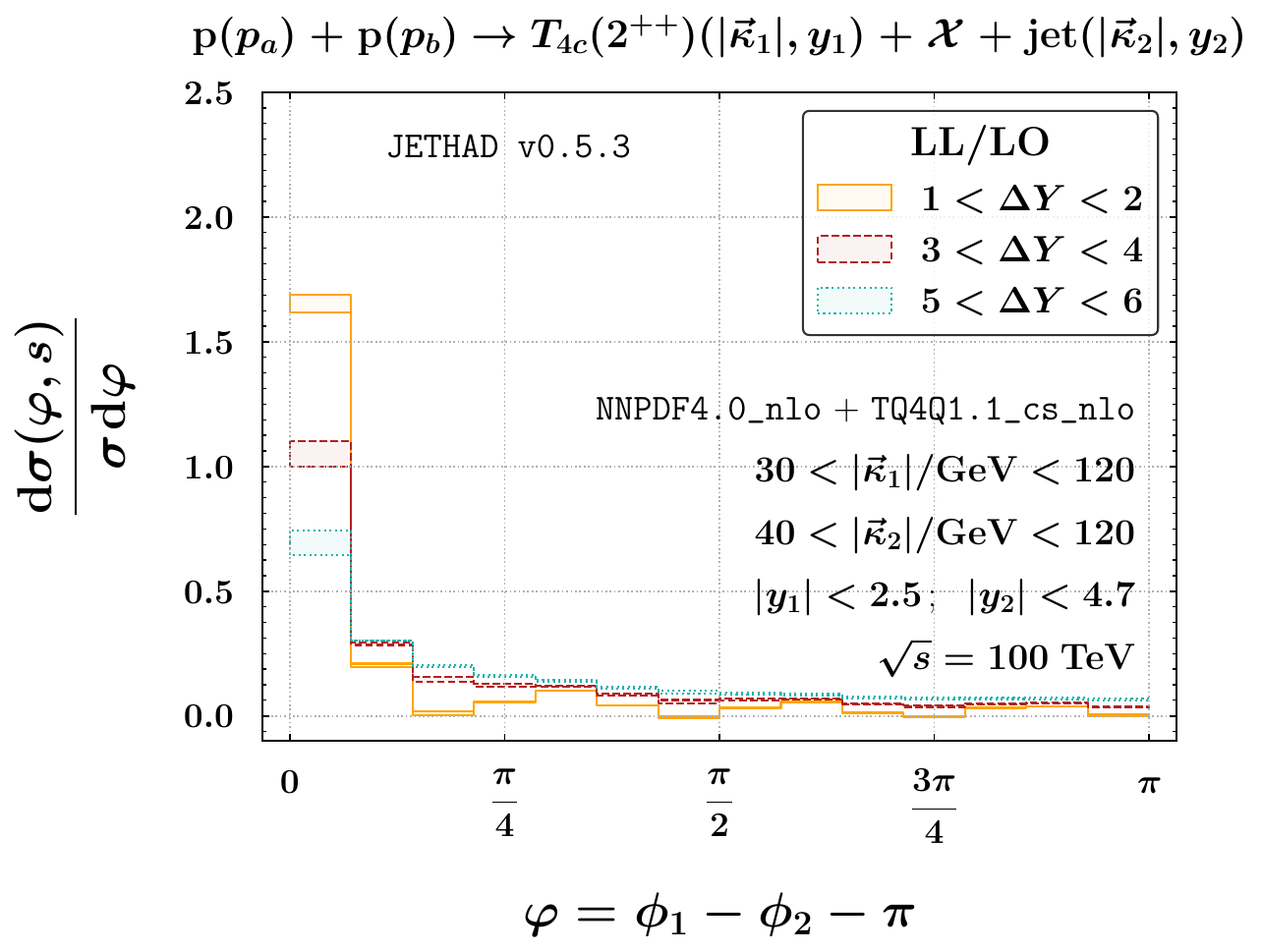}

 \includegraphics[scale=0.395,clip]{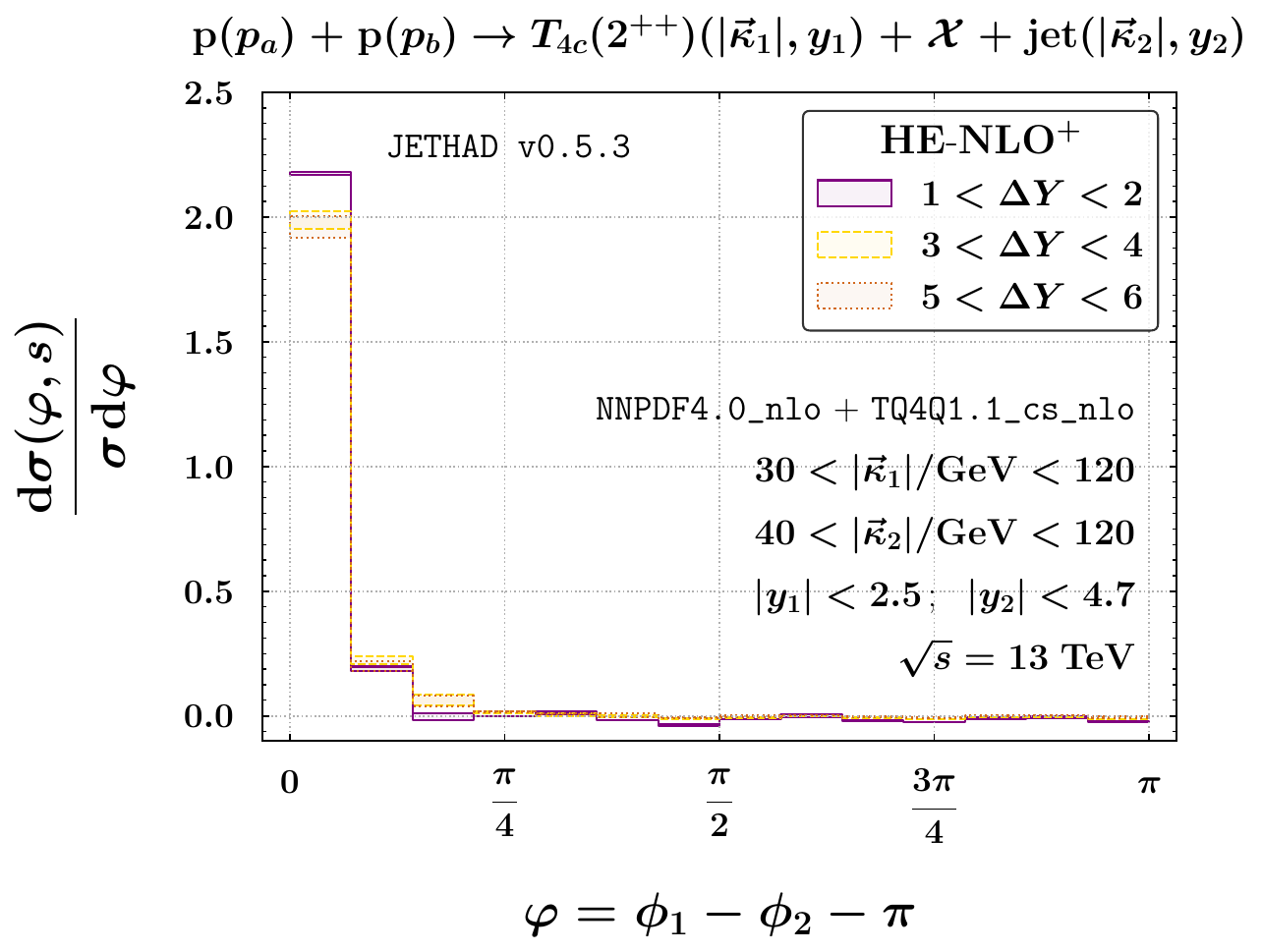}
 \hspace{-0.00cm}
 \includegraphics[scale=0.395,clip]{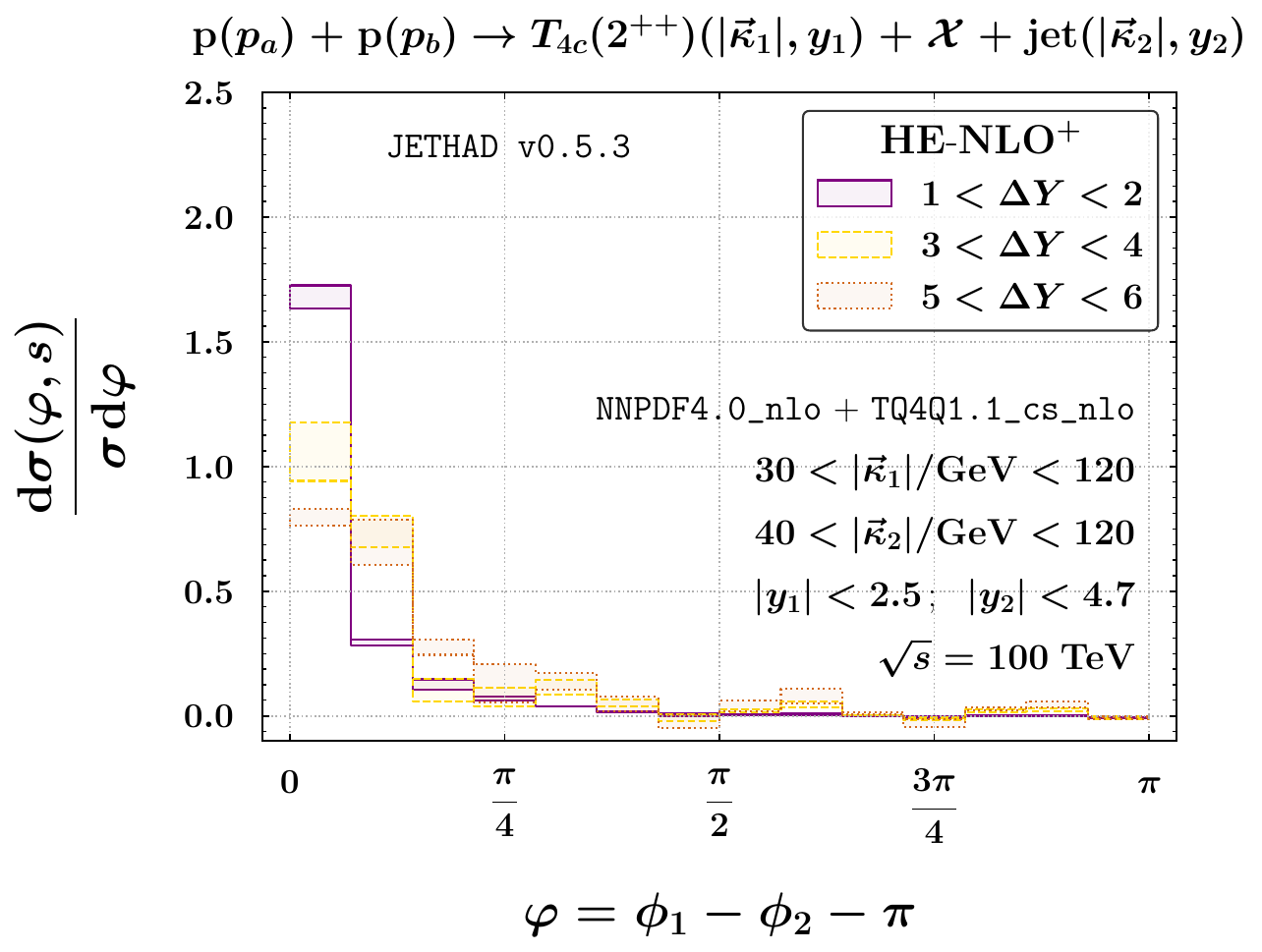}

\caption{
\justifying
\noindent
Angular multiplicities for tensor tetraquarks $\TQcTpp$ produced in association with a jet at $\sqrt{s} = 13$ TeV (HL-LHC, left) and $100$ TeV (nominal FCC, right). 
Predictions shown in the upper, central, and lower panels are computed at $\NLLp$, $\LL$, and $\HENLOp$ accuracy, respectively, within the HyF framework.
Filled bands in the main panels indicate the uncertainty coming from H-MHOUs.
}
\label{fig:I-phi_Tc2}
\end{figure*}

In Fig.~\ref{fig:I-phi_Tc2} we compare HyF predictions within $\NLLp$ accuracy (upper plots) with the corresponding $\LL$ accuracy (central plots) and the $\HENLOp$ background (lower plots).
Left and right columns show multiplicities taken at $13$~TeV HL-LHC and $100$~TeV FCC, respectively.
The rapidity distance $\DY$ is scanned over the interval from three to six units, and results are shown for three representative, nonoverlapping bins of unit width.

Due to the intrinsic symmetry of the azimuthal distributions defined in Eq.~\eqref{angular_multiplicity} under the transformation [$-\varphi \to \varphi$], we display them only within the range $0 < \varphi < \pi$.
To facilitate future experimental comparisons, the multiplicities are averaged over fixed-size $\varphi$ bins.
Since these distributions are defined as normalized ratios of cross sections, uncertainties from LDMEs and from the multidimensional phase-space integration effectively cancel out.
We have also numerically verified that F-MHOUs have negligible, if any, impact on our distribution ratios.
As a result, the error bands displayed in the plots reflect only the impact of H-MHOUs.

Our multiplicities display a pronounced peak located at $\varphi = 0$, corresponding to nearly back-to-back configurations between the final-state tetraquark and the jet.
This back-to-back enhancement is a standard feature of semihard final states produced with large rapidity separation because it reflects momentum conservation and the suppression of hard-gluon emissions at lowest order.
At the same time, the $\DY$-shape of these multiplicities offers insight into the interplay between collinear and high-energy logarithmic dynamics.

For all panels and all accuracy levels, we observe that the height of the peak at $\varphi = 0$ diminishes as $\DY$ grows.
This reduction is particularly pronounced in the $\NLLp$ and $-\LL$ case.
Such a trend reveals the increasing influence of BFKL-driven decorrelation effects, which become more visible as the rapidity span enlarges and more phase space opens for the emission of undetected gluons.
These secondary emissions tend to balance the transverse momentum between the final-state objects, thereby degrading their azimuthal alignment.

In contrast, $\HENLOp$ predictions show a comparatively weaker sensitivity to $\DY$.
The corresponding curves display a sharper peak at small $\varphi$, with only a mild flattening at larger separations.
This behavior is consistent with what is expected from a pure DGLAP-like evolution, where gluon radiation is collinear and less capable of inducing large-angle decorrelation.
This $\LL$ versus $\NLLp$ and $\HENLOp$ separation mirrors the classical ``BFKL versus DGLAP'' dichotomy, which has been extensively analyzed in the context of Mueller-Navelet jets, hadron-jet and dihadron systems~\cite{Celiberto:2015yba,Celiberto:2020wpk,Celiberto:2022rfj,Celiberto:2021dzy,Celiberto:2022dyf}.
In particular, according to BFKL dynamics, increasing $\DY$ leads to a broadening of the angular distribution and a suppression of the back-to-back peak, whereas in DGLAP-like scenarios such effects are more localized and less dependent on the total rapidity gap.

This dichotomic behavior is robust and persists at both 13~TeV and 100~TeV.
However, FCC energies offer additional benefits.
In all predictions at 100~TeV, uncertainty bands appear less sensitive to MHOUs, remaining relatively flat or even slightly shrinking.
This observation reflects the mitigation of soft-threshold logarithms that can affect the high-$\varphi$ tail.
Such threshold effects originate from Sudakov-like double logarithms that survive in configurations where the longitudinal momentum fractions $x_{1,2} \to 1$, a region that is approached at LHC but avoided at FCC.
Indeed, at fixed rapidities and transverse momenta, increasing $\sqrt{s}$ pushes the relevant $x$ values further from unity (see Eq.~\eqref{y-vs-x}).
Therefore, FCC kinematics helps suppress the \emph{threshold} region, leading to better stability and reduced sensitivity to logarithmic contributions not resummed in our HyF formalism.

Finally, we highlight that all our azimuthal multiplicities remain positive over the entire $\varphi$-range, within the estimated theoretical uncertainties.
This is a nontrivial feature.
In previous studies of different final states, such as vector quarkonia~\cite{Celiberto:2022dyf} and $B_c$ mesons~\cite{Celiberto:2022keu}, oscillating behavior and negative bins were occasionally observed at large $\varphi$, due to residual \emph{threshold} logarithmic instabilities~\cite{Celiberto:2022kxx}.
In contrast, the fragmentation of the tensor tetraquark exhibits a stabilizing influence that smoothens the distributions, even at the $\LL$ level, reinforcing the robustness of the observable.
This qualitative difference underscores the beneficial role played by multiple heavy-flavor fragmentation in taming theoretical instabilities, making tetraquark-jet systems a reliable probe for the analysis of BFKL effects in exotic-hadron production.

\section{Summary and new directions}
\label{sec:conclusions}

Guided by a hadron-structure-oriented vision, we have advanced the exploration of exotic matter production via the derivation and public release of the {\tt TQ4Q1.1} set of collinear FFs for a family of fully heavy tetraquarks~\cite{Celiberto:2025_TQ4Q11}.
Our analysis spans the three accessible channels characterized by quantum numbers $J^{PC} = 0^{++}$ (scalar), $1^{+-}$ (axial vector), and $2^{++}$ (tensor), and targets both charm and bottom flavors.
The construction of these FFs relies on the leading-power, single-parton fragmentation picture embedded in a NRQCD framework, which incorporates the contribution of distinct color-spin Fock-state configurations through their associated LDMEs.

SDCs are calculated at the initial scale for both gluon and heavy-quark fragmentation channels.  
These inputs are then evolved through a DGLAP evolution in the VFNS, implemented via the threshold-matched {\HFNRevo} scheme~\cite{Celiberto:2025euy,Celiberto:2024mex,Celiberto:2024bxu,Celiberto:2024rxa}.  
The {\tt TQ4Q1.1} sets represent the first release to include a detailed treatment of theoretical uncertainties from three sources: $(i)$ nonperturbative LDMEs $(ii)$ hard-scattering scale dependence (H-MHOUs), and $(iii)$ fragmentation-scale inputs at the initial condition (F-MHOUs).  
Each of these contributions has been analyzed separately and then combined, allowing us to disentangle their relative weight and to provide collider-level predictions with a transparent and systematic uncertainty budget.

As a direct phenomenological application, we have employed the {\psymJethad} multimodular interface~\cite{Celiberto:2020wpk,Celiberto:2022rfj,Celiberto:2023fzz,Celiberto:2024mrq,Celiberto:2024swu} to evaluate high-energy cross sections for semi-inclusive tetraquark plus jet production at the HL-LHC and the FCC, operating at full $\NLLp$ accuracy within the HyF factorization formalism, that consistently combines collinear factorization with BFKL resummation.

To enhance the phenomenological impact of our predictions, we have complemented the cross section analysis with a quantitative estimate of expected event yields. 
These are obtained by integrating the NLL-resummed rapidity distributions over a HyF fiducial rapidity range ($1.5 < |\Delta Y| < 6.5$), and multiplying by the integrated luminosity collected by CMS during Run~2 at $\sqrt{s} = 13$~TeV. 
The same value is conservatively adopted at 100~TeV to allow for a baseline comparison of energy scaling. 
A second set of yields, integrated over the extended range ($|\Delta Y| < 6.5$), is also provided to assess the maximal statistical reach. 
Our event-yield analysis offers a concrete reference for ongoing and future searches for exotic multiquark states at the LHC and FCC, and is expected to guide both phenomenological developments and experimental strategies in upcoming high-luminosity collider programs.

We have computed differential distributions in the rapidity interval between the final-state hadrons, $\DY = y_1 - y_2$, as well as angular multiplicities in the relative azimuthal angle $\varphi = \phi_1 - \phi_2 - \pi$.
These observables are known to be highly sensitive to the structure of logarithmic resummation at high energies and to the interplay between perturbative and nonperturbative mechanisms of hadron formation.

Our rapidity-interval analyses revealed characteristic trends across the scalar, axial-vector, and tensor channels, with cross sections that decrease as $\DY$ increases and uncertainty bands that display a consistent stabilization pattern, especially at FCC energies.
We found that the HyF signal becomes increasingly distinguishable from fixed-order backgrounds in the large-$\DY$ regime, particularly in the scalar and tensor channels.

Notably, axial-vector configurations, despite yielding lower absolute cross sections due to fragmentation suppression, exhibit enhanced stability and tightly constrained uncertainty bands, rooted in the strong $\mu_F$ evolution of the gluon FF and in the reduced propagation of LDME ambiguities.
This \emph{natural stability}~\cite{Celiberto:2022grc} highlights the potential of this channel as a benchmark for future precision studies.

Angular multiplicities, on the other hand, provide complementary diagnostic power by isolating genuine resummation effects at fixed $\DY$.
Thanks to their definition as ratios of differential cross sections, these observables are largely insensitive to LDME modeling and phase-space integration effects, allowing for unambiguous comparison and discrimination between NLL resummed and high-energy NLO predictions.

The phenomenological richness arising from this multichannel investigation endows the {\tt TQ4Q1.1} functions with a twofold role.
On the one hand, they serve as precision tools to probe high-energy QCD dynamics through stable, infrared-safe observables.
On the other hand, they enable an in-depth exploration of exotic matter, providing access to the color-spin composition and hadronization pathways of fully heavy tetraquarks in different quantum states.
This synergy between structural and dynamical insights constitutes a cornerstone for future studies at both current and next-generation hadron colliders.

Looking ahead, the framework and tools developed in this study may also support a broader program of validation for the underlying methodology used to describe exotic-hadron production at high energies. 
While direct data on fully heavy tetraquarks are not yet available, related systems involving heavy quarkonia, such as double-$\Jpsi$ or $\Jpsi$-plus-jet final states, can provide valuable testing grounds for key components of our approach. 
In particular, measurements of quarkonium production at large transverse momentum and wide rapidity separation would help assess the accuracy of the high-energy resummation strategy and the shape of the corresponding FFs.

Although current experimental analyses, such as those by LHCb on double-$\Jpsi$ production~\cite{LHCb:2023ybt}, focus on the forward region and on moderate transverse-momentum regimes, future extensions to harder kinematics could offer meaningful comparisons with our predictions. 
In this sense, studies of heavy quarkonium at high transverse momentum and large rapidity separation represent an essential step toward validating the theoretical machinery that underpins our predictions for exotic tetraquark signals. 
They would not only reinforce the robustness of the HyF factorization formalism, but also strengthen its phenomenological relevance for upcoming high-luminosity collider programs.

Future developments will aim to consolidate the bridge between exotic hadron structure and the high-energy regime of QCD, building upon the multifaceted potential offered by the new {\tt TQ4Q1.1} functions. 
These tools open promising avenues for investigating not only the fragmentation mechanisms responsible for the emergence of fully heavy tetraquarks, but also the underlying dynamics of hadronization and gluon radiation in the forward domain. 
Advancing our grasp of exotic matter production will require increasingly accurate modeling of perturbative and nonperturbative effects, calling for a broader and more refined treatment of the many theoretical components involved.

A central priority is a further refinement of uncertainty quantification, especially for F-MHOUs.  
As shown in recent studies~\cite{Kassabov:2022orn,Harland-Lang:2018bxd,Ball:2021icz,McGowan:2022nag,NNPDF:2024dpb,Pasquini:2023aaf}, new methodologies for the propagation and statistical treatment of theoretical uncertainties are increasingly being established as standard practice.  
These developments should soon be extended to the case of exotic-hadron fragmentation.  
In our formalism, the explicit role of LDMEs as hadron-specific, nonperturbative parameters, together with their interplay with F-MHOUs, enables dedicated strategies to trace and reduce theoretical errors channel by channel.

Another crucial upgrade will come from the inclusion of color-octet configurations in the NRQCD expansion for tetraquark states. 
Although the present {\tt TQ4Q1.1} release focuses on the color-singlet channels at LO, the inclusion of higher Fock-state components is expected to affect the magnitude and shape of the FFs, especially for the scalar and tensor states. 
Once reliable LDME estimates for octet configurations become available, we plan to incorporate them into a consistent framework, extending the predictive power of our approach.

In the long term, progress will also involve an expansion of the HyF framework into a \emph{multilateral} scheme integrating complementary resummation techniques. 
Specifically, establishing connections with soft-gluon~\cite{Hatta:2020bgy,Hatta:2021jcd,Caucal:2022ulg,Taels:2022tza} and jet-radius~\cite{Dasgupta:2014yra,Dasgupta:2016bnd,Banfi:2012jm,Banfi:2015pju,Liu:2017pbb} resummations represents a compelling path forward. 
The investigation of jet angularities and their interplay with fragmentation~\cite{Luisoni:2015xha,Caletti:2021oor,Reichelt:2021svh} is equally promising, especially as experimental access to such observables improves. These enhancements will render our hybrid $\NLLp$ approach increasingly suitable for precision physics at high energies.

From a phenomenological perspective, the rare and suppressed character of fully bottomed tetraquarks provides an ideal laboratory to test the limits of the fragmentation formalism. 
Although current predictions of $\TQb$ cross sections remain small, the evolution of bottom-flavored FFs at high scales, as described by {\tt TQ4Q1.1}, will allow us to probe more exclusive observables and eventually extract structural information even in low-statistics environments. 

The upcoming data from the FCC~\cite{FCC:2025lpp,FCC:2025uan,FCC:2025jtd}, as well as from other planned facilities~\cite{Chapon:2020heu,LHCspin:2025lvj,Anchordoqui:2021ghd,Feng:2022inv,AlexanderAryshev:2022pkx,LinearCollider:2025lya,LinearColliderVision:2025hlt,Arbuzov:2020cqg,Accettura:2023ked,InternationalMuonCollider:2024jyv,MuCoL:2024oxj,Black:2022cth,InternationalMuonCollider:2025sys,Accardi:2023chb,Bose:2022obr,Gessner:2025acq,Altmann:2025feg}, will play a decisive role in driving these developments. 
Complementary programs at future lepton-hadron colliders such as the EIC~\cite{AbdulKhalek:2021gbh,Khalek:2022bzd,Hentschinski:2022xnd,Amoroso:2022eow,Abir:2023fpo,Allaire:2023fgp} will offer further ways to investigate exotic states in cleaner environments, where gluon-initiated fragmentation may be isolated and studied more directly.

Among the next targets of our program stands the phenomenological investigation of the enigmatic $Z_c(3900)$ state~\cite{Guo:2013ufa}, whose prompt production has not yet been observed.
Tetraquark production channels, powered by the high-energy and high-luminosity capabilities of new-generation machines, may help clarify its production mechanisms and reveal its inner composition.

In parallel, a deeper understanding of hadron structure will emerge from a progressive refinement of our knowledge of the underlying dynamics governing quarkonium and exotic-matter formation.
This effort will benefit from data collected at future colliders, which will serve as a promising environment for probing the existence of intrinsic heavy-quark components.
In particular, the case of intrinsic charm (IC)~\cite{Brodsky:1980pb,Brodsky:2015fna,Jimenez-Delgado:2014zga,Ball:2016neh,Hou:2017khm,Ball:2022qks,Guzzi:2022rca} has recently received renewed attention, with compelling evidence for its valence distribution in the proton~\cite{NNPDF:2023tyk}.
This opens the possibility of using exotic-hadron production, including heavy tetraquarks, as indirect probes of IC dynamics.
Reciprocally, the modeling of exotic hadrons containing heavy quarks may benefit from improved knowledge of IC, potentially suggesting a two-way portal between proton structure and exotic spectroscopy~\cite{Vogt:2024fky}.

An additional frontier involves exploring the sensitivity of tetraquark production to the \emph{dead-cone} effect~\cite{Dokshitzer:1991fd}, a unique prediction of QCD for heavy-quark fragmentation.
This effect, recently confirmed by ALICE~\cite{ALICE:2021aqk}, may manifest itself in observables involving fully heavy tetraquarks and their distribution inside jets.
Tetraquark-in-jet observables, when analyzed within the HyF factorization formalism, may provide a clean environment to quantify and constrain the dynamics of gluon radiation suppression at small angles, thereby enriching our understanding of the mass-dependent features of QCD radiation.

Finally, we stress the importance of treating individual tetraquark quantum states as distinct diagnostic tools. 
Each spin configuration examined here, namely scalar, axial-vector, and tensor, exhibit specific features in the behavior of FFs, their evolution, and the resulting production rates. 
Their differential impact on angular and rapidity observables enriches the phenomenological picture and offers valuable access to both the hadronization mechanisms at work and the core QCD dynamics in high-energy collisions.

In this context, the recent CMS spin-parity determination of fully charmed tetraquarks~\cite{CMS:2023owd,CMS:2025fpt} marks a major step forward in the exploration of exotic hadrons.
The {\tt TQ4Q1.1} functions serve as a case study for a broader, evolution-consistent fragmentation framework that encompasses different families of fully heavy states.
Thanks to its modular structure, our approach lends itself naturally to future extensions toward \emph{multimodal} descriptions, where distinct classes of initial-scale inputs can be systematically included.
Such a theoretical infrastructure provides versatile support to ongoing experimental efforts and enhances data-driven investigations of exotic multiquark systems.

The release of {\tt TQ4Q1.1}, now extended to include scalar, axial-vector, and tensor configurations---and now providing an estimate of uncertainties from the nonperturbative sector---lays the groundwork for a new generation of studies on the formation of exotic matter via collinear fragmentation.
This work represents a timely step toward decoding the structure and formation of fully heavy hadrons, connecting state-of-the-art QCD methods to experimental programs at the energy and precision frontiers.

\section*{Acknowledgments}
\label{sec:Acknowledgments}

We acknowledge the use of calculations from Refs.~\cite{Feng:2020riv,Bai:2024ezn}, which were independently rederived using {\psymJethad}~\cite{Celiberto:2020wpk,Celiberto:2022rfj,Celiberto:2023fzz,Celiberto:2024mrq,Celiberto:2024swu}, and employed as proxies for the fragmentation process at the initial scale.
We are grateful to Marco Bonvini, Angelo Esposito, Luca Maxia, Alessandro Papa, Fulvio Piccinini, and Alessandro Pilloni for valuable discussions on the physics of exotic hadronic states.
This work is supported by the Atracción de Talento Grant No. 2022-T1/TIC-24176 from the Comunidad Autónoma de Madrid, Spain.

\section*{Data availability}
\label{sec:data}

The data that support the findings of this article are openly available~\cite{Celiberto:2025_TQ4Q11}.
The {\tt TQ4Q1.1} fragmentation [functions for fully heavy tetraquarks $\TQQ(J^{PC})$~\cite{Celiberto:2025_TQ4Q11} are available at: ~\cite{Celiberto:2025_TQ4Q11_url}.  
For convenience, we provide three replica sets that encode perturbative fragmentation-scale MHOUs (F-MHOUs): replica~0 corresponds to the central input without variations, while replica~1 and replica~2 implement, respectively, a variation of the evolution-ready scale $Q_0$ to one half and to twice its central value, $4m_Q$. 
As the functions scale linearly with the LDME (see Eqs.~\eqref{TQQ_FF_initial-scale}), there is no need to provide separate error sets; users can straightforwardly rescale the FFs by varying the LDME within the uncertainty range specified in Tables~\ref{tab:T4c_LDMEs} and~\ref{tab:T4b_LDMEs}.  
This setup allows users to consistently disentangle perturbative and nonperturbative sources of uncertainty.

To further support reproducibility and independent verification, we provide a stand-alone \textsc{Mathematica} notebook, extracted from our internal {\symJethad} codebase~\cite{Celiberto:2020wpk,Celiberto:2022rfj,Celiberto:2023fzz,Celiberto:2024mrq,Celiberto:2024swu} and containing symbolic expressions for all short-distance coefficients as implemented in this work. 
This self-contained notebook requires no external dependencies and can be directly used for numerical testing or symbolic manipulation. 
It is available as Supplemental Material~\cite{Celiberto:2025_TQ4Q11_CaseStudy_suppl} and hosted in the same GitHub repository as the {\tt TQ4Q1.1} FF grids.

\clearpage

\appendix
\onecolumngrid

\counterwithin*{equation}{section}
\renewcommand\theequation{\thesection\arabic{equation}}

\hypertarget{app:A}{
\section{Analytic expressions for the SDCs}
}

In this appendix, we provide the functional form of all the nonvanishing dimensionless SDCs for our $\TQQ$ fragmentation channels.

\vspace{1em}
\noindent
\textbf{Scalar channel ($0^{++}$).}

The $[g \to \TQQ(0^{++})]$ SDCs read~\cite{Feng:2020riv}
\begin{equation}
\begin{split}
 \label{Dg_FF_SDC_0pp_33}
\hspace{-0.00cm}
 \tilde{\cal D}^{(0^{++})}_g&(z,[3,3]) \,=\, 
 \frac{\pi^{2} \alpha_{s}^{4}(4m_Q)}{497664 \, d^{\TQQ}_g(z)}\left[186624-430272 z+511072 z^2-425814 z^3\right. \\
 & +\, 217337 z^4-61915 z^5+7466 z^6+42(1-z)(2-z)(3-z)(-144+634 z\\
 & \left.-\, 385 z^2+70 z^3\right) \ln (1-z)+36(2-z)(3-z)\left(144-634 z+749 z^2-364 z^3\right. \\
 & \left.+\, 74 z^4\right) \ln \left(1-\frac{z}{2}\right)+12(2-z)(3-z)\left(72-362 z+361 z^2-136 z^3+23 z^4\right) \\
 & \left.\times\, \ln \left(1-\frac{z}{3}\right)\right]
 \;,
\end{split}
\end{equation}
\\[-0.35cm]
\begin{equation}
\begin{split}
 \label{Dg_FF_SDC_0pp_66}
\hspace{-0.00cm}
 \tilde{\cal D}^{(0^{++})}_g&(z,[6,6]) \,=\,  
 \frac{\pi^{2} \alpha_{s}^{4}(4m_Q)}{331776 \, d^{\TQQ}_g(z)}\left[186624-430272 z+617824 z^2-634902 z^3\right. \\
 & +\, 374489 z^4-115387 z^5+14378 z^6-6(1-z)(2-z)(3-z)(-144-2166 z\\
 & \left.+\, 1015 z^2+70 z^3\right) \ln (1-z)-156(2-z)(3-z)\left(144-1242 z+1693 z^2-876 z^3\right. \\
 & \left.+\, 170 z^4\right) \ln \left(1-\frac{z}{2}\right)+300(2-z)(3-z)\left(72-714 z+953 z^2-472 z^3+87 
 z^4\right) \\
 & \left.\times\, \ln \left(1-\frac{z}{3}\right)\right]
 \;,
\end{split}
\end{equation}
\\[-0.35cm]
\begin{equation}
\begin{split}
 \label{Dg_FF_SDC_0pp_36}
\hspace{-0.00cm}
 \tilde{\cal D}^{(0^{++})}_g&(z,[3,6]) \,=\,  
 \frac{\pi^{2} \alpha_{s}^{4}(4m_Q)}{165888 \, d^{\TQQ}_g(z)}\left[186624-430272 z+490720 z^2-394422 z^3\right. \\
 & +\, 199529 z^4-57547 z^5+7082 z^6+6(1-z)(2-z)(3-z)(-432+3302 z \\
 & \left.-\, 1855 z^2+210 z^3\right) \ln (1-z)-12(2-z)(3-z)\left(720-2258 z+2329 z^2-1052 z^3\right. \\
 & \left.+\, 226 z^4\right) \ln \left(1-\frac{z}{2}\right)+12(2-z)(3-z)\left(936-4882 z+4989 z^2-1936 z^3+331 z^4\right) \\
 & \left.\times\, \ln \left(1-\frac{z}{3}\right)\right]
 \;,
\end{split}
\end{equation}
with $d^{\TQQ}_g(z) = z(2-z)^{2}(3-z)$.
Analogously, the $[Q \to \TQQ(0^{++})]$ SDCs read~\cite{Bai:2024ezn}
\begin{equation}
\begin{split}
 \label{DQ_FF_SDC_0pp_33}
 \hspace{-0.00cm}
 \tilde{\cal D}&^{(0^{++})}_Q(z,[3,3]) \,=\, 
 \frac{\pi^2 \as^4(5m_Q)}{559872 \, d^{\TQQ}_Q(z)} \left[ -264 (z-4) (11 z-12) (z^2-16 z+16) \right. \\
 & \times\, (13 z^4-57 z^3-656 z^2+1424z-512) (3 z-4)^5 \log (z^2-16 z+16) + 6 (11 z-12)(z^2-16 z+16) \\
 & \times\, (1273 z^5-16764 z^4+11840 z^3 + 247808z^2-472320 z+171008) (3 z-4)^5 \log (4-3 z) \\
 & -\, 3 (11 z-12)(z^2-16 z+16) (129 z^5 - 7172 z^4+49504 z^3-108416z^2 + 73984 z-9216) (3 z-4)^5  \\
 & \times\, \log\left[\left(4-\frac{11z}{3}\right)(4-z)\right] + 16 (z-1) (657763 z^{12}-10028192z^{11} + 188677968 z^{10}-2600899712 z^9 \\
 & +\, 18018056448 z^8-71685000192z^7 + 179414380544 z^6-294834651136 z^5 \\
 & +\, 321642168320z^4-229388845056 z^3 + 102018056192 z^2-25480396800z \left. + 2717908992)\right]
 \;,
\end{split}
\end{equation}
\\[-0.35cm]
\begin{equation}
\begin{split}
 \label{DQ_FF_SDC_0pp_66}
 \hspace{-0.00cm}
 \tilde{\cal D}&^{(0^{++})}_Q(z,[6,6]) \,=\, 
 \frac{\pi^2 \as^4(5m_Q)}{373248 \, d^{\TQQ}_Q(z)} \left[ -120 (z-4) (11 z-12) (z^2-16 z+16) \right. \\
 & \times\, (35 z^4-535 z^3+3472 z^2-4240z+512) (3 z-4)^5 \log (z^2-16 z+16)  \\
 & -\, 30 (11 z-12)(z^2-16 z+16) (3395 z^5-48020 z^4+126144 z^3 \\
 & -\, 75776^2-38656 z+62464) (3 z-4)^5 \log (4-3 z) + 75 (11 z-12)(z^2-16 z+16) \\
 & \times\, (735 z^5 - 10684 z^4+34208 z^3-44160z^2 + 20224 z+9216) (3 z-4)^5 \log\left[\left(4-\frac{11z}{3}\right)(4-z)\right]\\
 & +\, 16 (z-1) (7916587 z^{12}-263987840z^{11} + 3125201872 z^{10}-16993694336 z^9 \\
 & +\, 51814689024 z^8-99638283264^7 + 133459423232 z^6-140136398848 z^5 \\
 & +\, 127161204736z^4-96695746560 z^3 + 53372518400 z^2-17930649600z \left. + 2717908992)\right]
 \;,
\end{split}
\end{equation}
\\[-0.35cm]
\begin{equation}
\begin{split}
 \label{DQ_FF_SDC_0pp_36}
 \hspace{-0.00cm}
 \tilde{\cal D}&^{(0^{++})}_Q(z,[3,6]) \,=\, 
 \frac{\pi^2 \as^4(5m_Q)}{186624 \sqrt{6} \, d^{\TQQ}_Q(z)} \left[ 24 (z-4) (11 z-12) (z^2-16 z+16) \right. \\
 & \times\, (225 z^4-3085 z^3+17456 z^2 - 19760z+1536) (3 z-4)^5 \log (z^2-16 z+16)  \\
 & -\, 6 (11 z-12)(z^2-16 z+16) (555 z^5+52428 z^4-363328 z^3 + 616448z^2-270080 z+70656) \\
 & \times\, (3 z-4)^5 \log (4-3 z) + 75 (11 z-12)(z^2-16 z+16) (1245 z^5 -84308 z^4 \\
 & +\, 601696z^3-1333120z^2 + 914688 z-119808) (3 z-4)^5 \log\left[\left(4-\frac{11z}{3}\right)(4-z)\right]\\
 & +\, 16 (z-1) (1829959z^{12}-44960912 z^{11} + 285792656 z^{10}-1090093952z^9 \\
 & +\, 5123084544 z^8-24390724608 z^7 + 77450817536 z^6-153897779200z^5 \\
 & +\, 194102034432 z^4-155643543552 z^3 + 77091307520 z^2-21705523200z \left. + 2717908992)\right]
 \;,
\end{split}
\end{equation}
with $d^{\TQQ}_Q(z) = (4-3 z)^6(z-4)^2 z(11 z-12)(z^2-16 z+16)$.

\vspace{1em}
\noindent
\textbf{Axial-vector channel ($1^{+-}$).}

As discussed in Sec.~\ref{ssec:FFs_initial_scale}, Fermi-Dirac statistics, together with the symmetry constraints imposed by the $S$-wave configuration, allow only the $[3,3]$ color-spin channel to contribute to the axial-vector state.
Furthermore, the $[g \to \TQQ(1^{+-})]$ fragmentation channel is suppressed at LO due to the Landau-Yang selection rule.

\begin{equation}
\begin{split}
\label{sDQ_FF_SDC_1pm_33}
 \tilde{\cal D}^{(1^{+-})}_Q&(z,[3,3])
 \,=\, \frac{\pi^{2}\left[\alpha_{s}^4(5m_Q)\right]}{279936 \, d^{\TQQ}_Q(z)}
 \left[ 480 (z-4)
 (11 z-12)\left(z^2-16 z+16\right) \right.
 \\
 \,&\times\, 
 \left(4 z^4+115 z^3-316 z^2+112z+64\right) (3 z-4)^5 \log \left(z^2-16 z+16\right)
 \\
 \,&+\, 6 (11 z-12)\left(z^2-16 z+16\right) (4825 z^5-56232 z^4+378480z^3
 \\
 \,&-\, 942528 z^2+672768 z-60416) (3 z-4)^5 \log (4-3 z)-3 (11z-12) \left(z^2-16 z+16\right) (5465 z^5
 \\
 \,&-\, 40392 z^4+254320z^3-722368 z^2+611328 z-101376) (3 z-4)^5 \log\left[\left(\frac{11z}{3}-4\right)(z-4)\right]
 \\
 \,&+\, 16 (z-1) z(476423 z^{11}+32559240 z^{10}-934590720 z^9+8015251776z^8
 \\
 \,&-\, 35393754624 z^7+94265413632 z^6-160779010048 z^5+177897046016z^4
 \\
 \,&-\,  \left. 124600254464 z^3 + 51223461888 z^2-10217324544z+490733568) \right] \;,
\end{split}
\end{equation}

\vspace{1em}
\noindent
\textbf{Tensor channel ($2^{++}$).}

As noted in Sec.~\ref{ssec:FFs_initial_scale}, the combined effect of Fermi-Dirac statistics and the combined effect and the $S$-wave structure restricts the surviving SDCs for tensor states to the $[3,3]$ structures only.
In the $[g \to \TQQ(2^{++})]$ channel, one finds~\cite{Feng:2020riv}
\begin{equation}
\begin{split}
 \label{Dg_FF_SDC_2pp_33}
\hspace{-0.00cm}
 \tilde{\cal D}^{(2^{++})}_g&(z,[3,3]) \,=\, 
 \frac{\pi^{2} \alpha_{s}^{4}(4m_c)}{622080 \, z \, d^{\TQQ}_g(z)}\left[\left(46656-490536 z+1162552 z^2-1156308 z^3\right.\right. \\
 & \left.+\, 595421 z^4-170578 z^5+21212 z^6\right) 2z+3(1-z)(2-z)(3-z)(-20304-31788 z) \\
 & \left.\left.\times\, (1296+1044 z + 73036 z^2-36574 z^3+7975 z^4\right)\right. \\
 & \left.\times\, \ln (1-z)+33(2-z)(3-z)(1296+25)\right] \\
 & \left.\left.\, -9224 z^2+9598 z^3-3943 z^4+725 z^5\right) \ln \left(1-\frac{z}{3}\right)\right]
  \;.
\end{split}
\end{equation}
Analogously, the $[Q \to \TQQ(2^{++})]$ SDC reads~\cite{Bai:2024ezn}
\begin{equation}
\begin{split}
 \label{DQ_FF_SDC_2pp_33}
 \hspace{-0.00cm}
 \tilde{\cal D}&^{(2^{++})}_Q(z,[3,3]) \,=\, 
 \frac{\pi^2 \as^4(5m_Q)}{2799360 \, z \, d^{\TQQ}_Q(z)} \left[ 672 (z-4) (11 z-12) (z^2-16 z+16) \right. \\
 & \times\, (47z^5+12186 z^4-44608 z^3 + 40000 z^2 -7936 z+4608) \\
 & \times\, (3 z-4)^5 \log (z^2-16 z+16) + 6 (11 z-12)(z^2-16 z+16) \\
 & \times\, (107645 z^6-1088988 z^5+7805536 z^4 - 20734976 z^3 +8933504z^2 - 6013952 z+1695744) \\
 & \times\, (3 z-4)^5 \log (4-3 z) - 33 (11 z-12)(z^2-16 z+16) (3581 z^5 - 53216 z^4-326176 z^3+419456z^2 \\
 & -\, 6912 z+55296) (3 z-4)^6 \log\left[\left(4-\frac{11z}{3}\right)(4-z)\right]\\
 & +\, 16 (z-1) (96449507 z^{12} - 158520388z^{11} - 26228206896 z^{10} + 281743037888 z^9 \\
 & -\, 1355257362432 z^8 + 1355257362432 z^7 - 6637452959744 z^6 + 7595797282816 z^5 \\
 & -\, 5643951472640 z^4 + 2662988513280 z^3 - 788934950912 z^2 + 161828831232 z \left. - 24461180928)\right]
 \;.
\end{split}
\end{equation}

\hypertarget{app:B}{
\section{Tetraquark-jet systems in NLL/NLO$^+$ HyF}
}
\label{app:B}

\begin{figure*}[!t]
\includegraphics[width=0.575\textwidth]{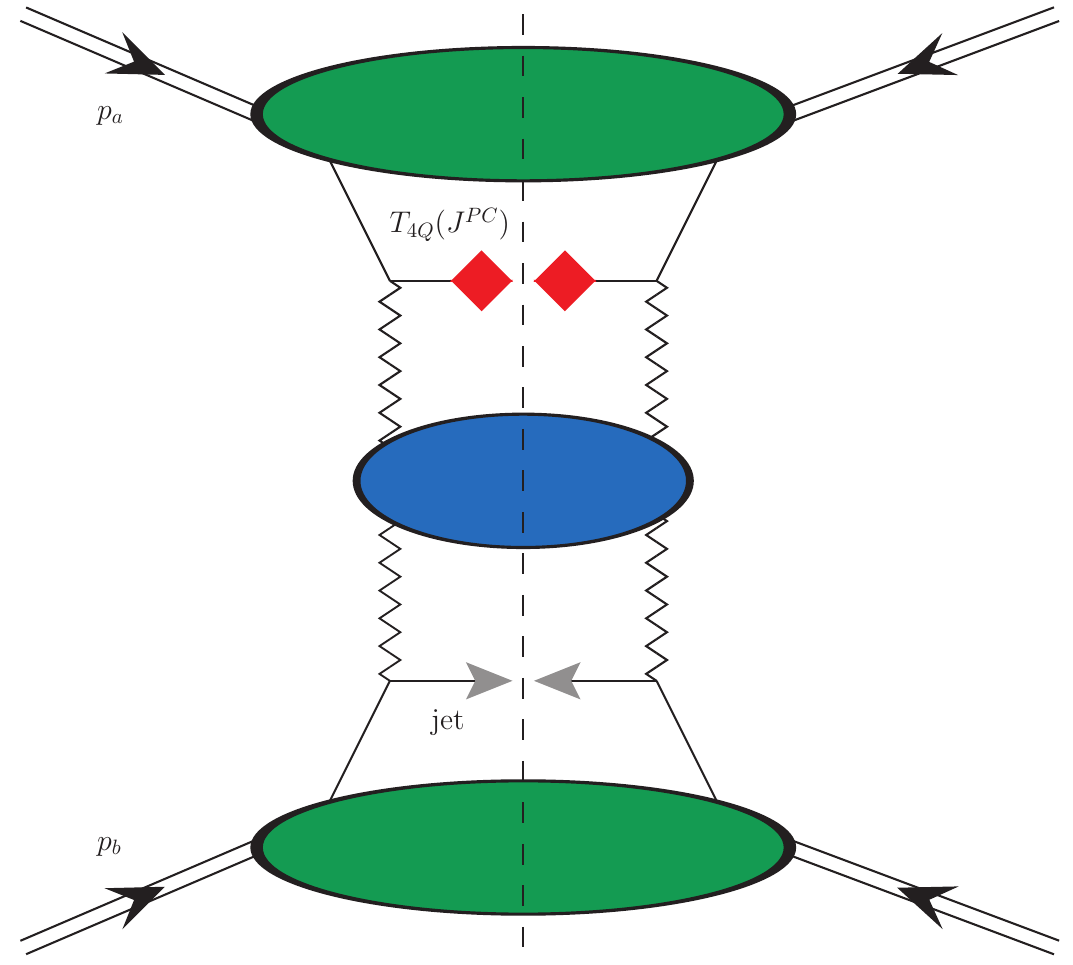}

\caption{
\justifying
\noindent
Pictorial representation of the semi-inclusive hadroproduction of a tetraquark-jet system within the hybrid collinear and high-energy factorization (HyF) framework.
Red rhombi indicate the collinear FFs of the fully heavy $\TQQ(J^{PC})$ tetraquark.
Black arrows represent the final-state jet, while green ovals correspond to proton collinear PDFs.
Blue blobs denote the exponentiated resummation kernel, which is connected to the two off shell gluon emission functions via zigzag Reggeon lines.
}
\label{fig:HyF_TQQ-plus-jet}
\end{figure*}

We consider the following reaction: (Fig.~\ref{fig:HyF_TQQ-plus-jet})
\begin{eqnarray}
\label{process}
 {\rm p}(p_a) + {\rm p}(p_b) \to \TQQ(\kappa_1, y_1) + {\cal X} + {\rm jet}(\kappa_2, y_2) \;,
\end{eqnarray}
where a fully heavy tetraquark, $\TQc$ or $\TQb$, is semi-inclusively identified together with a light jet, and an undetected gluon radiation cascade, ${\cal X}$.
The outgoing particles carry transverse momenta that fulfill the condition $|\vec \kappa_{1,2}| \gg \Lambda_{\rm QCD}$, where $\Lambda_{\rm QCD}$ denotes the QCD hadronization scale.
These objects are also separated by a large rapidity interval, $\DY = y_1 - y_2$.
We apply a Sudakov decomposition of the four-momenta $\kappa_{1,2}$ using the momenta of the incoming protons, $p_{a,b}$, which yields
\begin{eqnarray}
\label{sudakov}
\kappa_{1,2} = x_{1,2} p_{a,b} + \frac{\vec \kappa_{1,2}^{\,2}}{x_{1,2} s}p_{b,a} + \kappa_{1,2\perp} \, \qquad
\kappa_{1,2\perp}^2 = - \vec \kappa_{1,2}^{\,2} \;.
\end{eqnarray}
In the center-of-mass frame, the following relations hold between rapidities $y_{1,2}$ and longitudinal momentum fractions of the final-state particles
\begin{eqnarray}
\label{y-vs-x}
y_{1,2} = \pm \ln  \frac{x_{1,2} \sqrt{s}}{|\vec \kappa_{1,2}|} \;.
\end{eqnarray}

In a purely collinear factorization framework, the LO differential cross section for our processes would be expressed as a one-dimensional convolution involving the on-shell hard scattering kernel, the proton PDFs, and the $\TQQ$ FF
\begin{eqnarray}
\label{sigma_collinear}
\frac{\drv\sigma^{\rm LO}_{\rm [collinear]}}{\drv x_1\drv x_2\drv ^2\vec \kappa_1\drv ^2\vec \kappa_2}
= \hspace{-0.25cm} \sum_{i,j=q,{\bar q},g}\int_0^1 \hspace{-0.20cm} \drv x_a \!\! \int_0^1 \hspace{-0.20cm} \drv x_b\ f_i\left(x_a\right) f_j\left(x_b\right) 
\int_{x_1}^1 \hspace{-0.15cm} \frac{\drv \zeta}{\zeta} \, D^{\TQQ}_i\left(\frac{x_1}{\zeta}\right) 
\frac{\drv {\hat\sigma}_{i,j}\left(\hat s\right)}
{\drv x_1\drv x_2\drv ^2\vec \kappa_1\drv ^2\vec \kappa_2} \;.
\end{eqnarray}
Here, the indices $i,j$ run over all partons except for the top quark, which does not hadronize.
For compactness, the explicit dependence on the factorization scale $\mu_F$ is omitted in Eq.~\eqref{sigma_collinear}.
The functions $f_{i,j}(x_{a,b}, \mu_F)$ denote the proton PDFs, while $D^{\TQQ}_i(x_1/\zeta, \mu_F)$ represent the tetraquark FFs.
Here, $x{a,b}$ are the longitudinal momentum fractions of the incoming partons, and $\zeta$ is the momentum fraction carried by the outgoing parton that fragments into the $\TQQ$ hadron.
Finally, $\drv\hat\sigma_{i,j}(\hat s)$ refers to the partonic cross section, with $\hat s = x_a x_b s$ the partonic center-of-mass energy squared.

The differential cross section can be recast as a Fourier sum of angular coefficients, $C_{n \ge 0}$,
We write
\begin{eqnarray}
 \label{dsigmaFourier}
 \frac{\drv \sigma}{\drv \DY \, \drv \varphi \, \drv |\vec \kappa_1| \, \drv |\vec \kappa_2|} 
 =
 \frac{1}{\pi} \left[ \frac{1}{2} C_0 + \sum_{n=1}^\infty \cos (n \varphi)\,
 C_n \right]\;,
\end{eqnarray}
with $\varphi = \phi_1 - \phi_2 - \pi$ and $\phi_{1,2}$ standing for the azimuthal angles of the two outgoing particles.
Within the HyF framework, employing the $\MSb$ renormalization scheme~\cite{PhysRevD.18.3998}, we derive a master formula for the $C_n$ coefficients.
This formulation is valid at NLO accuracy and includes the NLL resummation of high-energy logarithms.
Explicitly, one obtains,
\begin{equation}
\begin{split}
\label{CnNLL}
 C_n^{\NLLp} \;&=\; 
 \int_{\kappa_1^{\rm min}}^{\kappa_1^{\rm max}} \drv |\vec \kappa_1|
 \int_{\kappa_2^{\rm min}}^{\kappa_2^{\rm max}} \drv |\vec \kappa_2|
 \int_{y_1^{\rm min}}^{y_1^{\rm max}}
 \drv y_1
 \int_{y_2^{\rm min}}^{y_2^{\rm max}} 
 \drv y_2
 \; \delta(y_1 - y_2 -\DY)
 \int_{-\infty}^{+\infty} \drv \nu \, e^{\bar \alpha_s \DY \chi^{\rm NLL}(n,\nu)}
\\[0.20cm]
 &\times \,
 \frac{e^{\DY}}{s}
 \alpha_s^2(\mu_R)
 \left\{ 
 \F_1^{\rm NLO}(n,\nu,|\vec \kappa_1|, x_1)[\F_2^{\rm NLO}(n,\nu,|\vec \kappa_2|,x_2)]^*
 + \bar \alpha_s^2 \frac{\beta_0 \DY}{4 N_c}\chi(n,\nu)\upsilon(\nu)
 \right\} \;,
\end{split}
\end{equation}
Here, $\bar \alpha_s(\mu_R) = \alpha_s(\mu_R) N_c/\pi$, where $N_c$ is the number of colors, and $\beta_0 = 11N_c/3 - 2 n_f/3$ is the leading coefficient of the QCD $\beta$-function.
We employ a two-loop running coupling with $\alpha_s\left(M_Z\right)=0.118$ and a dynamic flavor number, $n_f$.
The $\chi(n,\nu)$ function appearing in the exponent of Eq.~\eqref{CnNLL} denotes the BFKL kernel~\cite{Fadin:1975cb,Kuraev:1977fs,Balitsky:1978ic}, responsible for resumming NLL energy logarithms,
\begin{eqnarray}
 \label{chi}
 \chi^{\rm NLL}(n,\nu) = \chi(n,\nu) + \bar\alpha_s \hat \chi(n,\nu) \;,
\end{eqnarray}
where
\begin{eqnarray}
\chi\left(n,\nu\right) = -2\left\{\gamma_{\rm E}+{\rm Re} \left[\psi\left( (n + 1)/2 + i \nu \right)\right] \right\}
\label{chiLO}
\end{eqnarray}
are the LO BFKL eigenvalues.
Moreover, $\psi(z) = \Gamma^\prime(z)/\Gamma(z)$ represents the logarithmic derivative of the Gamma function, while $\gamma_{\rm E}$ denotes the Euler-Mascheroni constant.
The $\hat\chi(n,\nu)$ function stands for the NLO kernel correction
\begin{align}
\label{chiNLO}
\hat \chi\left(n,\nu\right) \;&=\; \bar\chi(n,\nu)+\frac{\beta_0}{8 N_c}\chi(n,\nu)
\left\{-\chi(n,\nu)+2\ln\left(\mu_R^2/\hat{\mu}^2\right)+\frac{10}{3}\right\} \;,
\end{align}
with $\hat{\mu} = \sqrt{|\vec \kappa_1| |\vec \kappa_2|}$.
The complete formula form the characteristic $\bar\chi(n,\nu)$ function is given, \emph{e.g.} in Sec.~2.1.1 of Ref.~\cite{Celiberto:2020wpk}.
The two quantities
\begin{eqnarray}
\label{EFs}
\F_{1,2}^{\rm NLO}(n,\nu,|\vec \kappa|,x) =
\F_{1,2}(n,\nu,|\vec \kappa|,x) +
\alpha_s(\mu_R) \, \hat \F_{1,2}(n,\nu,|\vec \kappa|,x)
\end{eqnarray}
portray the NLO singly off shell, transverse-momentum-dependent emission functions, commonly referred to in the BFKL context as forward-production impact factors.
Tetraquark emissions are captured by the NLO forward-hadron impact factor~\cite{Ivanov:2012iv}.
Although originally derived for light hadrons, this object remains valid within our VFNS-based framework~\cite{Mele:1990cw,Cacciari:1993mq}, provided that the relevant transverse-momentum regimes lie well above the DGLAP evolution thresholds associated with heavy-quark production.
At LO, we have
\begin{equation}
\begin{split}
\label{LOHEF}
\F_{\TQQ}(n,\nu,|\vec \kappa|,x) \;&=\; 2 \sqrt{\frac{C_F}{C_A}} \; |\vec \kappa|^{2i\nu-1}\int_{x}^1 \frac{\drv \zeta}{\zeta} \left( \frac{x}{\zeta} \right)^{1-2i\nu} 
\\
 &\times \, 
 \left[ \frac{C_A}{C_F} f_g(\zeta, \mu_F)D_g^{\TQQ}\left( \frac{x}{\zeta}, \mu_F \right)
 +\sum_{i=q,\bar q}f_i(\zeta, \mu_F)D_i^{\TQQ}\left( \frac{x}{\zeta}, \mu_F \right) \right] \;,
\end{split}
\end{equation}
with $C_F = (N_c^2-1)/(2N_c)$ and $C_A \equiv N_c$ being the Casimir invariants associated with gluon emissions from a quark and a gluon, respectively.
Here, $f_i\left(x, \mu_F \right)$ denotes the PDF of parton $i$ inside the parent proton, while $D_i^{\TQQ}\left(x/\zeta, \mu_F \right)$ is the FF describing the fragmentation of parton $i$ into the detected tetraquark, $\TQQ$.
The full NLO correction is detailed in~\cite{Ivanov:2012iv}.
At LO, the jet emission function is given by
\begin{eqnarray}
 \label{LOJEF}
 \hspace{-0.09cm}
 \F_J(n,\nu,|\vec \kappa|,x) = 2 \sqrt{\frac{C_F}{C_A}} \;
 |\vec \kappa|^{2i\nu-1}\,\hspace{-0.05cm} \left[ \frac{C_A}{C_F} f_g(x, \mu_F)
 +\hspace{-0.15cm}\sum_{j=q,\bar q}\hspace{-0.10cm}f_j(x, \mu_F) \right] \;.
\end{eqnarray}
while its NLO correction is derived from~\cite{Colferai:2015zfa}.
It bases on small-cone selection functions with the jet cone radius fixed to $R_J = 0.5$, in accordance with recent analyses at CMS~\cite{Khachatryan:2016udy,Khachatryan:2020mpd,CMS:2021maw}.
The remaining element in Eq.~\eqref{CnNLL} is the function $\upsilon(\nu)$, which reads
\begin{eqnarray}
 \upsilon(\nu) = \frac{1}{2} \left[ 4 \ln \hat{\mu} + i \frac{\drv}{\drv \nu} \ln\frac{\F_1(n,\nu,|\vec \kappa_1|, x_1)}{\F_2[(n,\nu,|\vec \kappa_1|, x_1)]^*} \right] \;.
\label{fnu}
\end{eqnarray}
Equations~\eqref{CnNLL} to~\eqref{LOJEF} provide a detailed characterization of our hybrid-factorization framework.
In line with the BFKL approach, the cross section exhibits high-energy factorization, represented as a convolution of the BFKL Green’s function with two singly off-shell emission functions.
These emission functions incorporate collinear inputs, namely the convolutions of the initial-state proton PDFs with the final-state hadron FFs.
The $\NLLp$ label denotes a fully resummed next-to-leading logarithmic treatment of energy logarithms within a next-to-leading order perturbative expansion.
The `$+$' superscript indicates the inclusion of subleading terms beyond NLL accuracy, arising from the product of NLO corrections to the emission functions in our computation of azimuthal coefficients.

For the sake of comparison, we also consider the pure LL limit within the $\MSb$ scheme, which is obtained by neglecting NLO corrections in both the resummation kernel (Eq.~\eqref{chi}) and the impact factors (Eq.~\eqref{EFs}).
This yields
\begin{eqnarray}
\label{CnLL}
 C_n^{\LL} \propto 
 \frac{e^{\DY}}{s} 
 \int_{-\infty}^{+\infty} \drv \nu \, 
 e^{\bar \alpha_s \DY \chi(n,\nu)} 
 \alpha_s^2(\mu_R) \, \F_{\TQQ}(n,\nu,|\vec \kappa_1|, x_1)[\F_J(n,\nu,|\vec \kappa_2|,x_2)]^* \;.
\end{eqnarray}
We note that integration over transverse momenta and rapidities of the final-state objects, explicitly displayed in the first line of Eq.~\eqref{CnNLL}, is omitted in Eq.~\eqref{CnLL} but remains implicitly understood.

A detailed comparison between high-energy resummation and fixed-order methods hinges on contrasting NLL-resummed predictions with purely fixed-order calculations.
However, to the best of our knowledge, no numerical framework currently allows for full NLO evaluations of observables sensitive to two-particle hadroproduction.
To provide a fixed-order baseline, we truncate the perturbative expansion of the $C_n$ coefficients in Eq.~(\ref{CnNLL}) at ${\cal O}(\alpha_s^3)$.
This procedure yields an effective high-energy fixed-order ($\HENLOp$) representation, suitable for phenomenological applications.
It retains the dominant high-energy logarithmic contributions found in a full NLO result, while systematically neglecting subleading terms suppressed by inverse powers of the partonic center-of-mass energy.
The resulting $\MSb$-scheme expression for the $\HENLOp$ angular coefficients reads
\begin{align}
\label{CnHENLO}
 C_n^{\HENLOp} &\propto 
 \frac{e^{\DY}}{s} 
 \int_{-\infty}^{+\infty} \drv \nu \, 
 \alpha_s^2(\mu_R) \,
 \left[ 1 + \bar \alpha_s(\mu_R) \DY \chi(n,\nu) \right] \,
 \F_{\TQQ}^{\rm NLO}(n,\nu,|\vec \kappa_1|, x_1)[\F_J^{\rm NLO}(n,\nu,|\vec \kappa_2|,x_2)]^* \;.
\end{align}
Here, the exponentiated kernel is expanded up to ${\cal O}(\alpha_s)$.
As before, for the sake of brevity, the integration over the transverse momenta and rapidities of the final-state objects is left implicit.

\twocolumngrid


\bibliographystyle{apsrev4-1} 
\bibliography{bibliography}

\end{document}